# Educational Intervention Re-Wires Social Interactions in Isolated Village Networks

Marios Papamichalis,[1,2]† Laura Forastiere,[1,3,4]† Edoardo M. Airoldi,[5,6] Nicholas A. Christakis[1,2,4,7,*]

[1] *Yale Institute for Network Science*
[2] *Human Nature Lab, Yale University*
[3] *Department of Biostatistics, Yale School of Public Health*
[4] *Department of Statistics and Data Science, Yale University*
[5] *Department of Statistics, Operations, and Data Science, Fox School of Business, Temple University*
[6] *Data Science Institute, Temple University*
[7] *Department of Sociology, Yale University*

†These authors shared equally in this work.
*Corresponding author: nicholas.christakis@yale.edu

**Social networks shape behavior, disseminate information, and undergird collective action within communities. Consequently, they can be very valuable in the design of effective interventions to improve community well-being[1–7]. But any exogenous intervention in networked groups, including ones that just involve the provision of information, can also possibly modify the underlying network structure itself, and some interventions are indeed designed to do so[8,9]. While social networks obey certain fundamental principles (captured by network-level statistics, such as the degree distribution or transitivity level)[10,11] they can nevertheless undergo change across time, as people form and break ties with each other within an overall population. Here, using a randomized controlled trial in 110 remote Honduran villages[1] involving 8,331 people, we evaluated the effects of a 22-month public health intervention on pre-existing social network structures. We leverage a two-stage randomized design, where a varying fraction of households received the intervention in each village. In low-dosage villages (5%, 10%, 20% and 30%) compared to untreated villages (0%), over a two-year period, individuals who received the intervention tended to sever both inbound and outbound ties with untreated individuals whom they previously trusted for health advice. Conversely, in high-dosage villages (50%, 75% and 100%), treated individuals increased both their inbound and outbound ties. Furthermore, although the intervention was health-focused, it also reshaped broader friendship and financial ties. In aggregate, the imposition of a novel health information regime in rural villages (as a kind of social institution) led to a significant rewiring in individuals' particular connections, but it still had a limited effect on the overall global structure of the village-wide social networks.**

In many small-scale traditional or rural societies, social connections provide important benefits otherwise typically available from formal institutions in large-scale or more developed societies[12,13]. In rural environments, people rely on their friends and family for loans rather than on banks, for security rather than on police or courts, and for health care advice and help rather than on doctors or clinics. For instance, women often turn to their social networks (e.g., to older female friends residing in the village) for information about, and assistance with, the delivery and care of newborns, rather than seek care from providers in the formal health care sector (e.g., clinics).

This raises an important issue: if formal health care, or health information, were to be made available, and be used, one motivation that people had for maintaining social connections to people who previously filled such a need might fall away. That is, the introduction of formal health care institutions (e.g., in the form of monthly educational sessions offered by visiting outsiders) might modify or replace traditional close-knit social connections, leading to potential declines in face-to-face interactions or to a rewiring of social ties. People who previously relied on friends and kin for health information might cut ties to some of their contacts or form new ties to people newly in possession of the novel information provided by the visiting health care workers. When formal institutions are present, in other words, an individual might turn for information and assistance to *different* community members invested with a possibly new role by the formal institution, even if the individual's number of connections stayed the same. Indeed, many public health and development initiatives that are ostensibly of a purely informational nature may be inadvertently re-wiring social connection in locales around the world, for better or worse. Here, we sought to quantify these changes in several ways.

**Honduras Village Cohort**

We explored the possible transformative impact of exogenous education interventions on social interactions in the setting of a randomized trial, examining the connections that individuals drop, form, or change, as well as the overall structure of village-wide networks that might arise as a result. We studied 110 remote villages in an isolated region of Honduras to which a novel intervention was delivered over 22 months. Our design improves on the design of a few prior studies that might support such an exploration of the broad effects of policy interventions on the network structure itself in a number of ways [2,5,6,14–18]. Specifically, we carried out a 2-stage randomized experiment in the 110 villages, and we modulated the fraction of households that received the intervention in each village (0%, 5%, 10%, 20%, 30%, 50%, 75%, and 100%), which, among other things, allows us to probe deeply the particular types of rewiring taking place across multiple types of social ties within the villages.

We conducted our experiment in Honduras's Copan Department, covering over 200 square miles of challenging mountainous terrain. We started with a census in 176 villages (**Supplementary Fig. 2**). Recruitment was strong: government data found approximately 32,500 eligible individuals, and N=30,815 (94.8%) participated in our census. From them, 24,702 (80.2%) individuals (in 10,013 households in 176 villages) enrolled in our longitudinal study. They committed to randomization, a 22- month intervention (with monthly visits from a health worker, if selected), as well as annual or biannual surveys and other data collection.

Unlike in the parent RCT (where the changes in the diffusion of outcomes for both "random targeting" and "friendship targeting" are examined),[1] in the present study on network structural dynamics, we use data only from the 110 villages that were assigned to the "random targeting" arms (66 villages which were partially treated, at dosages of 5%, 10%, 20%, 30%, 50% and 75%), 22 of which were fully treated (at a dosage of 100%), and 22 of which were assigned to a control arm (0% dosage). From the individuals in these 110 villages, we also drop any people who were not present for both waves of the study and those who moved between households; as a result, we analyzed 8,331 individuals who resided in 4,513 households (we evaluated some of these constraints in robustness checks in the **Supplementary Information**). The village populations in our data ranged from 11 to 313 adults, and the average village size was 75.7 (SD 50).

At baseline, the average age of the 8,331 participants was 37 (SD=18; range: 12-91); 37% were below 35 years old; 63% were women; and 41.4% were married, 22.5% had a civil union, 0.2% were divorced, and 2.1% were widowed.

The 22-month health education intervention was delivered to the villages from roughly November 2016 to August 2018. We mapped the village-wide social networks before and after the intervention using custom software which is publicly available (see Methods)[19]. That is, the social networks are the main outcomes of interest, and we measure them prior to intervention in 2016 ("wave 1") and afterwards in 2018 ("wave 3"). For a timeline of data collection, see **Supplementary Fig. 2**. We discerned health-related connections by asking all village residents: "Who would you ask for advice about health-related matters?" and "Who comes to you for health advice?" To ascertain financial connections, we asked people from whom they might borrow a small sum of money or to whom they might lend it. To ascertain general friendship connections, we asked people to identify with whom they (1) "spent free time," (2) discussed "personal or private" matters, or (3) were "close friends."

On average, at baseline, when we include intra-household social connections (e.g., when a person says they get health advice from their spouse), individuals had 13 (SD 9 range: 0-82) total close social connections, of which 2.9 (SD 3.1 range: 0-66) are health-advice connections, 7.4 (SD 4.9, range: 0-40) are friendship connections, and 2.7 (SD 2.5, range: 0-26) are financial connections. **Fig. 1** illustrates such a health network for an example village in wave 1 and how it changed by wave 3. Comparable analyses to those below that exclude intra-household connections yield similar and consistent results (see **Supplementary Tables 13-20**).

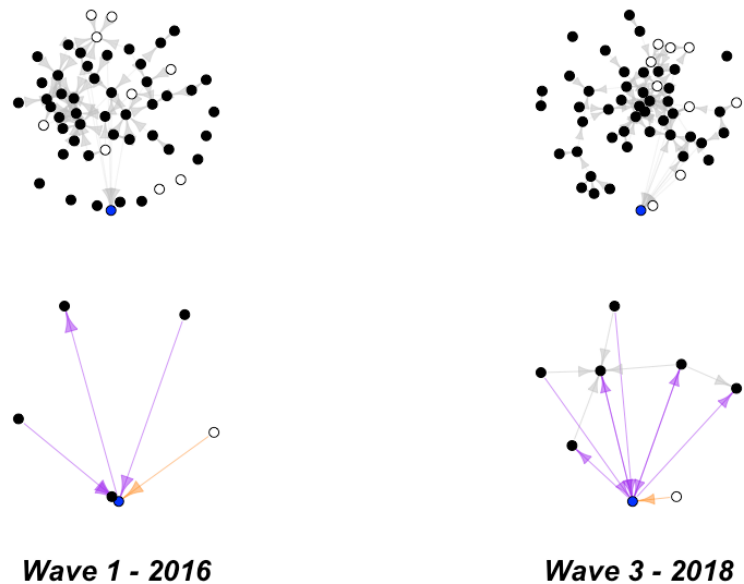

**Fig. 1: Illustrative health network relationships for a village with 54 individuals.** A total of 75% of the individuals were treated (black nodes) in wave 1 (left) and are also marked in wave 3 (right). In the bottom panels, we zoom in on the connections of a single treated individual who is highlighted in blue in both waves. Connections from the blue ego to its alters are color-coded: to/from treated individuals in purple, and to/from untreated individuals in yellow. Connections among the alters are displayed in gray. The effects of the intervention on the social structure of the blue node are apparent in this instance. Even though, visually speaking, the overall network structure does not seem to change too much (comparing top left to top right), the pattern of connections around the blue individual changed quite a bit (comparing bottom left to bottom right).

**Statistical Approach**

We estimate the average causal effects of the intervention assigned at the individual and village level on network properties as they changed between 2016 and 2018 – including overall, total, direct, and spillover effects – using the concepts and terminology established in prior work (see Methods) [20–25]. Such effects are defined under the framework of "partial interference,"[22] which permits the intervention's impact on an individual to propagate through their entire village's social network (but not to other villages – which is a reasonable assumption given the isolation of the villages and the paucity of inter-village ties here, which we also confirmed empirically). When total and spillover effects closely mirror the overall effects, we concentrate our discussion on the overall effects for conciseness (see Methods).

Permutation tests developed for randomized experiments with interference[23–25] are employed to test the average pooled effects (i.e., averaged across dosages) for different network for total degree, in-degree, and out-degree (in the **Supplementary Information**, we report effects on other network statistics, such as centrality and transitivity). As a test statistic, we use a difference-in-differences estimate of the change in the mean of each individual's network property, comparing treated and untreated networks between wave 1 and wave 3. We report percentage changes by dividing the estimates by the wave 1 mean of the relevant property in the control villages.

In addition, we conducted a link formation analysis to assess which types of links (treated to treated, untreated to treated, treated to untreated, untreated to untreated) already existing in Wave 1 are more likely to be maintained or broken in Wave 3, as well as which types of non-existing links are more likely to be formed in fully or partially treated villages, compared to fully untreated villages. For this analysis, we employ dyadic logistic regressions rather than permutation tests used for the individual-level analyses (see Methods).

Because the foregoing effects on social ties (including the interference effects) likely vary with the "dosage" of the intervention (in the sense of the fraction of village residents who were treated), further analysis by dosage level is needed to clarify the nature and magnitude of these changes. Hence, we stratify many of the analyses by low (5%, 10%, 20%, and 30%) and high (50%, 75% and 100%) dosages (see **Supplementary Fig. 4** for further justification of this sub-group analysis).

**Effects on health-advice networks**

We first estimate the average causal effects of the intervention on changes in specific village-wide network properties – such as degree distribution – across village-level networks focusing on health-advice ties. Although the intervention did not shift overall village-wide statistics, such as the shape of the degree distribution (see **Supplementary Fig. 3**), it did result in alterations to specific node-level metrics. To capture these changes, we examine how the intervention changed network properties such as mean total degree, in-degree, and out-degree (and also other metrics, such as clustering, closeness, and betweenness, reported in the **Supplementary Tables 4, 5 and 9**).

Initial comparisons of total degree, in-degree, and out-degree between partially and fully treated villages, on the one hand, and fully untreated villages, on the other hand, showed no significant statistical differences, with negligible overall effects (Panel A in **Fig. 2**, where the lines overlap). That is, the change in average total degree (Panel A in **Fig. 2**), in-degree, and out-degree in fully and partially treated villages, for treated and untreated individuals, compared with untreated individuals in fully untreated villages, differ not significantly by 0.06% (p-values 0.977, **Supplementary Table 12**). These results indicate a non-significant overall effect of the educational intervention on degree changes across dosages and *on average* for treated and untreated individuals, between wave 1 and wave 3.

However, treated individuals exhibit relatively *increased* connectivity, with an increase in degree, in-degree, and out-degree of 4.18%, 4.77%, and 3.59% (p-values = 0.143, 0.226, 0.295, respectively), as shown in **Fig. 2D** and **Supplementary Table 4**. In contrast, untreated individuals in treated villages show a relatively *decreased* connectivity, with a decrease in degree, in-degree, and out-degree of -3.62%, -4.15%, and -3.09% (p-values 0.228, 0.338 and 0.363) as shown in **Fig. 2G** and **Supplementary Table 4**. This divergence between treated and untreated nodes indicates that total effects (on treated individuals) and spillover effects (on untreated individuals) move in opposite directions at the individual level. The resulting direct effect, i.e., the difference in connectivity changes between treated and untreated individuals in partially or fully treated villages, as shown in **Fig. 2J** and **Supplementary Table 4**, is 7.78% (p-value < 0.001). To be clear, these computations are the summary findings, irrespective of dosage.

**Effects on health-advice networks at low dosages**

How does the dosage of the treatment at the village level matter? We focus on what happens in low-dosage and high-dosage villages separately. Our analysis reveals that, in low-dosage scenarios, where only a few individuals in a village receive treatment, the intervention appears to diminish connections, with overall, total, spillover, and direct effects showing similar decreasing patterns (second column of **Fig. 2** and **Supplementary Fig. 4**). Specifically, in low-dosage villages, we observe a significant overall reduction in network connectivity, characterized by decreases in total degree, in-degree, and out-degree by 6.36% for all three metrics, with a p-value of 0.042, compared to individuals in fully untreated villages. That is, in villages where only a small fraction of the households was treated, health care information can be viewed as a scare resource, and the effects of such information on changes in the social structure of these villages are bigger. (The reason the numbers are all the same is that, in a directed network, each edge adds one to both a node's in-degree and out-degree, making their averages equal and causing total degree (their sum) to change by the same percentage as each component.)

Treated individuals (in fully and partially treated villages) reduced their degree by 12.86% (p-value = 0.007); their in-degree by 13.35% (p-value = 0.042); and their out-degree by 12.38% (p-value = 0.0287), compared to the control villages. In contrast, untreated individuals in partially treated villages reduced their degree by 5.09%; their in-degree by 5.00%; and their out-degree by 5.18%; however, these effects are not statistically significant (p-values = 0.116, 0.266, and 0.123, respectively). The direct effects (difference between treated and untreated in partially treated villages) for the total degree, in-degree, and out-degree are -7.24 %, -8.18 and -7.07% (p-values = 0.119, 0.245 and 0.170). These results for the total degree are shown in **Fig. 2** (second column), depicting the change in this measure between wave 1 and wave 3 for the two comparison groups corresponding to each effect; results for in-degree and out-degree are reported in S**upplementary Table 5**.

Building on these results for spillover effects on individual-level node changes, we next examine how such changes in connectivity unfold among the immediate social circles of treated individuals – namely, untreated individuals with at least one treated one-hop-away network neighbor – compared with untreated individuals in control villages. This comparison is referred to as "first-order spillover effects," which represents the spillover effect on network metrics of being in a partially treated village and being connected to at least one treated individual, compared to being in a control village. This effect is also contrasted to that of untreated individuals who are in partially treated villages but who do not have any initial connections to treated individuals in wave 1.

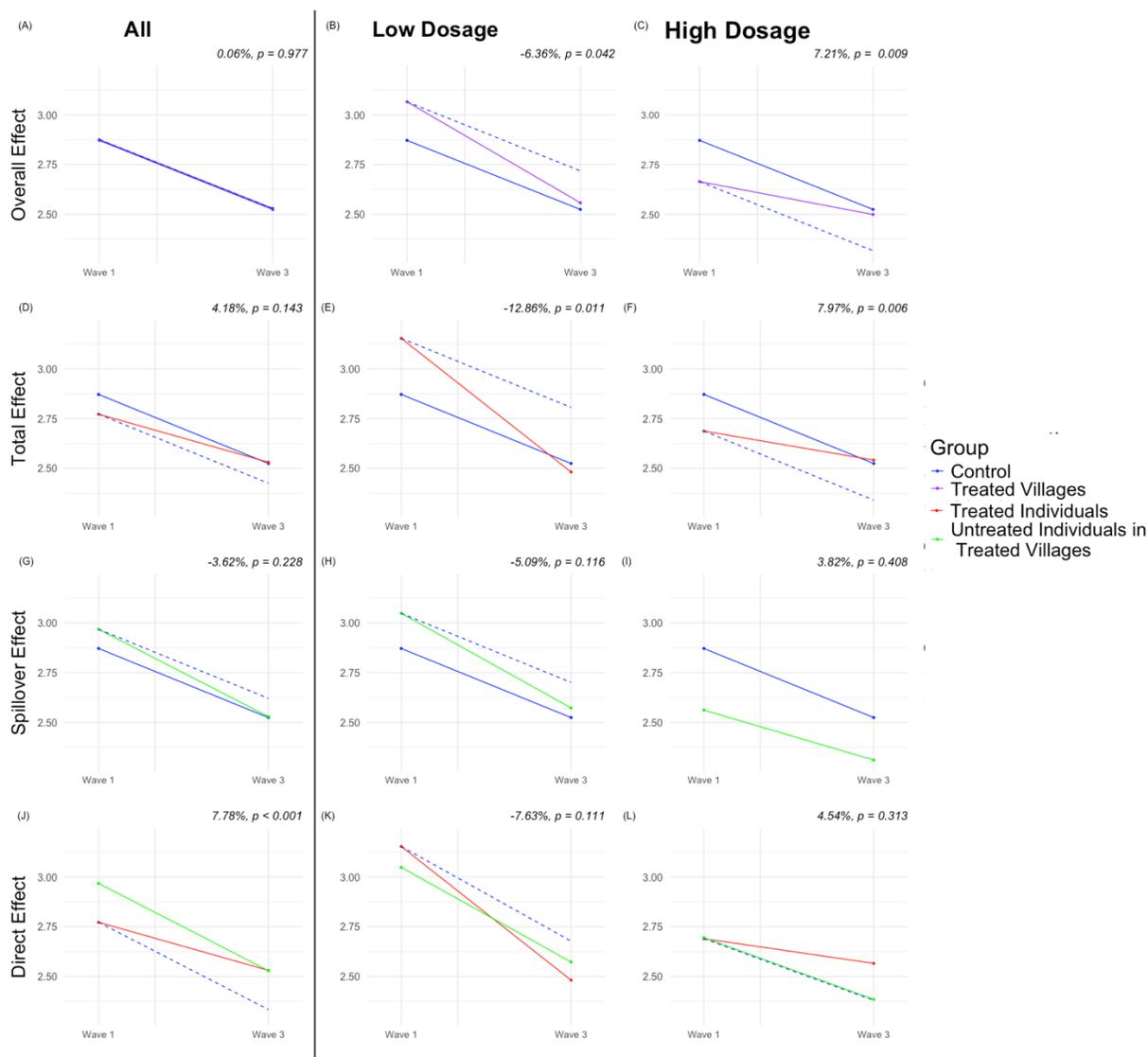

**Fig. 2: Overall, total, spillover, and direct effects on mean degree changes as assessed in all villages combined, and in low- and high-dosage villages separately.** The X-axis represents time (waves 1 and 3), and the Y-axis displays the mean degree for different groups: purple includes all individuals in fully or partially treated villages (Treated villages); red includes treated individuals in both partially and fully treated villages (Treated Individuals); green includes untreated individuals in partially treated villages (Untreated in Treated Villages); and blue includes untreated individuals in fully untreated villages (Control). Dashed blue lines represent expected mean degree for the group of interest without treatment under a parallel trend assumption, and serve as the counterfactual comparison. At the top of each panel, we report the estimated effect and the p-value.

Using permutation tests (see **Supplementary Table 61** for details), an examination of untreated one-hop-away neighbors of treated individuals reveals a pronounced decline in network degree compared with untreated individuals in control villages. Specifically, the first-order spillover effects, in low dosage, among 661 one-hop-away neighbors of treated individuals (both inbound and outbound) shows a significant reduction of the total degree, in-degree, and out-degree of 54.09%, 57.32%, and 50.85%, respectively (p-values < 0.001), suggesting that these untreated individuals are losing connections. That is, one-hop-away neighbors experienced a reduction of their inbound connections possibly because treated individuals became either self-reliant or more selective, choosing not to trust their untreated neighbors for health advice following the intervention. Moreover, a plausible explanation for the reduction in their outbound connections is that these untreated individuals not only lose some of their credibility regarding health issues, but also experience the potential diffusion of information within the villages[1] and also becoming more self-reliant. Overall, this further emphasizes that first-order neighbors to treated individuals experience a much greater reduction compared to the average spillover effects on all the untreated individuals.

We next examine the "higher-degree spillover" effect, comparing 2,280 untreated individuals only *indirectly* connected to treated individuals in wave 1 (i.e., through more than one degree of separation) to untreated individuals in control villages. For these individuals with a more-than-one-hop-away connection to the treated, we see an increase of 9.11% in degree (p-value = 0.002), 10.17% in in-degree (p-value = 0.006), and 8.05% in out-degree (p-value = 0.026), compared with control villages. This increase can be explained by the fact that individuals more than one hop away from any treated ego begin the study with ties exclusively to other untreated villagers – and, as we will show in the link analysis below, such ties between untreated individuals rarely dissolve. After the intervention, they become eligible to add entirely new ties to treated alters, so their degree can either remain unchanged or rise. Hence, the observed growth is an outcome of stable untreated-untreated links plus newly possible treated links, and not likely a substantive spillover effect of being indirectly connected to a treated individual (see **Supplementary Table 83** for further explanation).

Overall, the reduction in social connections observed in low-dosage villages compared to fully untreated villages is driven by the spillover effects on untreated individuals who are directly connected to treated neighbors in wave 1 (see **Supplementary Table 61** for detailed analyses). That is, the intervention appears to prompt treated people to have some churn in their social ties.

We next conduct a link formation analysis in low-dosage villages to examine which *types* of links already existing in Wave 1 are more likely to be maintained or broken in Wave 3, compared to fully untreated villages (**Supplementary Table 21**). Directed links from treated to untreated individuals (TU links henceforth) and directed links from untreated to treated individuals (UT links henceforth) break significantly more often compared to links between untreated individuals in untreated villages (UoUo links henceforth). The reduction of TU aligns with the observation that treated individuals either became self-reliant or more selective, choosing not to connect with individuals they trusted for health-related matters in wave 1 – after the intervention. Specifically, TU and UT links have 26.07% and 38.58% lower odds of persisting than the baseline in Wave 3 (p = 0.004 and p < 0.001, respectively), based on a 64.4% baseline persistence rate in control villages. In contrast, directed links between treated individuals (TT links henceforth) have 43.44% higher odds of persisting compared to UoUo links, making them less likely to disappear. Directed links from untreated to untreated individuals (UU links henceforth) have 7.29% higher odds of remaining intact compared to UoUo links; however, this difference is not statistically significant (p-value = 0.193). As mentioned above, this suggests that the observed stability of UU links helps preserve the baseline network structure of higher-order untreated individuals, who initially have only UU ties, and contributes to the observed increase in their average degree.

Also, we investigated the effect of low dosage on link formation by analyzing the probability that individuals unconnected in Wave 1 establish new ties by Wave 3 (**Supplementary Table 23**). Links between untreated individuals (UU) have 17.72% lower odds of forming compared to the relevant reference group (UoUo), making them less likely to be created, based on a 0.77% baseline formation rate in control villages (p < 0.001). Similarly, links that are formed between treated and untreated people in partially treated villages – that is, TU and UT links – have 36.86% and 34.53% lower odds of forming, respectively (both p < 0.001), compared to the reference group. The reduction of TU arises because treated individuals dissolve ties formed during Wave 1 with previously trusted untreated alters, adopting self-reliance or preferentially forming ties with other treated individuals. Correspondingly, TT links exhibit 43.92% higher odds (p<0.001) of formation compared to the baseline group, making them more likely to be created.

**Effects on health-advice networks at high dosages**

Next, we performed analogous analyses to the foregoing in the high-dosage villages (50%, 75% and 100%). In high-dosage situations, being in a partially or fully treated village results in an average relative increase of total degree, in-degree, and out-degree by 7.21%, for all three metrics, with p-value = 0.009, compared to an untreated village (**Supplementary Table 9** and third column in **Fig. 2**). For treated individuals, there is a relative increase in total degree, in-degree, and out-degree of 7.97%, 8.80% and 7.14 % (p-values = 0.006, 0.025 and 0.038), compared with untreated individuals in fully untreated villages (**Supplementary Table 9**). In contrast, for untreated individuals, the spillover effect shows no change in total degree, in-degree, and out-degree of 3.82%, 0.14%, and 7.49 % (p-values 0.408, 0.976 and 0.149), compared to untreated individuals in fully untreated villages. Overall, the treatment encourages treated individuals to expand their networks. These results for the total degree are shown in **Fig. 2** (third column), while results for in-degree and out-degree are reported in **Supplementary Table 9.**

Using permutation tests (see **Supplementary Table 65** for details), an examination of untreated one-hop-away neighbors of treated individuals in high-dosage villages reveals a continued, though less extreme, decline in network connectivity. Among the 314 untreated one-hop-away neighbors of treated individuals, we observe significant reductions of 19.05%, 20.06%, and 18.04% in total degree, in-degree, and out-degree, respectively (p-values = 0.002, 0.018, and 0.011). This suggests that, even in high-dosage contexts, untreated individuals adjacent to treated ones lose both incoming and outgoing ties – likely due to the same mechanisms seen in low-dosage villages: decreased credibility and increased self-reliance among the treated.

We next examine the "higher-degree spillover" effect in high-dosage villages, comparing 268 untreated individuals more than one hop away from any treated individual to their counterparts in control villages. For these indirectly exposed individuals, we observe significant increases in degree (30.60%, p < 0.001), in-degree (23.80%, p = 0.004), and out-degree (37.40%, p < 0.001). As in low-dosage villages, this growth is likely due to the stability of untreated-untreated ties and the post-intervention availability of new ties to treated individuals, rather than a direct behavioral spillover effect.

Once again, we examined which pre-existing connections are more likely to break or not, in the high-dosage villages, focusing on the set of links existing in Wave 1 (**Supplementary Table 37**). As in the low-dosage villages, TU links and UT links exhibit significantly higher odds of breaking compared to UoUo links. Specifically, TU and UT links have 43.50% and 28.47% lower odds of persisting than the reference group in Wave 3 ($p < 0.001$ and $p = 0.0109$, respectively), based on a 64.4% baseline persistence rate in control villages. Also similar to low-dosage villages, TT links have 8.02% higher odds of persisting compared to the reference group, indicating a greater likelihood of remaining intact, though with a not statistically significant p-value = 0.173. The same applies to UU links, which have an 11.83% higher probability, compared with UoUo of not breaking; however, this difference is not statistically significant ($p = 0.308$).

Furthermore, we consider nodes that were not connected in Wave 1 and find similar patterns to low-dosage villages (**Supplementary Table 39**). TU links have 12.99% lower odds of forming compared to the reference group, making them less likely to appear, based on a 0.77% baseline formation rate in control villages ($p = 0.061$). UU, TU, and TT links have non-significant p-values compared to the baseline, indicating no strong evidence of a preference or increased likelihood of connection formation among these links

**Overall effects on health-advice networks**

In summary, the foregoing analyses of network properties reveals significant trends and critical thresholds in connection behavior. In villages overall (with either high or low fractions of households treated) the treated households tend to cut connections with their untreated neighbors whom they previously trusted for health advice, and subsequently tend to opt for self-reliance. In low-dosage villages, because so few individuals are treated, treated individuals have limited opportunities to connect with others with similar knowledge (acquired either from the intervention or from other treated individuals through diffusion) [1] – thus leading to a decline in their network ties. However, once enough people (around 30%) receive the educational treatment, the treated group may reach a critical mass, and it becomes much easier for them to form or reinforce connections with individuals they already know who were also treated. This empirically observed phase-transition-like phenomenon increases the overall connections among the treated group and can even lead to stronger, more cohesive social networks for those who have been treated. Moreover, first-degree untreated neighbors of the treated can sever connections with the treated, possibly because they are perceived as less reliable following the intervention or opt for self-reliance. However, with their enduring UU ties intact and the experimental treatment newly permitting links to treated alters, higher order untreated neighbors (i.e., those who are more than one hop away from treated individuals and not otherwise connected to any treated person) display degree growth that reflects selection (whereby untreated individuals connected to other untreated individuals keep their links), rather than spillover from being indirectly connected to any treated persons (whereby network rewiring flows through the network, see **Supplementary Table 83** for details).

Finally, robustness tests that exclude intra-household connections remain consistent with all the foregoing results (see **Supplementary Table 13** for low dosages and **Supplementary Table 17** for high dosages).

**Effects on aggregated, friendship, and financial networks**

The intrinsic multiplexity of village social networks (where, for instance, an ego can have all three kinds of ties to the same alter – friendship, financial, and health-advice), means that some of the friendship and financial ties might be rewired as a consequence of the health-advice tie rewiring. We first test whether aggregated ties reconfigure in the same way as health-advice ties, and then disentangle whether these shifts arise from multiplex overlap among layers or from a direct intervention effect on the friendship and financial layers (a distinction examined more rigorously via the "residual-network" analysis in **Supplementary Tables 79-82**). Thus, we estimated the average causal effects of the intervention on network properties within aggregated undirected village-level networks (comprising the health, friendship, and financial relationships combined), and we also analyzed these other sorts of social ties separately, as distinct "layers" (and also as quasi-independent layers, whereby we assess connections solely of each kind, as discussed in the **Supplementary Tables 6-8** for low-dosage and **Supplementary Tables 10-12** for high dosage). The analysis reveals consistent impacts across these other network layers, mirroring the patterns observed in the health-advice network (**Fig. 3,** panels A, D, G and J and **Supplementary Fig. 4**).

For villages with a low fraction of households treated (5%, 10%, 20% and 30%), the *aggregated* networks (that concatenate health-advice, friendship, and financial ties) undergo similar changes as the health-advice network. Specifically, individuals in partially and fully treated villages experience a 5.12% decrease in total degree (p-value 0.004), compared to untreated individuals in fully untreated villages. Among treated individuals in particular, total degree decreases by 7.06% (p-value = 0.011). For untreated individuals in untreated villages, total degree decreases by 4.10% (p-value < 0.009), relative to control villages. First-order untreated neighbors of treated individuals experience a 20.9% reduction (p-value < 0.001). In contrast, untreated neighbors located more than one hop away exhibit a 7.50% increase (p-value < 0.001), compared to untreated individuals in untreated villages.

For villages with a high fraction of households treated (50%, 75% and 100%), we find no significant overall, total, direct and spillover effects, meaning that individuals in partially and fully treated villages do not exhibit significant changes in total degree, compared to fully untreated villages. However, one-hop-away untreated neighbors of the treated exhibit a 6.38% decrease (p-value = 0.035) in total degree compared to individuals in control villages. In contrast, more-than-one-hop-away untreated neighbors of the treated show a 23.67% increase in degree (p-value < 0.001).

In the friendship networks layer, independently of the fraction of households treated in a village, we do not observe significant overall, total, spillover, and direct causal effects. However, for low dosages, as with the aggregate networks, first-order untreated neighbors of treated individuals reduce their total degree, in-degree, and out-degree by 17.83%, 18.77%, and 16.89%, respectively (p-values < 0.001). In contrast, untreated neighbors located more than one hop away increase their total degree, in-degree, and out-degree by 8.14%, 9.13%, and 7.15%, respectively (p-values < 0.001), compared to untreated individuals in untreated villages. For high dosages (50%, 75% and 100%), one-hop-away untreated neighbors of the treated exhibit a 9.09%, 7.94 and 10.24% decrease (p-values 0.035, 0.07, and 0.017, respectively) in total degree, in-degree and out-degree, compared to individuals in control villages. In contrast, more-than-one-hop-away untreated neighbors of the treated show a 22.48%, 24.21% and 20.75% increase in degree (p-value < 0.001, <0.001, and 0.003).

In the financial networks, in villages with a low dosage, untreated individuals who are first-order neighbors of the treated reduce their degree, in-degree, and out-degree by 39.34%, 40.03%, and 38.65%, respectively (p-values < 0.001). In contrast, untreated individuals more than one hop away from the treated increase their degree, in-degree, and out-degree by 6.79%, 6.74%, and 6.83%, respectively (p-values = 0.02, 0.068, and 0.036), compared to untreated individuals in untreated villages. In villages with a high dosage, one-hop-away untreated neighbors of the treated exhibit a 20.93%, 11.64% and 30.21% decrease (p-values <0.001, 0.119, and <0.001, respectively) in total degree, in-degree and out-degree compared to individuals in control villages. In contrast, more-than-one-hop-away untreated neighbors of the treated show a 23.92%, 16.54% and 31.31% increase in degree (p-value < 0.001, 0.022, and <0.001).

Across every network layer, we observe a marked decline in first-order neighbors: the intervention not only rewired friendship and financial ties – eroding the status, connections, and trust of adjacent individuals – but also marginalized the untreated neighbors who were linked to treated participants in wave 1, making them less relevant to the treated for health, friendship, and financial exchanges alike. Moreover, across all examined networks (including friendship and financial layers), the same patterns regarding ties are observed. Specifically, TT links are less likely to break and more likely to form in both low and high dosages (**Supplementary Tables 21-52**). In contrast, TU and UT links are more likely to break and less likely to form, especially at low dosages.

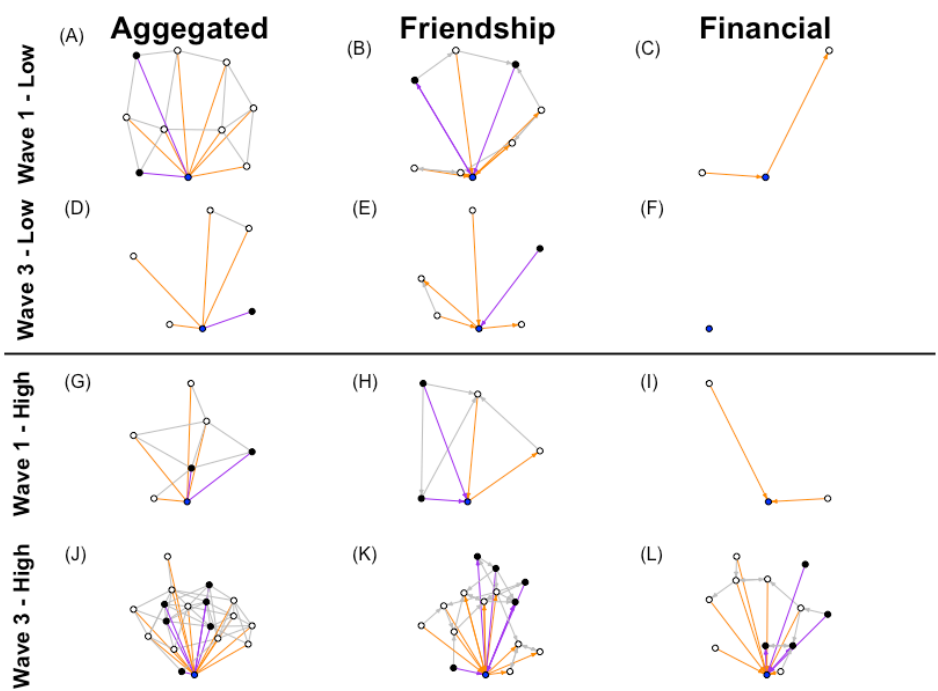

**Fig. 3: Representation of the aggregated (left), friendship (middle), and financial (right) relationships of selected individuals in two illustrative villages.** The first village (top rows) belongs to the low-dosage group, where 20% of the individuals are treated, and consists of 31 individuals (Panels A-F). The second village (bottom rows) belongs to the high-dosage group, with 75% of its 54 individuals treated (bottom panels G-L). The ties of a *single* treated individual in each village (in blue) are shown, in wave 1 (top row of each panel) and in wave 3 (bottom row). Treated individuals are represented as black nodes and untreated people with white nodes, and a single treated individual is indicated with a blue node. The aggregated networks are undirected (and concatenate friendship, health-advice, and financial ties), whereas friendship and financial networks are directed. Purple and yellow connection lines represent connections from treated and untreated individuals, respectively. Grey lines represent the rest of the links among other individuals. The treated individuals acquire more connections of all types.

Overall, although the intervention functions as an exogenous institution that delivers health information, it may also affect broader expectations of sociality, thereby reshaping interactions in other domains such as friendships and financial interactions, as illustrated in **Fig. 3** (see panels B, E, H, and K for friendship and panels C,F,I, and L for financial ties, and see also **Supplementary Fig. 5** and **Supplementary Fig. 6**). This underscores the broad-reaching impact of interventions on various relationships within rural village communities.

**Discussion**

Health education intervention in rural Honduran villages affects the dynamics of rural community social interactions and hence the structural details of village-wide social networks. Changes in the wiring patterns within networks, specifically regarding who is connected to whom, result in various structural properties of the overall health network being altered. In villages with a low fraction of the households treated, ties are rewired and there is some erosion of traditional social structures. But when a high fraction of the households treated, and villages are saturated with the exogenous institution, the intervention seems to bolster local networks, enhancing community-based knowledge sharing and decision-making. This nuanced effect is particularly evident in the differential impacts on treated individuals and their one-hop-away neighbors who were untreated, insofar as we observe that treated individuals sever connections with untreated individuals to whom they were initially linked and newly link to each other. Moreover, these re-wiring effects concern not only health and aggregated networks, but also extend to networks of friendship and financial interactions. Nevertheless, at both low and high dosages, while changes to the internal wiring within the villages are observed, there is no statistical evidence that the overall village-wide network properties change.

The structure of a social network clearly has relevance to its collective properties, as indicated by both mathematical models[26–32] and prior empirical work with online networks[8,9,33]. Still, situations where ensembles of many face-to-face networks have been collected (say in collections of villages or schools) remain rare,[1–7,34,35] and interventions to explicitly foster rewiring of naturally occurring face-to-face networks are rarer still. But whether advertently or inadvertently targeted at modifying network structure, it is clear that some interventions can still have such an effect, as shown here. However, the structure of human social networks is not coincidental, and it likely has evolutionary antecedents and implications[10,36–38]. Hence, social networks may be structurally consistent even when perturbed, to some extent. The equilibrium perspective suggests that despite local churning in social ties under new influences, overall network structures can remain remarkably stable.[39]

Many interventions seek to exploit social influence to enhance their impact. For instance, in face-to-face networks, phenomena as diverse as microfinance uptake in India[40], iron supplementation in India[2], vitamin use in Honduras[5], agricultural technology in Malawi[3], and bullying in American schools[6], have been shown experimentally to spread by social contagion. But such interventions, especially if substantial or sustained, are likely to also affect the underlying social ties in the populations at hand. Consequently, our study not only sheds light on the transformative impact of a health education intervention in rural Honduran villages, but also presents a broader blueprint for development interventions in other domains, such as security, finance, and healthcare.

A prior study of microcredit demonstrated that while microcredit increases borrowing, its impact on profits and developmental outcomes is not significant; but notably, this intervention altered the network dynamics by influencing social relationships and financial interactions within communities, as seen in the reduction of informal borrowing and gift exchanges[16]. Similarly, interventions promoting eco-friendly practices or providing agricultural subsidies not only target specific outcomes, such as the adoption of particular pesticides or improved production, but also reshaped the underlying social networks by influencing information diffusion and community interactions.[18] These effects on the network structure and its dynamics underscore the broader potential and impact of community-based interventions. This is especially likely to be the case with interventions that leverage social connections to accelerate behavior change in the first place, illustrating the critical role of understanding network structure for achieving sustainable public health and developmental outcomes. Our experimental design allows particular clarity regarding such effects, since we varied the dosage of the treatment and studied detailed interactions among treated and untreated individuals and since we studied many different types of social connections.

The intricate dynamics of social networks and the cascading effects of information provision that we observe offer the prospect of creating more effective interventions that might foster positive change and social resilience within communities. Underneath the enduring stability of social networks lies ongoing change in who is connected to whom.

## METHODS

**The Exogenous "Institution" Intervention**

The 22-month intervention involved a series of counseling sessions delivered by trained community health workers through 15 educational modules aimed at maternal, child, and neonatal practices typical of public health or development economics interventions. These workers paid monthly visits to the households and spent an hour or more in the home, providing health education, including about topics that were previously generally unknown (e.g., the use of zinc to treat diarrhea).

In the parent randomized controlled trial (which involved 176 villages), the villages were randomized to several dosages, i.e. fractions of households receiving the educational intervention, and, according to the assigned dosage in each village, there was random assignment of households to the intervention. For the present analyses, we are explicitly contrasting low and high dosages (involving random targeting in 88 villages); that is, non-zero dosages are categorized into low (5%, 10%, 20%, and 30%) and high (50%, 75% and 100%). While the 100% dosage scenario could be examined separately since there are no untreated individuals, we have grouped it with the 50% and 75% dosage arm for two reasons: (1) the behavioral patterns in these high-dosage arms are similar, and (2) grouping them increases the statistical power of our results. This stratification is grounded in preliminary exploratory analysis using non-linear models (i.e., Loess regression and smoothing); a detailed explanation is provided in **Supplementary Fig. 4**.

For clarity, we categorize individuals based on their treatment status and the treatment distribution within their villages. We merge treated individuals in fully treated villages (100% dosage) with treated individuals in partially treated villages (5%, 10%, 20%, 30%, 50% and 75% dosages) under the single term "treated individuals." We retain distinct categories for "untreated individuals in partially treated villages" (5%, 10%, 20%, 30%, 50% 75% dosages) and "untreated individuals in fully untreated villages" (0%).

**Network Ascertainment**

Using "name generator" questions, we recorded diverse social relationships, including health advice, friendship, and financial relationships within each village in waves 1 and 3 (the full list of name generators is available in **Supplementary Table 1**). For the main analyses, we first analyzed the health network (which consists of two questions: "Who would you ask for advice about health-related matters?" and "Who comes to you for health advice?"), and we allowed for directional ties.

In addition, in the main analyses, when referring to the "aggregated" social network, we aggregate the health-advice, friendship, and financial relationships into an undirected network that concatenated multiplex relationships for one level of analysis (i.e., treating any tie of any kind in either direction as a binary indicator of a tie between an ego and an alter). Note also that each of these three relationship types is in turn the result of the concatenation of several name generators (e.g., we treat any tie between two people – free time, important matters, and close friends – as indicating a "friendship" tie, possibly from the ego to the alter, from the alter to the ego, or both). Lastly, we analyzed multiplex layers (e.g., financial ties) allowing for directional ties in this layer-level analysis, unlike the concatenated analysis. To ensure the validity and robustness of the changes in the friendship and financial networks, we further examine the residual networks by removing the health ties from waves 1 and 3, as the treatment is health oriented. For instance, as part of the robustness checks, we exclude health-advice ties from Waves 1 and 3 from the friendship and financial networks. This ensures that observed changes are not driven by the health network ties per se.

**Permutation Tests, Test Statistics, and Causal Effects Under Partial Interference**

We ran a two-stage, cluster-randomized controlled trial: villages were first assigned to dosage arms and, within treated villages, households were randomly selected for the 22-month health-education program. [1] To test whether the intervention altered node-level structure (degree, in-degree, out-degree) we applied permutation tests designed for experiments with interference. The test statistic is a percentage-scaled difference-in-differences estimator: for each degree metric, we take the change from wave 1 to wave 3 in the treated group, subtract the corresponding change in the relevant control group, and then divide this difference by the wave 1 mean of that metric in control villages. Expressing all changes on a percentage basis using (Wave 3 Treated – Parallel Control Wave 3) / (Control Wave 3), as illustrated in **Fig. 2,** yields interpretable percentage effects that are assessed against the randomization distribution via the permutation test. This framework provides exact p-values[41] under the sharp null of no treatment effect while respecting the village-level randomization and the possibility of spillovers.

An ego's network in our experiment can change because the ego was treated, because others in their village (both near and far, geodesically) were treated, or both. Disentangling these effects under the two-stage design depends on proper comparisons. "Overall effects" were determined by comparing all individuals in the partially and fully treated villages (5%, 10%, 20%, 30%, 50%, 75% and 100% dosages) with untreated individuals in the 22 fully untreated villages (dosage 0%). This captures the average effect of assigning different treatment dosages to people in a village on an individual's network structure, regardless of whether they or anyone else in the village is treated. "Total effects" were determined by comparing *treated* individuals in treated villages (dosage greater than 0%) with untreated individuals in the fully untreated villages, representing the average effect on changes in the network properties of being treated and being in a treated village versus being untreated in a fully untreated village (dosage 0%). By comparing *untreated* individuals in treated villages with untreated individuals in untreated villages, "spillover effects" were established; these represent the average effect on network changes in being untreated and being exposed to a village dosage greater than 0%, versus being untreated in fully untreated villages (0%). Finally, the difference between the total effect and the spillover effect results in the "direct effect," which compares treated individuals to untreated individuals in partially treated villages.

**Link Analysis**

Although we can show a one-to-one correspondence between dyadic effects and causal effects on individual-level network properties (in particular, in-degree from treated individuals, in-degree from untreated individuals, out-degree to treated individuals, and out-degree to untreated individuals), in the main text we report results from the dyadic logistic regressions as they may be more intuitive (see **Supplementary Tables 22-53** for methodological details). Results from the corresponding permutation tests for the individual-level analysis show similar conclusions. **Supplementary Tables 22-53** also includes unconditional analyses for link formation.

**Acknowledgments**

We thank our field team in Honduras; the Honduras Ministry of Health; and our implementation partners, the InterAmerica Development Bank, World Vision Honduras, Child Fund Honduras, and Dimagi. Rennie Negron supervised the overall execution of the field data collection and Liza Nicoll managed the data. Mark McKnight and Wyatt Israel developed the Trellis software platform to map networks and collect the data. We are also grateful to Tom Keegan for superb project management.

**Funding**

This research was supported by the Bill and Melinda Gates Foundation, with additional support from the RWJ Foundation, the NOMIS Foundation, the Pershing Square Foundation, Paul Graham, and R01AG062668 from the National Institute on Aging. Additional support was also partly provided by NIH award R01MH134715, NSF awards CAREER IIS-1149662 and IIS-1409177, and by ONR awards YIP N00014-14-1-0485 and N00014-17-1-2131.

**Author contributions**

Conceptualization: MP, LF, EMA and NAC; Funding acquisition: EMA and NAC; Methodology: MP, LF, EMA, and NAC; Statistical analysis performance: MP and LF; Writing: MP, LF, EMA, and NAC.

**Competing interests**

The authors declare that they have no competing interests.

**Data and materials availability**

Data that complies with human subjects regulations will be made available at a URL upon publication.

**Code availability**

Source code for data analysis and data for reproduction of figures is available on GitHub (https://github.com/human-nature-lab/network_changes) and (at publication) permanently deposited at Zenodo (https://doi.org/10.5281/zenodo.***).

**Ethics**

The underlying field trial was registered with ClinicalTrials.gov, number NCT02694679. The Yale IRB and the Honduran Ministry of Health approved all data collection procedures (Protocol # 1506016012) and all participants provided informed consent.

**Supplementary Information**

# Educational Intervention Re-Wires Social Interactions in Isolated Village Networks

Marios Papamichalis,[1,2][†] Laura Forastiere,[1,3,4] [†] Edoardo M. Airoldi,[5,6] Nicholas A. Christakis[1,2,4,7,*]

[1] *Yale Institute for Network Science*
[2] *Human Nature Lab, Yale University*
[3] *Department of Biostatistics, Yale School of Public Health*
[4] *Department of Statistics and Data Science, Yale University*
[5] *Department of Statistics, Operations, and Data Science, Fox School of Business, Temple University*
[6] *Data Science Institute, Temple University*
[7] *Department of Sociology, Yale University*
†These authors shared equally in this work.
*Corresponding author: nicholas.christakis@yale.edu

**TABLE OF CONTENTS**

# 19. Explanation of Higher Order Untreated Respondent Increase in Degree Centralities

## 1. Experimental Design

In the parent study, we executed a multi-stage randomized controlled trial within 176 villages in the isolated Copan region of western Honduras, categorized into 16 treatment groups. The initial stage involved organizing the villages into 16 groups of 11, with each group being balance for household count and average family size. Subsequently, these clusters were randomly allocated to cells of a 2 x 8 factorial design, integrating two household selection approaches (random and friendship-nomination) and eight distinct household targeting ratios (0%, 5%, 10%, 20%, 30%, 50%, 75%, and 100%). The 0% and 100% targeting levels formed unique groups, with their 22 villages analyzed collectively. Our focus was on the villages under the random selection criteria, which removed 66 villages, for a final sample (for this project) of 110 villages.

The balancing strategy, informed by preliminary analyses, considered variables such as geographical location, elevation, healthcare access, population demographics, household characteristics, education, and economic factors. This was validated using simple linear regression.

For the intervention, villages in the random selection category had households selected as per their designated targeting percentage.

Social network analyses within these villages assessed healthcare, friendship, and financial connections, using criteria detailed in **Supplementary Table 1**. Networks were directed with allowance for multiplex links.

*Post-randomization, household lists were provided to our partners, the Inter-American Development Bank (IDB) and its local affiliates, who were blinded to the village categorization for deliver of the intervention.*

## 2. Honduras Dataset

The study employs a two-level randomization design in the Copan region of Honduras, spatially distributed across villages. *Supplementary Fig. 1a* depicts the allocation of treatment at both levels, while Figure 1b provides a geographic representation of a single village, incorporating observed network ties to illustrate the local social structure.

The data collection process (*Supplementary Fig. 2*) spanned multiple survey waves, tracking longitudinal changes in social and demographic variables. With an over 80% retention rate, the study systematically followed individuals while accounting for new entrants, births, deaths, and emigration. The MCNH intervention was implemented post-baseline (wave 1), with waves 3 and 4 conducted after the intervention, ensuring a structured follow-up to assess intervention effects over time.

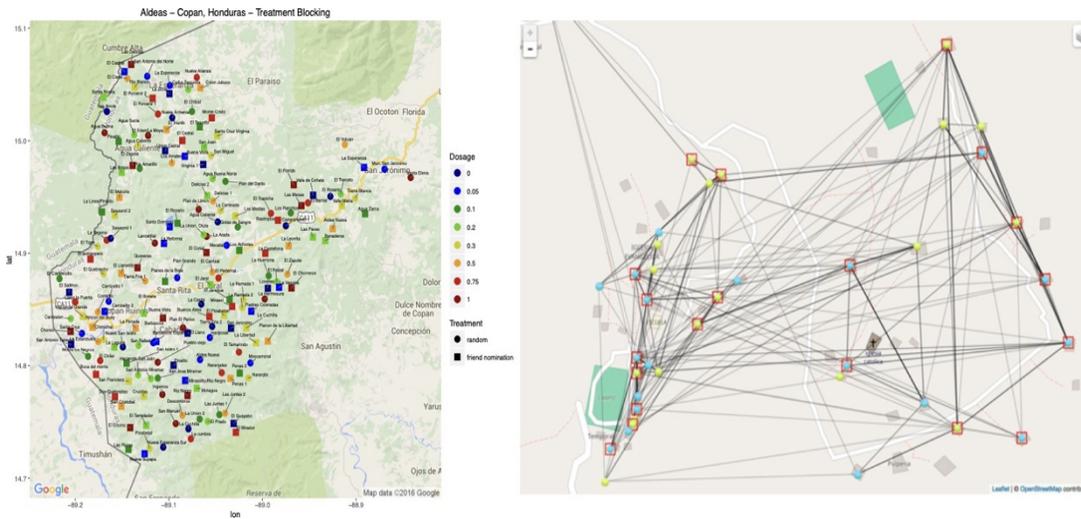

**Supplementary Figure 1 | Honduras Villages:** (a) A visualization of how the two-level randomization is spatially distributed in the Copan region of Honduras. (b) Geographic map of one village, with network ties also shown.

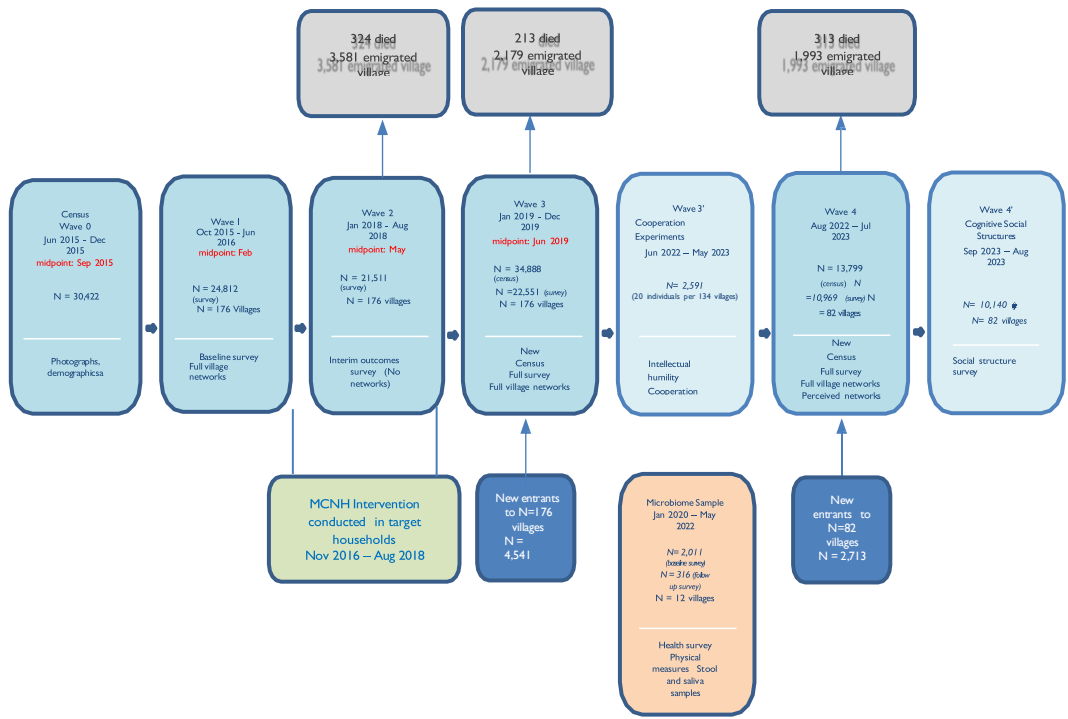

**Supplementary Figure 2 | Data Collection Overview:** Details waves, dates, sample sizes, and variable types. Highlights over 80% retention rate across waves, with tracking of new entrants, births, deaths, and emigration. Emphasizes the MCNH intervention timing, post-baseline (wave 1), with subsequent follow-up surveys (waves 3 and 4) conducted after intervention and previous wave completion, respectively.

### 3. Network Ascertainment

| Connection | User |
|---|---|
| Health | Who would you ask for advice about health-related matters? |
| Health | Who comes to you for health advice? (inverted) |
| Friendship | Who do you trust to talk to about something personal or private? |
| Friendship | With whom do you spend free time? |
| Friendship | Besides your partner, parents or siblings, who do you consider to be your closest friends? |
| Financial | Who would you feel comfortable asking to borrow 200 lempiras from if you needed them for the day? |
| Financial | Who do you think would be comfortable asking you to borrow 200 lempiras for the day? (inverted) |

**Supplementary Table 1 | Questionnaire with Connection Type**

*Supplementary Table 1* presents the questionnaire items ("name generators") used to construct four distinct social networks—health, financial, friendship, and aggregate social—based on self-reported interpersonal connections. The first three networks are derived from directed survey responses, with some ties inverted to capture reciprocal interactions.

The health network is constructed as the union of responses to two questions: one asking whom the respondent seeks for health advice and another (inverted) identifying those who seek advice from the respondent. The friendship network follows a similar approach, integrating responses about trust, shared free time, and close friendships. The financial network is built from questions regarding financial assistance, where both borrowing and lending relations are considered, with one direction inverted to maintain consistency. Finally, the aggregated social network aggregates all connection types into a single undirected network, ensuring no multiple edges between individuals.

To ensure consistency in directionality, inversion is applied to certain questions, particularly those related to advice-seeking and financial borrowing, allowing for a more coherent analysis of link formation and dissolution. This approach is crucial for examining changes in in-degree and out-degree patterns, such as whether treated individuals reduce their outgoing ties to untreated individuals. For example, in the health network, inversion ensures that a decrease in out-degree from treated to untreated individuals reflects a reduction in seeking health advice rather than a change in how often they are consulted. Similarly, in the financial network, inversion allows for the examination of whether treated individuals reduce their borrowing requests from untreated individuals. The inversion did not affect social and friendship networks.

## 4. Inclusion criteria

Respondents were included in our study if they meet the following criteria: (a) all applicable survey forms were finished by the respondent (b) they were enrolled in the census and participated in the survey for both waves 1 and 3 and (c) their status in terms of household and village in both wave 1 and 3 did not change.

## 5. Wasserstein Distance for Degree Distribution

We quantified each village's change in degree distribution from wave 1 (pre-treatment) to wave 3 (post-treatment) using the 1-Wasserstein distance[1–5] (Supplementary Fig. 3), yielding 22 distances for the control arm and 44 each for the low- and high-dosage arms. To test whether these changes differed between arms, we ran Welch two-sample t-tests: control vs low, t = –0.523, df = 43.1, p = 0.603, 95 % CI [–0.575, 0.338]; control vs high, t = 0.889, df = 28.1, p = 0.381, 95 % CI [–0.229, 0.581]. Both comparisons are non-significant, indicating no evidence that the degree-distribution change differs across treatment arms.

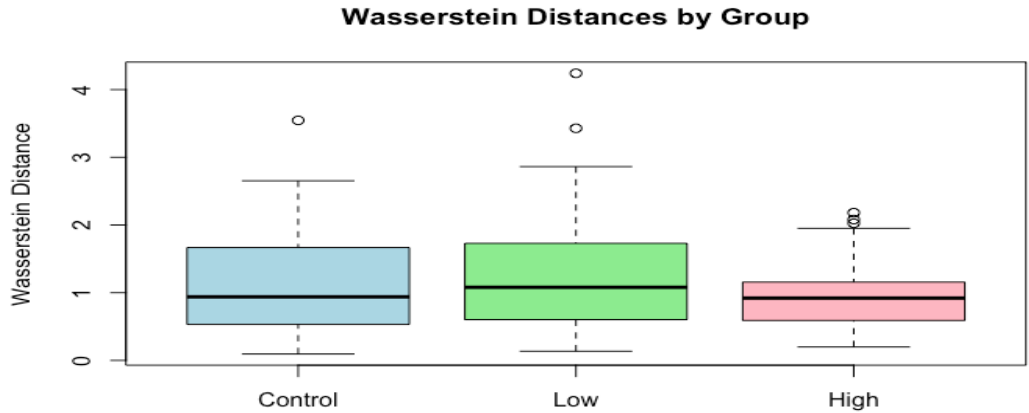

**Supplementary Figure 3** | Boxplot of Wasserstein distance comparisons between wave 1 (pre-treatment) and wave 3 (post-treatment) degree distributions. Welch Two Sample t-tests found no significant differences between groups, suggesting no evidence that the degree-distribution change differs across treatment arms.

## 6. Causal effects and test statistics in randomized experiments with interference

In our analysis, we consider both treated and untreated villages. Within these villages, individuals are also categorized as either treated or untreated. To analyze the impact of the interventions, we define overall, total, spillover and direct effects per standard practices in randomized experiments with interference[6].

To assess the impact of the intervention on network metrics, we used a percentage-scaled difference-in-differences estimator as our permutation-test statistic. For each individual, we first calculated how much a given metric (such as degree, in-degree, or out-degree) changed between wave 1 and wave 3. We then averaged these changes within the treatment group of interest and separately within the corresponding comparison group. The raw difference between these two averages was then scaled by dividing it by the wave-1 mean of the same metric in fully untreated (control) villages, yielding a percentage-based effect size. This percentage-scaled value served as the test statistic that was shuffled across randomization draws in the permutation test. The p-value was then computed based on how extreme the observed statistic was relative to the null distribution. The scaling simply facilitates interpretation as a percent change.

## 7. Low-dosage and High-dosage Effects on Network Connectivity

We noted that *low-dosage* villages were associated with a reduction in degree centrality. This decrease was more prominent among treated individuals but even significant in untreated individuals within these villages (*Supplementary Figure 4*). Conversely, *high-dosage* treatments were linked with an increase in degree centrality, suggesting enhanced network connectedness. This increase was less pronounced in both treated and untreated individuals within treated villages (*Supplementary Figure 4).*

## 8. LOESS Regression on Network Changes

Here, we include a series of figures (*Supplementary Figure 5 - Supplementary Figure 11*) that illustrate the Dosage-Mean Degree Change Relationship across different network types. In those figures, the 50% dosage group is included (from 11 villages that were not used in the main analysis), illustrating that there are no changes in the 50% dosage villages. Each figure presents three panels: (A) overall mean degree change, (B) mean degree change for treated individuals, and (C) mean degree change for untreated individuals in treated villages. The dose-response curves reveal bimodal patterns, supporting the distinction between "high" and "low" dosage categories to capture varying behavioral responses.

- *Supplementary Figure 4* focuses on health advice networks, showing how social ties related to health information shift in response to treatment.
- *Supplementary Figure 5* presents the aggregated network, which combines multiple social interactions to assess broad structural changes.
- *Supplementary Figure 6* examines friendship networks, highlighting how trust and companionship evolve post-treatment.
- *Supplementary Figure 7 investigates* financial networks, detailing how monetary exchange relationships adapt over time.
- *Supplementary Figure 8* and *Supplementary Figure 9* analyze residual friendship and financial networks, excluding health ties to isolate non-health-related social dynamics.

Across all networks, the observed bimodal patterns provide insights into how treatment exposure influences social connectivity, reinforcing the need to differentiate between varying dosage levels.

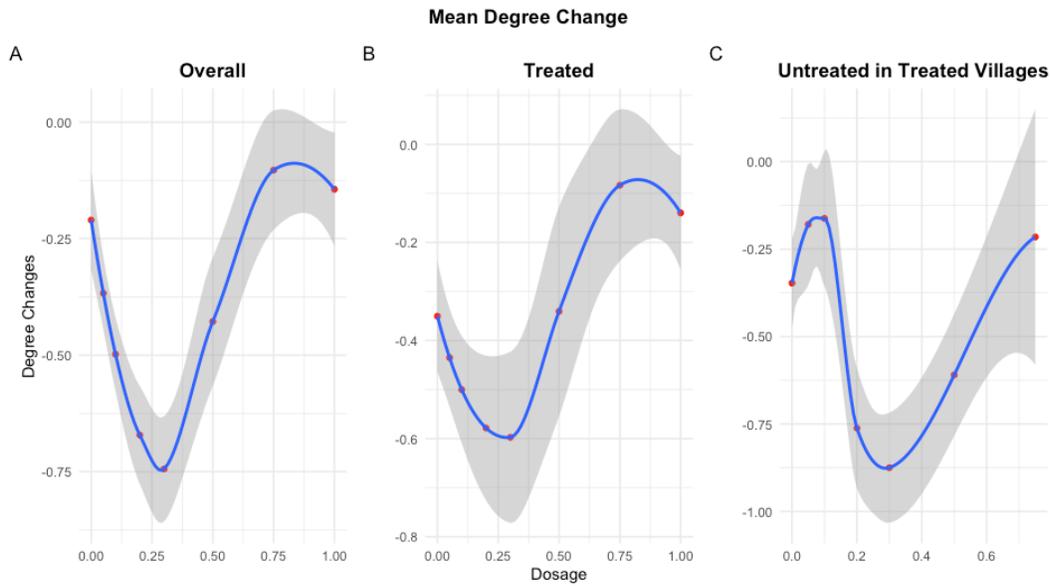

**Supplementary Figure 4 |** Dosage-Mean Degree Change Relationship in Health advice networks. A) Represents overall mean degree change, B) Shows mean degree change for treated individuals, and C) Depicts mean degree change for untreated in treated villages. The dose-response curves illustrate bimodal patterns, justifying the categorization into 'high' and 'low' dosages to reflect distinct behavioral responses across dosage ranges.

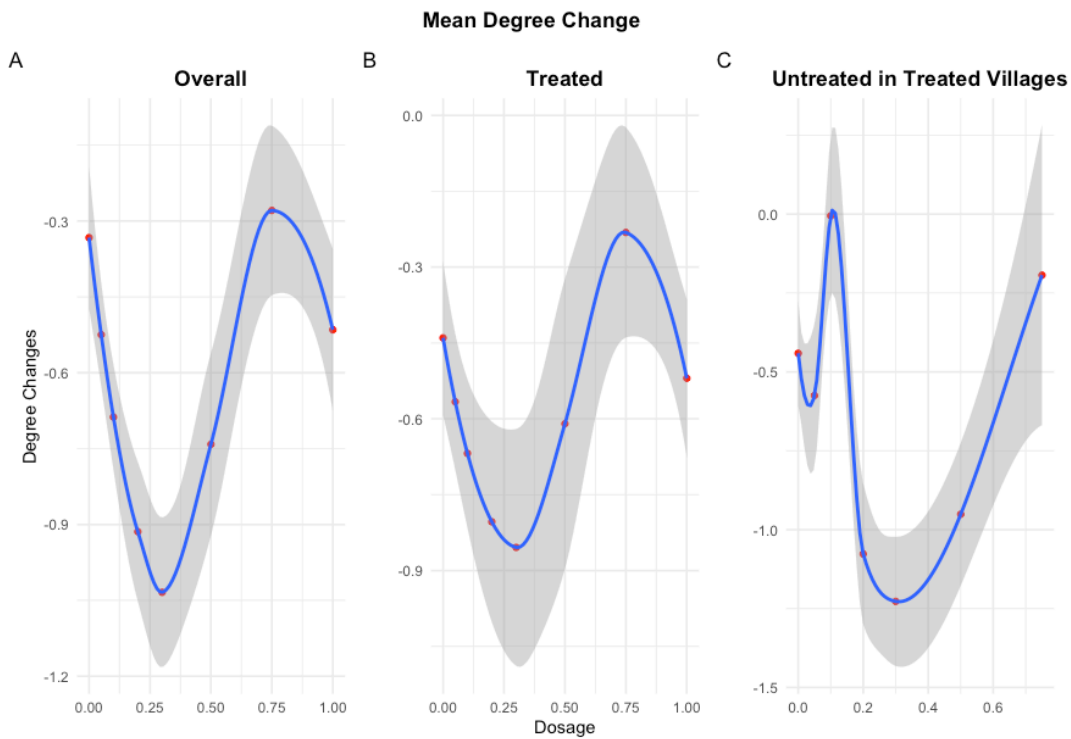

**Supplementary Figure 5 |** Dosage-Mean Degree Change Relationship in Aggregated networks. A) Represents overall mean degree change, B) Shows mean degree change for treated individuals, and C) Depicts mean degree change for untreated in treated villages. The dose-response curves illustrate bimodal patterns, justifying the categorization into 'high' and 'low' dosages to reflect distinct behavioral responses across dosage ranges.

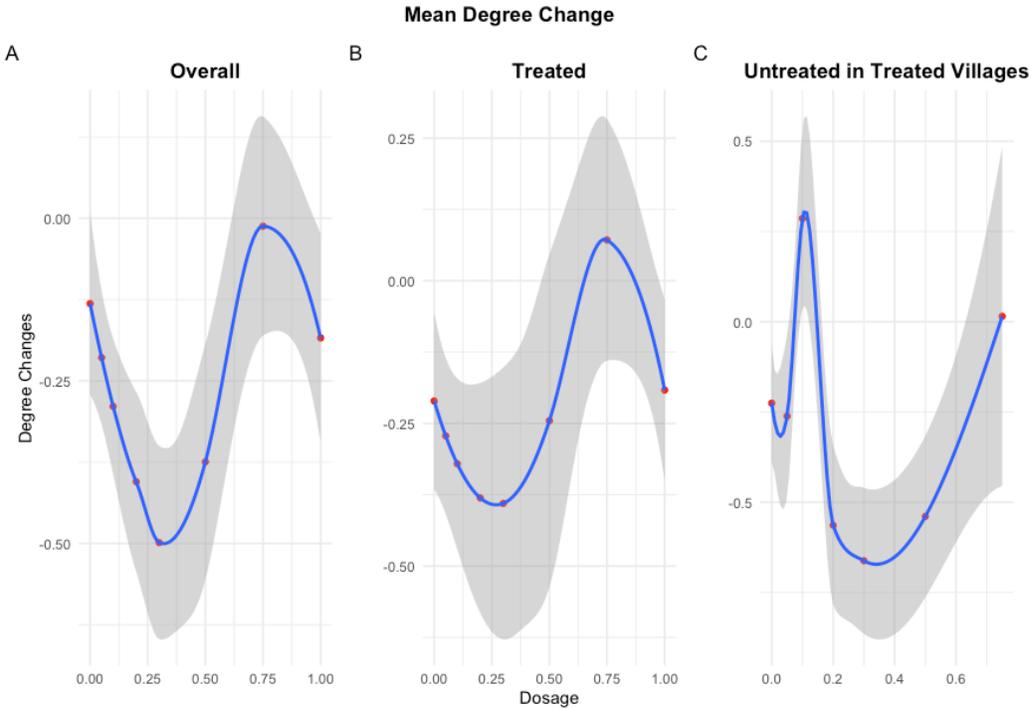

**Supplementary Figure 6** | Dosage-Mean Degree Change Relationship in Friendship networks. A) Represents overall mean degree change, B) Shows mean degree change for treated individuals, and C) Depicts mean degree change for untreated in treated villages. The dose-response curves illustrate bimodal patterns, justifying the categorization into 'high' and 'low' dosages to reflect distinct behavioral responses across dosage ranges.

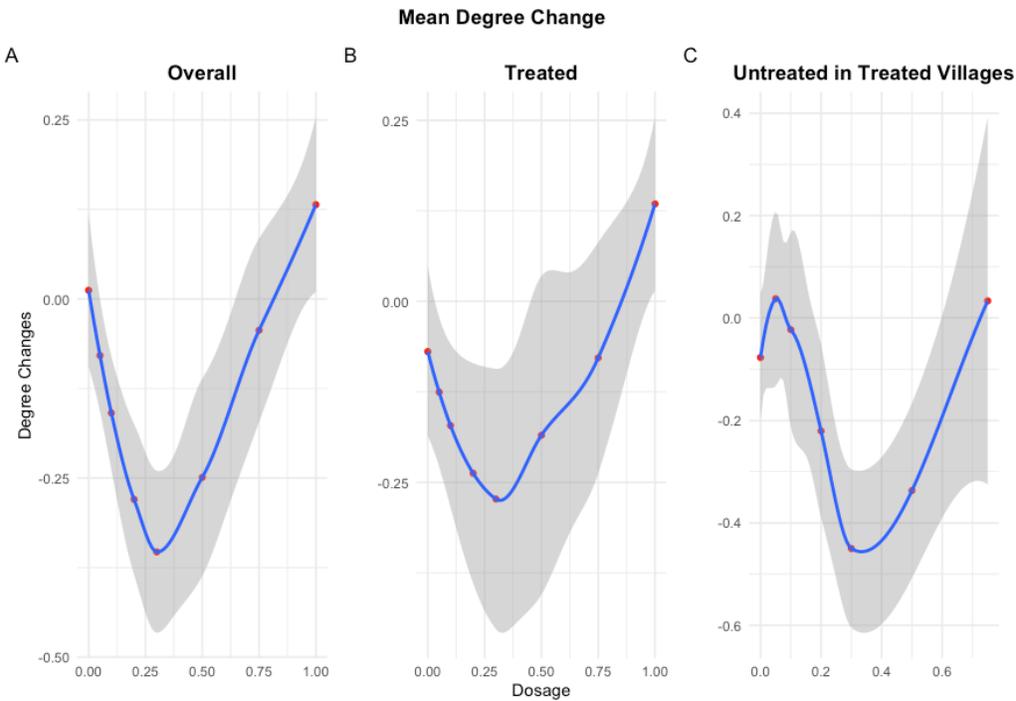

**Supplementary Figure 7** | Dosage-Mean Degree Change Relationship in Financial networks. A) Represents overall mean degree change, B) Shows mean degree change for treated individuals, and C) Depicts mean degree change for untreated in treated villages. The dose-response curves illustrate bimodal patterns, justifying the categorization into 'high' and 'low' dosages to reflect distinct behavioral responses across dosage ranges.

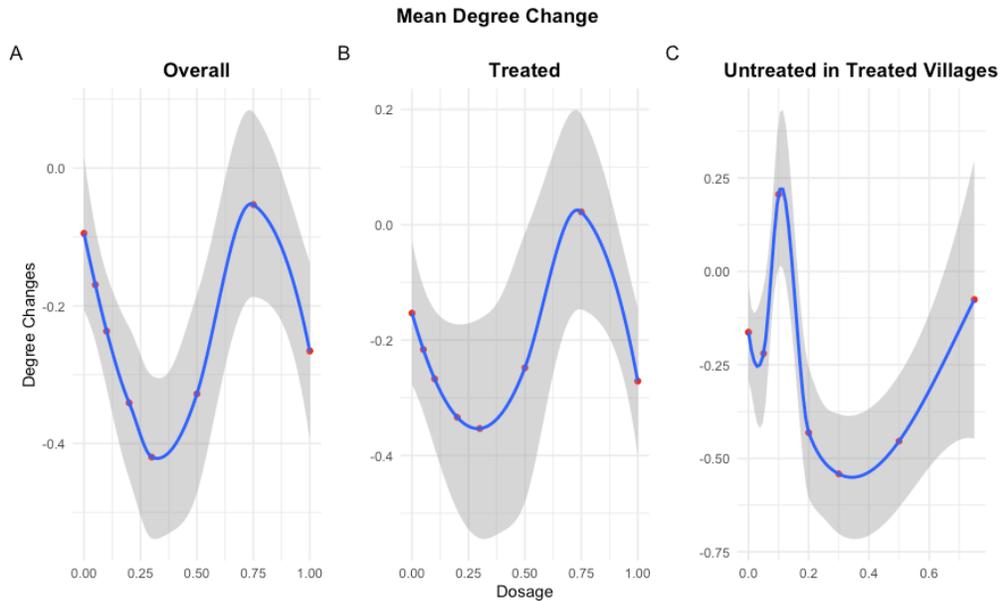

**Supplementary Figure 8** | Dosage-Mean Degree Change Relationship in residual friendship networks (excluded the health ties in wave 1 and 3). A) Represents overall mean degree change, B) Shows mean degree change for treated individuals, and C) Depicts mean degree change for untreated in treated villages. The dose-response curves illustrate bimodal patterns, justifying the categorization into 'high' and 'low' dosages to reflect distinct behavioral responses across dosage ranges

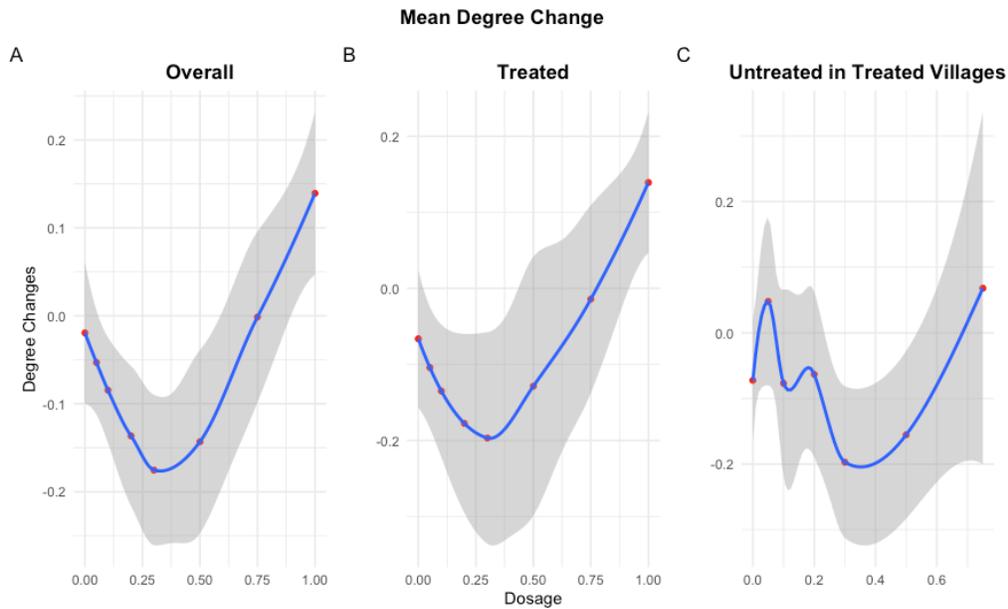

**Supplementary Figure 9** | Dosage-Mean Degree Change Relationship in residual financial networks (excluded the health ties in wave 1 and 3). A) Represents overall mean degree change, B) Shows mean degree change for treated individuals, and C) Depicts mean degree change for untreated in treated villages. The dose-response curves illustrate bimodal patterns, justifying the categorization into 'high' and 'low' dosages to reflect distinct behavioral responses across dosage ranges.

## 9. Analysis of Dosage Influence on Network Change

Our study concentrated on the impact of varying intervention dosages on the social network structures within Honduran villages, particularly focusing on changes in network centralities. We applied the *loess()* function from the stats package in *R* to flexibly model the non-linear relationship between intervention dosage and changes in degree centrality. By visually inspecting the smoothed data in *Supplementary Figure 4*, we identified a bimodal distribution of dosages, which prompted categorizing them into "low" and "high" groups. We used *Supplementary Figure 4* to select a threshold of 0.5 to distinguish between the two dosage categories. The same distinct patterns emerged when using non-parametric regression (via the *gam()* function from the *mgcv* package), supporting the conclusion that degree centrality changes behave differently at these two dosage levels.

Thus, we divided the intervention dosages into low (5%, 10%, 20%, 30%) and high (50%, 75% and 100%). We have also investigated the presence of an heterogenous effect for low and high dosage using causal trees[7]. More specifically, we observe that the partition distinguishes individuals in low and high dosage villages, supporting our claim that the mean degree of individuals in control villages, low-dosage villages, and high-dosage villages changed differently.

## 10. Number of Overall, Treated, Untreated and Control Individuals

Control villages had on average 75.4±43.7 individuals and treated villages 75.8±51.7 individuals. The total number of treated individuals is 3,149. The total number of untreated individuals in treated villages is 3,523 and untreated individuals in untreated villages is 1,659. Total effects were estimated among 4,808 individuals, direct effects were estimated among 6,672 individuals, and spillover effects were conducted among 5,182 individuals (*Supplementary Table 2*).

|  | Overall Individuals | Treated Individuals | Untreated Individuals | Control Individuals |
|---|---|---|---|---|
| All Dosages | 8331 | 3149 | 3523 | 1659 |
| Low Dosages | 5085 | 573 | 2941 | 1659 |
| High Dosages | 4755 | 2576 | 582 | 1659 |

**Supplementary Table 2** | Number of individuals in control and treated villages for all, low and high dosages.

|  | Overall Individuals | Overall Effect | Total Effect | Spillover Effect |
|---|---|---|---|---|
| All Dosages | 8331 | x | + | - |
| Low Dosages | 5085 | - | - | - |
| High Dosages | 4755 | + | + | + |

**Supplementary Table 3** | A plus sign (+) indicates that the degree of individuals in treated villages decreases less than in control villages, while a minus sign (–) indicates a larger decrease. An (x) denotes that no statistically significant difference in degree change was observed between treated and control villages.

## 11. Node-level and Network-level Statistical Results

In this section, we present the results of changes in low and high dosages for overall, total, spillover, and direct effects, as discussed in the section 'Average Effects on Health Networks.' All changes are calculated on a percentage basis using (Wave 3 Treated−Parallel Control Wave 3)/Control Wave 3, as illustrated in **Fig. 2.** We employ permutation tests with a difference-in-differences test statistics to analyze the overall, total, spillover, and direct effects on mean changes in degree, in-degree, out-degree, betweenness, closeness and clustering. These analyses are summarized in the tables below for all networks—health, friendship, financial and social—when including intra-household connections.

In directed networks, graph theoretical measures adapt to account for these complexities. This section discusses key measures for such networks: betweenness and closeness centrality. Betweenness centrality in directed networks with multiple edges quantifies the extent to which a node acts as a bridge along the shortest path between two other nodes. It is calculated as:

$B(v) = \sum_{\{s, t \in V\}} (\sigma_{\{st\}}(v) / \sigma_{\{st\}})$

where $\sigma_{\{st\}}$ is the number of shortest paths from node s to node t, and $\sigma_{\{st\}}(v)$ is the number of those paths passing through v.

The normalized betweenness centrality is given by:

$B_{norm}(v) = B(v) / ((n - 1) * (n - 2))$.

Closeness centrality in these networks measures how close a node is to all other nodes, considering the direction and multiplicity of edges. It is defined as:

$C(v) = \sum_{\{u \in V\}} (1 / d(v, u))$

where d(v, u) represents the shortest directed path length from v to u.

The normalized closeness centrality is given by:

$C_{norm}(v) = ((n - 1) * C(v)) / \sum_{\{u \neq v\}} d(v, u)$

For low dosages, the reduction in connectivity at the level of node degree in health advice networks also contributed to changes in other higher-order properties. We observed an increase in betweenness centrality by 29.92% (p-value < 0.001) and a diminution in closeness centrality by 11.99% less than that observed in control villages, (p-value < 0.001). Notably, treated individuals increased their betweenness by 32.18% (p-value = 0.013), while untreated individuals increased theirs by 29.48% (p-value < 0.001). Meanwhile, treated individuals decreased their closeness by 22.48% (p-value < 0.001), and untreated individuals decreased theirs by 9.94% (p-value < 0.001), with a direct effect of -12.69% (p-value = 0.004).

Betweenness for individuals in high-dosage treated villages increases by 22.92%; for only treated individuals by 21.87%; and for only untreated individuals by 27.58% – compared with individuals in fully untreated villages, with a p-values 0.004, 0.009 and 0.034 respectively. Finally, the clustering coefficient for individuals in treated villages increases by 33.70% and for only treated individuals increase by 32.65% compared with individuals in fully untreated villages, with both p-values < 0.001, indicating increased triangular connections among individuals.

All tables are presented below:

**All dosages**

|  | Overall | Total | Spillover | Direct |
| --- | --- | --- | --- | --- |
| Degree | 0.06% (0.977) | 4.18% (0.143) | -3.62% (0.228) | 7.78% (< 0.001) |
| In Degree | 0.06% (0.983) | 4.77% (0.226) | -4.15% (0.338) | 8.92% (0.010) |
| Out Degree | 0.06% (0.974) | 3.59% (0.295) | -3.09% (0.363) | 6.65% (0.017) |
| Betweenness | -22.51% (0.201) | 0.43% (0.976) | -43.01% (0.026) | 54.83% (0.005) |
| Closeness | -8.18% (0.005) | -8.30% (0.006) | -8.07% (0.005) | -0.22% (0.927) |
| Clustering | 19.80% (0.010) | 24.84% (0.004) | 15.30% (0.065) | 9.05% (0.155) |

**Supplementary Table 4** | Each cell presents the percentage change and p-values in treated groups (all individuals in fully and partially treated villages for the overall effect, treated individuals in fully and partially treated villages for the total and direct effects, untreated individuals in in fully and partially treated villages for the spillover effects) relative to control groups (all individuals in untreated villages for the overall, total and spillover effects, and untreated individuals in fully and partially treated villages for the direct effects) at wave 3 within the health all-dosages network, including intra-household connections, calculated using permutation tests with the difference-in-differences test statistics.

**Low dosages**

|  | Overall | Total | Spillover | Direct |
| --- | --- | --- | --- | --- |
| Degree | -6.36% (0.046) | -12.86% (0.007) | -5.09% (0.115) | -7.63% (0.119) |
| In Degree | -6.36% (0.149) | -13.35% (0.042) | -5.00% (0.268) | -8.18% (0.245) |
| Out Degree | -6.36% (0.072) | -12.38% (0.027) | -5.18% (0.157) | -7.07% (0.170) |
| Betweenness | 29.92% (<0.001) | 32.18% (0.013) | 29.48% (<0.001) | 2.72% (0.795) |
| Closeness | -11.99% (<0.001) | -22.48% (<0.001) | -9.94% (<0.001) | -12.69% (0.004) |
| Clustering | 7.31% (0.386) | -10.28% (0.456) | 10.74% (0.211) | -19.57% (0.105) |

**Supplementary Table 5** | Each cell presents the percentage change and p-values in treated groups (all individuals in fully and partially treated low-dosage villages for the overall effect, treated individuals in fully and partially treated low-dosage villages for the total and direct effects, untreated individuals in in fully and partially treated low-dosage villages for the spillover effects) relative to control groups (all individuals in untreated villages for the overall, total and spillover effects, and untreated individuals in fully and partially low-dosage treated villages for the direct effects) at wave 3 within the health low-dosage network, including intra-household connections, calculated using permutation tests with the difference-in-differences test statistics.

|  | Overall | Total | Spillover | Direct |
| --- | --- | --- | --- | --- |
| Degree | -10.77% (<0.001) | -11.07% (0.018) | -10.67% (0.003) | -0.40% (0.933) |
| In Degree | -10.77% (0.010) | -9.63% (0.125) | -11.16% (0.015) | 1.55% (0.824) |
| Out Degree | -10.77% (0.006) | -12.50% (0.033) | -10.19% (0.010) | -2.32% (0.726) |
| Betweenness | -11.96% (0.105) | -12.83% (0.280) | -11.66% (0.130) | -1.45% (0.917) |
| Closeness | -6.23% (0.111) | -12.14% (0.050) | -4.24% (0.302) | -8.11% (0.210) |
| Clustering | -4.36% (0.635) | 10.08% (0.510) | -9.22% (0.378) | 20.79% (0.196) |

**Supplementary Table 6** | Each cell presents the percentage change and p-values in treated groups (all individuals in fully and partially treated low-dosage villages for the overall effect, treated individuals in fully and partially treated low-dosage villages for the total and direct effects, untreated individuals in in fully and partially treated low-dosage villages for the spillover effects) relative to control groups (all individuals in untreated villages for the overall, total and spillover effects, and untreated individuals in fully and partially low-dosage treated villages for the direct effects) at wave 3 within the friendship low-dosage network, including intra-household connections, calculated using permutation tests with the difference-in-differences test statistics.

|  | Overall | Total | Spillover | Direct |
|---|---|---|---|---|
| Degree | -6.73% (0.027) | -8.15% (0.076) | -6.36% (0.053) | -1.86% (0.714) |
| In Degree | -6.73% (0.082) | -8.02% (0.177) | -6.40% (0.103) | -1.69% (0.773) |
| Out Degree | -6.73% (0.059) | -8.29% (0.121) | -6.33% (0.082) | -2.04% (0.731) |
| Betweenness | -8.14% (0.225) | -8.05% (0.461) | -8.17% (0.255) | 0.15% (0.992) |
| Closeness | -6.94% (0.059) | -17.11% (0.004) | -4.32% (0.267) | -12.86% (0.033) |
| Clustering | 4.44% (0.610) | 14.28% (0.310) | 1.89% (0.848) | 13.65% (0.372) |

**Supplementary Table 7** | Each cell presents the percentage change and p-values in treated groups (all individuals in fully and partially treated low-dosage villages for the overall effect, treated individuals in fully and partially treated low-dosage villages for the total and direct effects, untreated individuals in in fully and partially treated low-dosage villages for the spillover effects) relative to control groups (all individuals in untreated villages for the overall, total and spillover effects, and untreated individuals in fully and partially low-dosage treated villages for the direct effects) at wave 3 within the financial low-dosage network, including intra-household connections, calculated using permutation tests with the difference-in-differences test statistics.

|  | Overall | Total | Spillover | Direct |
|---|---|---|---|---|
| Degree | -5.12% (0.004) | -7.06% (0.011) | -4.74% (0.009) | -2.43% (0.404) |
| Betweenness | 4.12% (0.343) | -3.09% (0.643) | 5.53% (0.228) | -8.68% (0.192) |
| Closeness | -3.47% (<0.001) | -5.36% (0.001) | -3.10 (< 0.001) | -2.38% (0.106) |
| Clustering | 1.35% (0.627) | -2.04% (0.638) | 2.01% (0.481) | -4.14% (0.332) |

**Supplementary Table 8** | Each cell presents the percentage change and p-values in treated groups (all individuals in fully and partially treated low-dosage villages for the overall effect, treated individuals in fully and partially treated low-dosage villages for the total and direct effects, untreated individuals in in fully and partially treated low-dosage villages for the spillover effects) relative to control groups (all individuals in untreated villages for the overall, total and spillover effects, and untreated individuals in fully and partially low-dosage treated villages for the direct effects) at wave 3 within the social low-dosage network, including intra-household connections, calculated using permutation tests with the difference-in-differences test statistics.

**High dosages**

|  | Overall | Total | Spillover | Direct |
|---|---|---|---|---|
| Degree | 7.21% (0.009) | 7.97% (0.006) | 3.82% (0.408) | 4.54% (0.313) |
| In Degree | 7.21% (0.063) | 8.80% (0.025) | 0.14% (0.976) | 9.68% (0.137) |
| Out Degree | 7.21% (0.028) | 7.14% (0.038) | 7.49% (0.149) | –0.37% (0.945) |
| Betweenness | 22.92% (0.004) | 21.87% (0.009) | 27.58% (0.034) | –5.51% (0.626) |
| Closeness | –3.94% (0.176) | –5.14% (0.087) | 1.37% (0.770) | –6.43% (0.154) |
| Clustering | 33.70% (¡0.001) | 32.65% (¡0.001) | 38.35% (0.005) | –5.97% (0.654) |

**Supplementary Table 9** | Each cell presents the percentage change and p-values in treated groups (all individuals in fully and partially treated high-dosage villages for the overall effect, treated individuals in fully and partially treated high-dosage villages for the total and direct effects, untreated individuals in in fully and partially treated high-dosage villages for the spillover effects) relative to control groups (all individuals in untreated villages for the overall, total and spillover effects, and untreated individuals in fully and partially treated high-dosage villages for the direct effects) at wave 3 within the health high-dosage network, including intra-household connections, calculated using permutation tests with the difference-in-differences test statistics.

|  | Overall | Total | Spillover | Direct |
|---|---|---|---|---|
| Degree | 1.38% (0.458) | 1.92% (0.321) | -1.01% (0.712) | 3.42% (0.280) |
| In Degree | 1.38% (0.560) | 1.62% (0.517) | 0.29% (0.942) | 1.54% (0.704) |
| Out Degree | 1.38% (0.561) | 2.21% (0.399) | -2.31% (0.560) | 5.32% (0.208) |
| Betweenness | -2.39% (0.634) | -2.66% (0.606) | -1.20% (0.868) | -1.30% (0.858) |
| Closeness | 0.40% (0.771) | 0.67% (0.667) | -0.75% (0.751) | 1.50% (0.578) |
| Clustering | 7.59% (0.109) | 8.42% (0.073) | 3.93% (0.599) | 4.95% (0.529) |

**Supplementary Table 10** | Each cell presents the percentage change and p-values in treated groups (all individuals in fully and partially treated high-dosage villages for the overall effect, treated individuals in fully and partially treated high-dosage villages for the total and direct effects, untreated individuals in in fully and partially treated high-dosage villages for the spillover effects) relative to control groups (all individuals in untreated villages for the overall, total and spillover effects, and untreated individuals in fully and partially treated high-dosage villages for the direct effects) at wave 3 within the friendship high-dosage network, including intra-household connections, calculated using permutation tests with the difference-in-differences test statistics.

|  | Overall | Total | Spillover | Direct |
|---|---|---|---|---|
| Degree | 3.52% (0.200) | 4.21% (0.135) | 0.50% (0.915) | 4.41% (0.366) |
| In Degree | 3.52% (0.303) | 3.91% (0.276) | 1.82% (0.747) | 2.45% (0.672) |
| Out Degree | 3.52% (0.255) | 4.51% (0.175) | –0.82% (0.863) | 6.42% (0.289) |
| Betweenness | –6.59% (0.311) | –3.63% (0.590) | –19.71% (0.085) | 22.57% (0.078) |
| Closeness | –3.38% (0.340) | –2.68% (0.474) | –6.50% (0.257) | 3.72% (0.491) |
| Clustering | 11.01% (0.195) | 10.21% (0.263) | 14.57% (0.267) | –5.23% (0.735) |

**Supplementary Table 11** | Each cell presents the percentage change and p-values in treated groups (all individuals in fully and partially treated high-dosage villages for the overall effect, treated individuals in fully and partially treated high-dosage villages for the total and direct effects, untreated individuals in in fully and partially treated high-dosage villages for the spillover effects) relative to control groups (all individuals in untreated villages for the overall, total and spillover effects, and untreated individuals in fully and partially treated high-dosage villages for the direct effects) at wave 3 within the financial high-dosage network, including intra-household connections, calculated using permutation tests with the difference-in-differences test statistics.

|  | Overall | Total | Spillover | Direct |
|---|---|---|---|---|
| Degree | -0.31% (0.851) | -0.10% (0.949) | -1.21% (0.660) | 1.33% (0.648) |
| Betweenness | -8.27% (0.105) | -7.12% (0.173) | -13.35% (0.089) | 6.27% (0.453) |
| Closeness | 2.97% (0.003) | 2.88% (0.004) | 3.37% (0.012) | -0.52% (0.742) |
| Clustering | 5.02% (0.077) | 4.66% (0.094) | 6.61% (0.142) | -2.15% (0.642) |

**Supplementary Table 12** | Each cell presents the percentage change and p-values in treated groups (all individuals in fully and partially treated high-dosage villages for the overall effect, treated individuals in fully and partially treated high-dosage villages for the total and direct effects, untreated individuals in in fully and partially treated high-dosage villages for the spillover effects) relative to control groups (all individuals in untreated villages for the overall, total and spillover effects, and untreated individuals in fully and partially treated high-dosage villages for the direct effects) at wave 3 within the social high-dosage network, including intra-household connections, calculated using permutation tests with the difference-in-differences test statistics.

## 12. Analysis Excluding Intra-Household Connections

We conducted a robustness analysis, excluding intra-household connections. For low-dosage networks (5%, 10%, 20%, and 30%), in the situation excluding intra-household relationships, comparing waves 1 and 3, we observe a trend of overall declines in degree in-degree, and out-degree by 8.66% for all these three metrics, compared to individuals in fully untreated villages, with corresponding p-values of 0.002, 0.011, and 0.105, respectively. Additionally, a slight decrease in closeness by 2.74%, with a p-value of 0.005, further corroborates the pervasive nature of the treatment effect.

In high-dosage villages (50 %, 75 %, and 100 % treatment) and after removing intra-household ties, network connectivity rises only modestly and statistically not significantly. Relative to residents of fully untreated villages, the overall (total) effect is a 3.7 % gain in degree, in-degree, and out-degree, none of which reaches statistical significance (p = 0.302, 0.464, 0.385). Treated individuals in these partially treated villages show slightly larger—but still non-significant—gains: degree + 4.8 % (p = 0.204), in-degree + 5.9 % (p = 0.230), and out-degree + 3.6 % (p = 0.431). Spillover onto untreated peers is minimal: their degree and in-degree fall by 1.0 % and 6.1 % (p = 0.998 and 0.029, respectively), while their out-degree rises by 4.1 % (p = 0.540). Taken together, these results show that high-dosage interventions do not materially alter overall connectivity. The small upticks observed are driven largely by the treated group, with little evidence of meaningful spillover to untreated villagers.

Similar findings apply to the total and spillover effects. *Supplementary Table 13.* to *Supplementary Table 20* report the point estimates (percentages) and p-values for the node properties changes of the health, friendship, financial, and social networks, excluding intra-household connections. Moreover, direct effects are presented in *Supplementary Table 13* to *Supplementary Table 20*.

**Low dosages**

|  | **Overall** | **Total** | **Spillover** | **Direct** |
|---|---|---|---|---|
| Degree | -11.82% (0.003) | -16.21% (0.011) | -10.97% (0.009) | -5.36% (0.402) |
| In Degree | -11.82% (0.037) | -16.89% (0.034) | -10.84% (0.058) | -6.17% (0.490) |
| Out Degree | -11.82% (0.009) | -15.54% (0.031) | -11.10% (0.010) | -4.55% (0.507) |
| Betweenness | 24.61% (0.006) | 26.44% (0.078) | 24.26% (0.013) | 2.13% (0.879) |
| Closeness | -13.10% (< 0.001) | -13.26% (0.017) | -13.07% (< 0.001) | -0.21% (0.974) |
| Clustering | -7.09% (0.605) | 6.60% (0.756) | -9.75% (0.477) | 16.09% (0.391) |

**Supplementary Table 13** | Each cell presents the percentage change and p-values in treated groups (all individuals in fully and partially treated low-dosage villages for the overall effect, treated individuals in fully and partially treated low-dosage villages for the total and direct effects, untreated individuals in in fully and partially treated low-dosage villages for the spillover effects) relative to control groups (all individuals in untreated villages for the overall, total and spillover effects, and untreated individuals in fully and partially treated low-dosage villages for the direct effects) at wave 3 within the health low-dosage network, excluding intra-household connections, calculated using permutation tests with the difference-in-differences mean as test statistic.

|  | **Overall** | **Total** | **Spillover** | **Direct** |
|---|---|---|---|---|
| Degree | -9.56% (< 0.001) | -6.25% (0.150) | -10.67% (< 0.001) | 4.59% (0.330) |
| In Degree | -9.56% (0.008) | -7.81% (0.174) | -10.15% (0.008) | 2.43% (0.705) |
| Out Degree | -9.56% (0.012) | -4.70% (0.426) | -11.19% (0.004) | 6.74% (0.257) |
| Betweenness | 4.26% (0.454) | 2.08% (0.816) | 5.00% (0.401) | -3.50% (0.715) |
| Closeness | -0.40% (0.875) | -2.19% (0.570) | 0.21% (0.943) | -2.27% (0.563) |
| Clustering | -0.11% (0.986) | -3.77% (0.756) | 1.12% (0.885) | -4.74% (0.693) |

**Supplementary Table 14** | Each cell presents the percentage change and p-values in treated groups ((all individuals in fully and partially treated low-dosage villages for the overall effect, treated individuals in fully and partially treated low-dosage villages for the total and direct effects, untreated individuals in in fully and partially treated low-dosage villages for the spillover effects) relative to control groups (all individuals in untreated villages for the overall, total and spillover effects, and untreated individuals in fully and partially treated low-dosage villages for the direct effects) at wave 3 within the friendship low-dosage network, excluding intra-household connections, calculated using permutation tests with the difference-in-differences mean as test statistics.

|  | **Overall** | **Total** | **Spillover** | **Direct** |
|---|---|---|---|---|
| Degree | -9.30% (0.019) | -10.67% (0.068) | -8.85% (0.032) | -1.81% (0.777) |
| In Degree | -9.30% (0.062) | -8.79% (0.247) | -9.48% (0.070) | 0.68% (0.932) |
| Out Degree | -9.30% (0.034) | -12.54% (0.050) | -8.21% (0.083) | -4.24% (0.543) |
| Betweenness | -36.19% (< 0.001) | -40.06% (0.002) | -34.89% (< 0.001) | -7.46% (0.680) |
| Closeness | 5.00% (0.279) | -1.88% (0.783) | 7.32% (0.136) | -9.00% (0.193) |
| Clustering | -9.47% (0.477) | 8.77% (0.677) | -15.61% (0.274) | 26.04% (0.239) |

**Supplementary Table 15** | Each cell presents the percentage change and p-values in treated groups (all individuals in fully and partially treated low-dosage villages for the overall effect, treated individuals in fully and partially treated low-dosage villages for the total and direct effects, untreated individuals in in fully and partially treated low-dosage villages for the spillover effects) relative to control groups (all individuals in untreated villages for the overall, total and spillover effects, and untreated individuals in fully and partially treated low-dosage villages for the direct effects) at wave 3 within the financial low-dosage network, excluding intra-household connections, calculated using permutation tests with the difference-in-differences mean as estimator.

|  | Overall | Total | Spillover | Direct |
|---|---|---|---|---|
| Degree | -6.71% (< 0.001) | -8.79% (0.011) | -6.31% (0.007) | -2.62% (0.467) |
| Betweenness | 5.02% (0.283) | 4.22% (0.593) | 5.18% (0.282) | -1.03% (0.902) |
| Closeness | -2.74% (0.048) | -4.48% (0.031) | -2.40% (0.108) | -2.15% (0.352) |
| Clustering | 2.27% (0.603) | -5.65% (0.435) | 3.81% (0.419) | -9.53% (0.167) |

**Supplementary Table 16** | Each cell presents the percentage change and p-values in treated groups (all individuals in fully and partially treated low-dosage villages for the overall effect, treated individuals in fully and partially treated low-dosage villages for the total and direct effects, untreated individuals in in fully and partially treated low-dosage villages for the spillover effects) relative to control groups (all individuals in untreated villages for the overall, total and spillover effects, and untreated individuals in fully and partially treated low-dosage villages for the direct effects) at wave 3 within the social low-dosage network, excluding intra-household connections, calculated using permutation tests with the difference-in-differences mean astest statistics.

**High dosages**

|  | Overall | Total | Spillover | Direct |
|---|---|---|---|---|
| Degree | .71% (0.302) | 4.78% (0.204) | -1.02% (0.870) | 6.45% (0.279) |
| In Degree | .71% (0.464) | 5.94% (0.230) | -6.14% (0.445) | 13.90% (0.104) |
| Out Degree | .71% (0.385) | 3.62% (0.431) | 4.11% (0.540) | -0.52% (0.938) |
| Betweenness | 3.04% (0.019) | 22.03% (0.031) | 27.50% (0.088) | -4.75% (0.722) |
| Closeness | 7.29% (0.039) | -7.45% (0.044) | -6.61% (0.210) | -0.94% (0.876) |
| Clustering | 8.99% (0.023) | 26.26% (0.046) | 41.05% (0.051) | -16.19% (0.409) |

**Supplementary Table 17** | Each cell presents the percentage change and p-values in treated groups (all individuals in fully and partially treated high-dosage villages for the overall effect, treated individuals in fully and partially treated high-dosage villages for the total and direct effects, untreated individuals in in fully and partially treated high-dosage villages for the spillover effects) relative to control groups (all individuals in untreated villages for the overall, total and spillover effects, and untreated individuals in fully and partially treated high-dosage villages for the direct effects) at wave 3 within the health high-dosage network, excluding intra-household connections, calculated using permutation tests with the difference-in-differences mean as test statistics.

|  | Overall | Total | Spillover | Direct |
|---|---|---|---|---|
| Degree | 1.38% (0.458) | 1.92% (0.321) | -1.01% (0.712) | 3.42% (0.280) |
| In Degree | 1.38% (0.560) | 1.62% (0.517) | 0.29% (0.942) | 1.54% (0.704) |
| Out Degree | 1.38% (0.561) | 2.21% (0.399) | -2.31% (0.560) | 5.32% (0.208) |
| Betweenness | -2.39% (0.634) | -2.66% (0.606) | -1.20% (0.868) | -1.30% (0.858) |
| Closeness | 0.40% (0.771) | 0.67% (0.667) | -0.75% (0.751) | 1.50% (0.578) |
| Clustering | 7.59% (0.109) | 8.42% (0.073) | 3.93% (0.599) | 4.95% (0.529) |

**Supplementary Table 18**| Each cell presents the percentage change and p-values in treated groups (all individuals in fully and partially treated high-dosage villages for the overall effect, treated individuals in fully and partially treated high-dosage villages for the total and direct effects, untreated individuals in in fully and partially treated high-dosage villages for the spillover effects) relative to control groups (all individuals in untreated villages for the overall, total and spillover effects, and untreated individuals in fully and partially treated high-dosage villages for the direct effects) at wave 3 within the friendship low-dosage network, excluding intra-household connections, calculated using permutation tests with the difference-in-differences mean as test statistics.

|  | Overall | Total | Spillover | Direct |
|---|---|---|---|---|
| Degree | 3.52% (0.200) | 4.21% (0.135) | 0.50% (0.915) | 4.41% (0.366) |
| In Degree | 3.52% (0.303) | 3.91% (0.276) | 1.82% (0.747) | 2.45% (0.672) |
| Out Degree | 3.52% (0.255) | 4.51% (0.175) | –0.82% (0.863) | 6.42% (0.289) |
| Betweenness | –6.59% (0.311) | –3.63% (0.590) | –19.71% (0.085) | 22.57% (0.078) |
| Closeness | –3.38% (0.340) | –2.68% (0.474) | –6.50% (0.257) | 3.72% (0.491) |
| Clustering | 11.01% (0.195) | 10.21% (0.263) | 14.57% (0.267) | –5.23% (0.735) |

**Supplementary Table 19** | Each cell presents the percentage change and p-values in treated groups (all individuals in fully and partially treated high-dosage villages for the overall effect, treated individuals in fully and partially treated high-dosage villages for the total and direct effects, untreated individuals in in fully and partially treated high-dosage villages for the spillover effects) relative to control groups (all individuals in untreated villages for the overall, total and spillover effects, and untreated individuals in fully and partially treated high-dosage villages for the direct effects) at wave 3 within the financial high-dosage network, excluding intra-household connections, calculated using permutation tests with the difference-in-differences mean as test statistics.

|  | Overall | Total | Spillover | Direct |
|---|---|---|---|---|
| Degree | -0.31% (0.851) | -0.10% (0.949) | -1.21% (0.660) | 1.33% (0.648) |
| Betweenness | -8.27% (0.105) | -7.12% (0.173) | -13.35% (0.089) | 6.27% (0.453) |
| Closeness | 2.97% (0.003) | 2.88% (0.004) | 3.37% (0.012) | -0.52% (0.742) |
| Clustering | 5.02% (0.077) | 4.66% (0.094) | 6.61% (0.142) | -2.15% (0.642) |

**Supplementary Table 20** | Each cell presents the percentage change and p-values in treated groups (all individuals in fully and partially treated high-dosage villages for the overall effect, treated individuals in fully and partially treated high-dosage villages for the total and direct effects, untreated individuals in in fully and partially treated high-dosage villages for the spillover effects) relative to control groups (all individuals in untreated villages for the overall, total and spillover effects, and untreated individuals in fully and partially treated high-dosage villages for the direct effects)  at wave 3 within the social high-dosage network, excluding intra-household connections, calculated using permutation tests with the difference-in-differences mean as test statistics.

## 13. Result Tables for Link Formation and Link Dissolution

This section examines patterns of link formation and dissolution across health, friendship, financial, and aggregated social networks under both low- and high-dosage conditions. By comparing ties that persisted or dissolved between Wave 1 (pre-treatment) and Wave 3 (post-treatment), we identify how exposure to treatment influences social connections. Results indicate that existing links involving treated individuals are less likely to dissolve compared to fully untreated pairs, particularly in the health and financial networks. Moreover, new ties are more likely to form among individuals who were already connected to treated peers, suggesting that treatment fosters within-group cohesion but may also contribute to the fragmentation of mixed-treatment relationships.

In addition to examining overall link formation and dissolution, we analyze how network position and treatment exposure influence social connectivity by categorizing individuals into four distinct groups: U0, untreated individuals not directly connected to treated peers, U1, untreated individuals directly connected to at least one treated peer, T1 treated individuals connected to other treated peers, and T0, treated individuals who are not connected to other treated peers. This classification helps assess whether proximity to treated individuals affects tie persistence and new link formation.

**Low dosages**

**Health Network**

| Coefficient | Estimate | Std. Error | z value | Pr(>|t|) |
|---|---|---|---|---|
| (Intercept) | -0.59446 | 0.04283 | -13.880 | < 2e-16 *** |
| UU | 0.07033 | 0.05406 | 1.301 | 0.19329 |
| TU | -0.30137 | 0.10367 | -2.907 | 0.00365 ** |
| UT | -0.48684 | 0.10808 | -4.505 | 6.65e-06 *** |
| TT | 0.36137 | 0.11417 | 3.165 | 0.00155 ** |

**Supplementary Table 21** | Link dissolution in the health network, low dosage

Focusing on links that existed in Wave 1, we next examine whether they dissolved by Wave 3 under low-dosage conditions. Compared to the baseline category (UoUo: links between untreated individuals in untreated villages), several types of links were significantly more likely to break. In particular, U1Uh (untreated connected to treated in Wave 1 → untreated not connected to treated) links have 54.2% lower odds of persisting ($p < 0.0001$), while ToU1 and U1To links (treated not connected to treated → untreated connected to treated, and vice versa) show 30.6% and 30.8% lower odds of persisting, respectively (both $p < 0.01$). Additionally, U1T1 links (untreated connected to a treated individual who was itself linked to another treated person) have 58.3% lower odds of persisting ($p < 0.0001$). By contrast, UhUh (untreated not connected to treated → untreated not connected to treated) and UhU1 (untreated not connected to treated → untreated connected to treated) links remain more stable, with 15.8% ($p = 0.019$) and 18.8% ($p = 0.027$) higher odds of persisting, respectively. Finally, ToTo links (treated not connected to treated → treated not connected to treated) display a 43.5% higher chance of persisting ($p = 0.002$). These patterns suggest that once individuals have a prior connection to someone treated in Wave 1, their ties to other partially treated villagers are more likely to dissolve, whereas those without such connections—along with pairs of treated individuals—tend to maintain their existing links.

| Coefficient | Estimate | Std. Error | z value | Pr(>|t|) |
|---|---|---|---|---|
| (Intercept) | -0.59446 | 0.04283 | -13.880 | < 2e-16 *** |
| UhUh | 0.14715 | 0.06274 | 2.345 | 0.01900 * |
| U1Uh | -0.78136 | 0.13868 | -5.634 | 1.76e-08 *** |
| UhU1 | 0.17213 | 0.07783 | 2.212 | 0.02699 * |
| U1U1 | 0.07545 | 0.09549 | 0.790 | 0.42943 |
| ToU1 | -0.36471 | 0.12251 | -2.977 | 0.00291 ** |
| T1U1 | -0.16263 | 0.17191 | -0.946 | 0.34412 |
| U1To | -0.36802 | 0.11999 | -3.067 | 0.00216 ** |
| U1T1 | -0.87483 | 0.22169 | -3.946 | 7.94e-05 *** |
| ToTo | 0.36137 | 0.11417 | 3.165 | 0.00155 ** |

**Supplementary Table 22** | Link dissolution in the health network, low dosages

| Coefficient | Estimate | Std. Error | z value | Pr(>\|t\|) |
|---|---|---|---|---|
| (Intercept) | -4.85349 | 0.02845 | -170.589 | < 2e-16 *** |
| UU | -0.19657 | 0.03653 | -5.381 | 7.43e-08 *** |
| TU | -0.45546 | 0.07008 | -6.499 | 8.08e-11 *** |
| UT | -0.42208 | 0.06915 | -6.103 | 1.04e-09 *** |
| TT | 0.36558 | 0.08588 | 4.257 | 2.07e-05 *** |

**Supplementary Table 23** | Link formation in the health network, low dosages

We investigate new connections formed by Wave 3 among individuals who were unlinked in Wave 1. Again, using UoUo (untreated–untreated in untreated villages) as the reference category, T1T1 links (treated individuals already connected to other treated individuals) are 166.6% more likely to form (p < 0.0001), and T1U1 links (treated connected-to-treated → untreated connected-to-treated) exhibit 68.1% higher odds of forming (p < 0.001). Similarly, U1U1 (untreated connected-to-treated → untreated connected-to-treated) ties show 90.0% higher odds of emerging (p < 0.0001), and ToTo (treated not connected to treated → treated not connected to treated) links are 38.3% more likely to appear (p = 0.005). By contrast, pairs such as UhUh (untreated not connected to treated → untreated not connected to treated) are 31.9% less likely to form (p < 0.001), and UhT1 or T1Uh links (untreated not connected to treated ↔ treated connected to treated) are both more than 40% less likely to form (p < 0.001). These findings suggest that treated individuals who were already tied to other treated individuals are particularly inclined to build additional connections with similarly treated peers, whereas those who remained unconnected to treated individuals in Wave 1 tend to stay isolated or form fewer new links.

| Coefficient | Estimate | Std. Error | z value | Pr(>\|t\|) |
|---|---|---|---|---|
| (Intercept) | -4.85349 | 0.02845 | -170.589 | < 2e-16 *** |
| UhUh | -0.38369 | 0.04153 | -9.238 | < 2e-16 *** |
| UhU1 | 0.29217 | 0.05798 | 5.040 | 4.66e-07 *** |
| UhT1 | -0.52791 | 0.14188 | -3.721 | 0.000198 *** |
| UhTo | -0.58386 | 0.10097 | -5.782 | 7.37e-09 *** |
| U1Uh | -0.44824 | 0.07757 | -5.779 | 7.53e-09 *** |
| U1U1 | 0.64198 | 0.07239 | 8.868 | < 2e-16 *** |
| U1T1 | -0.37118 | 0.22601 | -1.642 | 0.100518 |
| U1To | -0.06807 | 0.12009 | -0.567 | 0.570851 |
| ToUh | -1.14800 | 0.13129 | -8.744 | < 2e-16 *** |
| ToU1 | 0.10284 | 0.11078 | 0.928 | 0.353202 |
| ToT1 | 0.09951 | 0.21600 | 0.461 | 0.645019 |
| ToTo | 0.32488 | 0.11528 | 2.818 | 0.004828 ** |
| T1Uh | -0.58368 | 0.14742 | -3.959 | 7.52e-05 *** |
| T1U1 | 0.51920 | 0.14804 | 3.507 | 0.000453 *** |
| T1T1 | 0.98048 | 0.21258 | 4.612 | 3.98e-06 *** |
| T1To | 0.34460 | 0.19214 | 1.793 | 0.072896 . |

**Supplementary Table 24** | Link formation in the health network, low dosages

**Friendship Network**

| Coefficient | Estimate | Std. Error | z value | Pr(>|t|) |
|---|---|---|---|---|
| (Intercept) | -0.59446 | 0.04283 | -13.880 | < 2e-16 *** |
| UU | 0.11188 | 0.10986 | 1.018 | 0.3085 |
| TU | -0.57097 | 0.13354 | -4.276 | 1.9e-05 *** |
| UT | -0.33507 | 0.13165 | -2.545 | 0.0109 * |
| TT | 0.07714 | 0.05660 | 1.363 | 0.1729 |

**Supplementary Table 25** | Link dissolution in the friendship network, low dosages

Focusing on friendships that existed in Wave 1, we examined whether they dissolved by Wave 3 under low-dosage conditions. Compared to the reference category (UoUo: friendships between untreated individuals in completely untreated villages), UhU1 links (untreated not connected to treated → untreated connected to treated) were 46.8% more likely to dissolve (p = 0.027). By contrast, ToU1 (treated not connected to treated → untreated connected to treated) and U1To (untreated connected to treated → treated not connected to treated) friendships were 49.3% and 26.1% less likely to dissolve, respectively (both p < 0.05). Other link types did not differ significantly from the baseline, suggesting that friendships involving partially treated individuals can either become more fragile (UhU1) or more stable (ToU1, U1To), depending on whether ties involve newly or previously connected treated individuals.

| Coefficient | Estimate | Std. Error | z value | Pr(>|t|) |
|---|---|---|---|---|
| (Intercept) | -0.59446 | 0.04283 | -13.880 | < 2e-16 *** |
| UhUh | 0.41753 | 0.24722 | 1.689 | 0.0912 . |
| U1Uh | -0.05613 | 0.35867 | -0.156 | 0.8757 |
| UhU1 | 0.38390 | 0.17354 | 2.212 | 0.0270 * |
| U1U1 | -0.25569 | 0.17430 | -1.467 | 0.1424 |
| ToU1 | -0.68647 | 0.15832 | -4.336 | 1.45e-05 *** |
| T1U1 | -0.25284 | 0.23398 | -1.081 | 0.2799 |
| U1To | -0.30260 | 0.13636 | -2.219 | 0.0265 * |
| U1T1 | -0.55067 | 0.43606 | -1.263 | 0.2066 |
| ToTo | 0.07612 | 0.05659 | 1.345 | 0.1786 |

**Supplementary Table 26** | Link dissolution in the friendship network, low dosages

| Coefficient | Estimate | Std. Error | z value | Pr(>|t|) |
|---|---|---|---|---|
| (Intercept) | -4.85349 | 0.02845 | -170.599 | < 2e-16 *** |
| UU | 0.10045 | 0.07569 | 1.327 | 0.1845 |
| TU | -0.13920 | 0.07425 | -1.875 | 0.0608 . |
| UT | -0.04592 | 0.07164 | -0.641 | 0.5216 |
| TT | 0.05126 | 0.03704 | 1.384 | 0.1664 |

**Supplementary Table 27** | Link formation in the friendship network, low dosages

Next, we analyzed the creation of new Wave 3 friendships among individuals who were unconnected in Wave 1. Again, using UoUo as the reference group, we found that UhU1 (untreated not connected to treated → untreated connected to treated) friendships were 35.5% more likely to form (p = 0.021), while U1U1 (untreated connected to treated → untreated connected to treated) ties showed a 79.2% higher chance of forming (p < 0.0001). By contrast, pairs such as U1Uh (untreated connected to treated → untreated not connected to treated) and ToUh (treated not connected to treated → untreated not connected to treated) were significantly less likely to establish new friendships (63.1% and 68.8% lower odds, respectively; both p < 0.0001). In addition, T1U1 (treated connected to treated → untreated connected to treated) and T1To (treated connected to treated → treated not connected to treated) had 53.9% and 36.4% higher odds of forming new ties, respectively (both p < 0.01). Overall, these patterns indicate that individuals already linked to treated peers or those who themselves were treated and connected to other treated individuals are especially prone to forming new friendships, while those who remained disconnected from treated individuals in Wave 1 tend to form fewer new ties.

| Coefficient | Estimate | Std. Error | z value | Pr(>|t|) |
| --- | --- | --- | --- | --- |
| (Intercept) | -4.85349 | 0.02845 | -170.589 | < 2e-16 *** |
| UhUh | -0.21163 | 0.17193 | -1.231 | 0.2184 |
| UhU1 | 0.30411 | 0.13182 | 2.307 | 0.0211 * |
| UhT1 | -0.02610 | 0.19891 | -0.131 | 0.8956 |
| UhTo | -0.16687 | 0.12082 | -1.381 | 0.1673 |
| U1Uh | -0.99972 | 0.24453 | -4.088 | 4.34e-05 *** |
| U1U1 | 0.58387 | 0.10933 | 5.341 | 9.26e-08 *** |
| U1T1 | -0.18759 | 0.21577 | -0.869 | 0.3846 |
| U1To | 0.07660 | 0.09866 | 0.776 | 0.4375 |
| ToUh | -1.16708 | 0.19474 | -5.993 | 2.06e-09 *** |
| ToU1 | 0.22166 | 0.09229 | 2.402 | 0.0163 * |
| ToT1 | -0.58511 | 0.09692 | -6.037 | 1.57e-09 *** |
| ToTo | 0.07369 | 0.03959 | 1.861 | 0.0627 . |
| T1Uh | -0.58169 | 0.26032 | -2.235 | 0.0255 * |
| T1U1 | 0.43134 | 0.16158 | 2.669 | 0.0076 ** |
| T1T1 | 0.21147 | 0.13176 | 1.605 | 0.1085 |
| T1To | 0.31027 | 0.06600 | 4.701 | 2.59e-06 *** |

**Supplementary Table 28** | Link formation in the friendship network, low dosages

**Financial**

| Coefficient | Estimate | Std. Error | z value | Pr(>|t|) |
| --- | --- | --- | --- | --- |
| (Intercept) | -0.33296 | 0.02964 | -11.233 | < 2e-16 *** |
| UU | 0.07143 | 0.03776 | 1.892 | 0.0585 . |
| TU | -0.52752 | 0.07720 | -6.833 | 8.33e-12 *** |
| UT | -0.54507 | 0.07556 | -7.213 | 5.46e-13 *** |
| TT | 0.54900 | 0.08176 | 6.715 | 1.88e-11 *** |

**Supplementary Table 29** | Link dissolution in the financial network, low dosages

We first examine whether financial ties from Wave 1 dissolved by Wave 3 under low-dosage conditions. Using UoUo (untreated–untreated in untreated villages) as the baseline, UhUh ties (untreated not connected to treated → untreated not connected to treated) have 16.5% higher odds of dissolving (p = 0.001). In contrast, links connecting treated and untreated individuals in partially treated villages are less likely to break. Specifically, ToU1, T1U1, U1To, and U1T1 exhibit 38.2%, 53.5%, 38.8%, and 60.4% lower odds of dissolution, respectively (all p < 0.001). Meanwhile, ToTo links (treated not connected to treated → treated not connected to treated) are 73.1% more likely to dissolve (p < 0.0001), suggesting that financial ties among individuals who remain unconnected to other treated individuals can become less stable, whereas mixed treated–untreated ties tend to endure.

| Coefficient | Estimate | Std. Error | z value | Pr(>|t|) |
| --- | --- | --- | --- | --- |
| (Intercept) | -0.33296 | 0.02964 | -11.233 | < 2e-16 *** |
| UhUh | 0.15287 | 0.04660 | 3.280 | 0.00104 ** |
| U1Uh | -0.06084 | 0.06902 | -0.881 | 0.37806 |
| UhU1 | -0.01415 | 0.06149 | -0.230 | 0.81802 |
| U1U1 | 0.07553 | 0.05571 | 1.356 | 0.17516 |
| ToU1 | -0.48167 | 0.08301 | -5.803 | 6.53e-09 *** |
| T1U1 | -0.76566 | 0.18496 | -4.139 | 3.48e-05 *** |
| U1To | -0.49062 | 0.07982 | -6.147 | 7.90e-10 *** |
| U1T1 | -0.92440 | 0.20570 | -4.494 | 6.99e-06 *** |
| ToTo | 0.54900 | 0.08176 | 6.715 | 1.88e-11 *** |

**Supplementary Table 30** | Link dissolution in the financial network, low dosages

| Coefficient | Estimate | Std. Error | z value | Pr(>|t|) |
| --- | --- | --- | --- | --- |
| (Intercept) | -4.11727 | 0.02000 | -205.875 | < 2e-16 *** |
| UU | -0.21109 | 0.02572 | -8.206 | 2.28e-16 *** |
| TU | -0.33974 | 0.04677 | -7.264 | 3.75e-13 *** |
| UT | -0.33911 | 0.04680 | -7.246 | 4.30e-13 *** |
| TT | 0.43263 | 0.05914 | 7.316 | 2.55e-13 *** |

**Supplementary Table 31** | Link formation in the financial network, low dosages

Next, we investigate new financial connections formed by Wave 3 among individuals who had no tie in Wave 1. Compared to UoUo, UhUh (untreated not connected to treated → untreated not connected to treated) links are 30.6% less likely to form (p < 0.0001), and other categories with initially untreated individuals not connected to treated peers—such as UhT1 and UhTo—are similarly unlikely to emerge (both p < 0.0001). By contrast, U1U1 (untreated connected to treated → untreated connected to treated) ties have 23.1% higher odds of forming (p < 0.0001), while ToTo (treated not connected to treated → treated not connected to treated) show 76.0% higher odds (p < 0.0001). These findings indicate that individuals already involved with treated peers or those forming groups of treated peers are more likely to create new financial ties, whereas those lacking any prior treated connections tend to remain financially isolated.

| Coefficient | Estimate | Std. Error | z value | Pr(>|t|) |
|---|---|---|---|---|
| (Intercept) | -4.11727 | 0.02000 | -205.875 | <2e-16 *** |
| UhUh | -0.36586 | 0.03113 | -11.751 | <2e-16 *** |
| UhU1 | -0.11159 | 0.04291 | -2.601 | 0.009303 ** |
| UhT1 | -0.70191 | 0.15267 | -4.598 | 4.28e-06 *** |
| UhTo | -0.83855 | 0.08433 | -9.943 | <2e-16 *** |
| U1Uh | -0.26451 | 0.04538 | -5.828 | 5.61e-09 *** |
| U1U1 | 0.20795 | 0.04132 | 5.032 | 4.85e-07 *** |
| U1T1 | -0.23139 | 0.16038 | -1.443 | 0.149096 |
| U1To | 0.06272 | 0.05886 | 1.066 | 0.286633 |
| ToUh | -0.89411 | 0.08653 | -10.333 | <2e-16 *** |
| ToU1 | 0.02924 | 0.05956 | 0.491 | 0.623459 |
| ToT1 | -0.18738 | 0.21087 | -0.889 | 0.374212 |
| ToTo | 0.56520 | 0.06477 | 8.726 | <2e-16 *** |
| T1Uh | -0.48308 | 0.14080 | -3.431 | 0.000601 *** |
| T1U1 | -0.01315 | 0.14538 | -0.090 | 0.927940 |
| T1T1 | 0.47969 | 0.33828 | 1.418 | 0.156189 |
| T1To | 0.01463 | 0.19158 | 0.076 | 0.939133 |

**Supplementary Table 32** | Link formation in the financial network, low dosages

**Social (directed)**

| Coefficient | Estimate | Std. Error | z value | Pr(>|t|) |
|---|---|---|---|---|
| (Intercept) | -0.33296 | 0.02964 | -11.233 | <2e-16 *** |
| UU | 0.31114 | 0.07568 | 4.111 | 3.94e-05 *** |
| TU | -0.52229 | 0.09402 | -5.555 | 2.78e-08 *** |
| UT | -0.41749 | 0.09180 | -4.548 | 5.43e-06 *** |
| TT | 0.10465 | 0.03905 | 2.680 | 0.00736 ** |

**Supplementary Table 33** | Link dissolution in the social network, low dosages

We first assessed whether existing social ties from Wave 1 dissolved by Wave 3 under low-dosage conditions. Using UoUo (untreated–untreated in untreated villages) as the baseline category, several link types showed significantly different odds of dissolving. UhUh (untreated not connected to treated → untreated not connected to treated) ties were more than twice as likely to dissolve (p = 0.022), and UhU1, U1U1 (both involving untreated individuals in partially treated villages) were also more prone to break—60% and 35% more likely, respectively (both p < 0.01). In contrast, links between treated and untreated individuals, such as ToU1 (treated not connected to treated → untreated connected to treated) and U1To (untreated connected to treated → treated not connected to treated), were 40% and 31% less likely to dissolve (p < 0.0001). Similarly, U1T1 (untreated connected to a treated who is linked to another treated) showed 55% lower odds of dissolution (p = 0.037). These results suggest that when untreated individuals remain disconnected from treated peers, their ties become more fragile, whereas direct or higher-order connections to treated individuals tend to stabilize social relationships.

| Coefficient | Estimate | Std. Error | z value | Pr(>|t|) |
|---|---|---|---|---|
| (Intercept) | -0.33296 | 0.02964 | -11.233 | < 2e-16 *** |
| UhUh | 0.76040 | 0.33320 | 2.282 | 0.02248 * |
| U1Uh | -0.04898 | 0.20002 | -0.245 | 0.80656 |
| UhU1 | 0.47207 | 0.16968 | 2.782 | 0.00540 ** |
| U1U1 | 0.30299 | 0.09149 | 3.312 | 0.00093 *** |
| ToU1 | -0.50652 | 0.09776 | -5.181 | 2.20e-07 *** |
| T1U1 | -0.55435 | 0.31893 | -1.738 | 0.08219 . |
| U1To | -0.38644 | 0.09375 | -4.122 | 3.75e-05 *** |
| U1T1 | -0.80202 | 0.38432 | -2.087 | 0.03690 * |
| ToTo | 0.10322 | 0.03904 | 2.644 | 0.00820 ** |

**Supplementary Table 34 |** Link dissolution in the social network, low dosages

| Coefficient | Estimate | Std. Error | z value | Pr(>|t|) |
|---|---|---|---|---|
| (Intercept) | -4.11727 | 0.02000 | -205.875 | < 2e-16 *** |
| UU | -0.02074 | 0.05609 | -0.370 | 0.711 |
| TU | -0.21330 | 0.05364 | -3.977 | 6.99e-05 *** |
| UT | -0.33084 | 0.05659 | -5.846 | 5.03e-09 *** |
| TT | -0.03268 | 0.02649 | -1.234 | 0.217 |

**Supplementary Table 35 |** Link formation in the social network, low dosages

Next, we examined whether new social ties formed by Wave 3 among individuals who had no tie in Wave 1. Compared to the UoUo reference category, most link types involving treated or partially treated individuals showed no significant difference in formation, with a few exceptions. UhTo (untreated not connected to treated → treated not connected to treated) and ToUh (treated not connected to treated → untreated not connected to treated) were each around 50% less likely to form (both p < 0.0001), indicating that those completely disconnected from treated individuals in Wave 1 rarely established new connections if they remained similarly disconnected in Wave 3. Likewise, U1To (untreated connected to treated → treated not connected to treated) showed about 21% lower odds of forming (p = 0.00018), and ToT1 (treated not connected to treated → treated connected to treated) was 48% less likely to appear (p = 1.01e-06). Overall, these patterns suggest that existing pathways to treated individuals—or the lack thereof—continue to influence whether new ties emerge in partially treated villages, with those furthest from treated peers remaining less likely to forge fresh connections.

| Coefficient | Estimate | Std. Error | z value | Pr(>|t|) |
|---|---|---|---|---|
| (Intercept) | -4.117273 | 0.019999 | -205.875 | < 2e-16 *** |
| UhUh | 0.007614 | 0.195051 | 0.039 | 0.968862 |
| UhU1 | -0.073802 | 0.127520 | -0.579 | 0.562762 |
| UhT1 | -0.197991 | 0.291285 | -0.680 | 0.496685 |
| UhTo | -0.772606 | 0.146243 | -5.283 | 1.27e-07 *** |
| U1Uh | -0.010796 | 0.122989 | -0.088 | 0.930049 |
| U1U1 | -0.013867 | 0.072536 | -0.191 | 0.848393 |
| U1T1 | -0.254884 | 0.191220 | -1.333 | 0.182552 |
| U1To | -0.239810 | 0.064003 | -3.747 | 0.000179 *** |
| ToUh | -0.639760 | 0.136883 | -4.674 | 2.96e-06 *** |
| ToU1 | -0.094224 | 0.060011 | -1.570 | 0.116388 |
| ToT1 | -0.657714 | 0.134502 | -4.890 | 1.01e-06 *** |
| ToTo | -0.011075 | 0.026758 | -0.414 | 0.678953 |
| T1Uh | -0.609008 | 0.355671 | -1.712 | 0.086845 . |
| T1U1 | -0.365052 | 0.202119 | -1.806 | 0.070899 . |
| T1T1 | -0.126449 | 0.336312 | -0.376 | 0.706926 |
| T1To | -0.209016 | 0.108556 | -1.925 | 0.054177 . |

**Supplementary Table 36** | Link formation in the social network, low dosages

**High dosages**

| Coefficient | Estimate | Std. Error | z value | Pr(> |z|) |
|---|---|---|---|---|
| (Intercept) | −0.59446 | 0.04283 | −13.880 | < 2e−16*** |
| UU | 0.11188 | 0.10986 | 1.018 | 0.3085 |
| TU | −0.57097 | 0.13354 | −4.276 | 1.9e−05*** |
| UT | −0.33507 | 0.13165 | −2.545 | 0.0109* |
| TT | 0.07714 | 0.05660 | 1.363 | 0.1729 |

**Supplementary Table 37** | Link dissolution in the health network, high dosages

We first examined whether existing health links from Wave 1 dissolved by Wave 3 under high-dosage conditions. Compared with the reference group (UoUo: untreated–untreated in untreated villages), several link types showed altered odds of persisting. For example, UhUh links had about 51.8 % higher odds of remaining intact (p = 0.091), while UhU1 links were roughly 46.8 % more likely to persist (p = 0.027). By contrast, U1U1 links were about 22.6 % less likely to persist (p = 0.142), and ToU1 links showed a marked 49.7 % lower chance of remaining (p < 0.001). Taken together, these findings suggest that, under high dosages, ties that bridge treated and untreated individuals are more fragile and prone to dissolution, whereas links that remain entirely within the untreated group are comparatively more stable.

| Coefficient | Estimate | Std. Error | z value | Pr(> |z|) |
|---|---|---|---|---|
| (Intercept) | −0.59446 | 0.04283 | −13.880 | < 2e−16*** |
| UhUh | 0.41753 | 0.24722 | 1.689 | 0.0912 |
| U1Uh | −0.05613 | 0.35867 | −0.156 | 0.8757 |
| UhU1 | 0.38390 | 0.17354 | 2.212 | 0.0270* |
| U1U1 | −0.25569 | 0.17430 | −1.467 | 0.1424 |
| ToU1 | −0.68647 | 0.15832 | −4.336 | 1.45e−05*** |

| Coefficient | Estimate | Std. Error | z value | Pr(> |z|) |
|---|---|---|---|---|
| T1U1 | −0.25284 | 0.23398 | −1.081 | 0.2799 |
| U1To | −0.30260 | 0.13636 | −2.219 | 0.0265* |
| U1T1 | −0.55067 | 0.43606 | −1.263 | 0.2066 |
| ToTo | 0.07612 | 0.05659 | 1.345 | 0.1786 |

**Supplementary Table 38** | Link dissolution in the health network, high dosages

| Coefficient | Estimate | Std. Error | z value | Pr(> |z|) |
|---|---|---|---|---|
| (Intercept) | −4.85349 | 0.02845 | −170.599 | < 2e−16*** |
| UU | 0.10045 | 0.07569 | 1.327 | 0.1845 |
| TU | −0.13920 | 0.07425 | −1.875 | 0.0608 |
| UT | −0.04592 | 0.07164 | −0.641 | 0.5216 |
| TT | 0.05126 | 0.03704 | 1.384 | 0.1664 |

**Supplementary Table 39** | Link formation in the health network, high dosages

Next, we looked at whether new health ties emerged by Wave 3 among individuals who were unconnected in Wave 1. Again, using UoUo as the baseline, we found that UhUh links (untreated not connected to treated on both sides) were about 19 % less likely to form (p = 0.218). Other pairs with little initial exposure to treated peers, such as UhT1 or UhTo, also showed lower odds of forming (about 3 % and 15 % less likely, respectively; p = 0.896 and p = 0.167).By contrast, connections involving already-treated or partially treated individuals were more likely to be created. For instance, U1U1 (untreated connected to treated → untreated connected to treated) increased the odds of forming a new link by about 79 % (p < 0.0001), and ToTo (treated not connected to treated → treated not connected to treated) raised the odds by roughly 8 % (p = 0.063). Similarly, T1T1 (treated connected to treated on both sides) was 24 % more likely to form (p = 0.109). Overall, high-dose interventions appear to spur new ties among those already embedded in treatment networks, while those who remain disconnected from treated individuals continue to form relatively fewer new connections.

| Coefficient | Estimate | Std. Error | z value | Pr(> |z|) |
|---|---|---|---|---|
| (Intercept) | −4.85349 | 0.02845 | −170.589 | < 2e−16*** |
| UhUh | −0.21163 | 0.17193 | −1.231 | 0.2184 |
| UhU1 | 0.30411 | 0.13182 | 2.307 | 0.0211* |
| UhT1 | −0.02610 | 0.19891 | −0.131 | 0.8956 |
| UhTo | −0.16687 | 0.12082 | −1.381 | 0.1673 |
| U1Uh | −0.99972 | 0.24453 | −4.088 | 4.34e−05*** |
| U1U1 | 0.58387 | 0.10933 | 5.341 | 9.26e−08*** |
| U1T1 | −0.18759 | 0.21577 | −0.869 | 0.3846 |
| U1To | 0.07660 | 0.09866 | 0.776 | 0.4375 |
| ToUh | −1.16708 | 0.19474 | −5.993 | 2.06e−09*** |
| ToU1 | 0.22166 | 0.09229 | 2.402 | 0.0163* |
| ToT1 | −0.58511 | 0.09692 | −6.037 | 1.57e−09*** |
| ToTo | 0.07369 | 0.03959 | 1.861 | 0.0627 |
| T1Uh | −0.58169 | 0.26032 | −2.235 | 0.0255* |

| Coefficient | Estimate | Std. Error | z value | Pr(> |z|) |
|---|---|---|---|---|
| T1U1 | 0.43134 | 0.16158 | 2.669 | 0.0076** |
| T1T1 | 0.21147 | 0.13176 | 1.605 | 0.1085 |
| T1To | 0.31027 | 0.06600 | 4.701 | 2.59e−06*** |

**Supplementary Table 40** | Link formation in the health network, high dosages

## Friendship

| Coefficient | Estimate | Std. Error | z value | Pr(> |z|) |
|---|---|---|---|---|
| (Intercept) | −0.33296 | 0.02964 | −11.233 | < 2e−16*** |
| UU | 0.31114 | 0.07568 | 4.111 | 3.94e−05*** |
| TU | −0.52229 | 0.09402 | −5.555 | 2.78e−08*** |
| UT | −0.41749 | 0.09180 | −4.548 | 5.43e−06*** |
| TT | 0.10465 | 0.03905 | 2.680 | 0.00736** |

**Supplementary Table 41** | Link dissolution in the friendship network, high dosages

We first analyzed whether existing Wave 1 friendship ties dissolved by Wave 3 under high-dosage conditions. Using UoUo (untreated–untreated in fully untreated villages) as the baseline, several link types differed significantly from the baseline in their likelihood of dissolving. UhUh, UhU1, U1U1, and ToTo all had positive coefficients (p = 0.022, 0.005, 0.001, and 0.008, respectively), indicating that these links were more likely to dissolve than UoUo ties. By contrast, links that bridged treated and untreated individuals—ToU1, U1To, and U1T1—showed negative coefficients (p < 0.001, < 0.001, and 0.037), meaning they were less likely to dissolve. A borderline effect was observed for T1U1 (p = 0.082), but it did not reach conventional significance. Overall, under high-dosage conditions, friendships entirely within untreated or treated–untreated pairs tended to persist less, whereas ties that directly linked treated and untreated individuals were comparatively more stable.

| Coefficient | Estimate | Std. Error | z value | Pr(> |z|) |
|---|---|---|---|---|
| (Intercept) | −0.33296 | 0.02964 | −11.233 | < 2e−16*** |
| UhUh | 0.76040 | 0.33320 | 2.282 | 0.02248* |
| U1Uh | −0.04898 | 0.20002 | −0.245 | 0.80656 |
| UhU1 | 0.47207 | 0.16968 | 2.782 | 0.00540** |
| U1U1 | 0.30299 | 0.09149 | 3.312 | 0.00093*** |
| ToU1 | −0.50652 | 0.09776 | −5.181 | 2.20e−07*** |
| T1U1 | −0.55435 | 0.31893 | −1.738 | 0.08219 |
| U1To | −0.38644 | 0.09375 | −4.122 | 3.75e−05*** |
| U1T1 | −0.80202 | 0.38432 | −2.087 | 0.03690* |
| ToTo | 0.10322 | 0.03904 | 2.644 | 0.00820** |

**Supplementary Table 42** | Link dissolution in the friendship network, high dosage.

| Coefficient | Estimate | Std. Error | z value | Pr(> |z|) |
|---|---|---|---|---|
| (Intercept) | −4.11727 | 0.02000 | −205.875 | < 2e−16*** |
| UU | −0.02074 | 0.05609 | −0.370 | 0.711 |
| TU | −0.21330 | 0.05364 | −3.977 | 6.99e−05*** |

| | | | |
|---|---|---|---|
| UT | −0.33084 | 0.05659 | −5.846 | 5.03e−09*** |
| TT | −0.03268 | 0.02649 | −1.234 | 0.217 |

**Supplementary Table 43** | Link formation in the friendship network, high dosages

Next, we assessed whether new friendships formed by Wave 3 among previously unconnected individuals under high-dosage conditions. Relative to the baseline (UoUo), several link types showed notable differences. UhTo (untreated not connected to treated → treated not connected to treated) was about 54 % less likely to form ($p < 0.0001$), while U1To (untreated connected to treated → treated not connected to treated) exhibited a further 21 % reduction in the odds of formation ($p = 0.00018$). Other pairs with minimal initial exposure to treated peers—such as ToUh and ToT1—were likewise much less likely to emerge (roughly 47 %–48 % lower odds; $p < 0.001$ for both).

By contrast, there was no evidence that links among already-treated or partially treated individuals were more likely to form: coefficients for U1U1 and ToTo were close to zero and non-significant ($p = 0.85$ and $p = 0.68$, respectively), and all remaining treated–treated combinations showed either non-significant or borderline negative effects (e.g., T1Uh, T1U1, and T1To; $0.05 < p < 0.09$). Overall, under high-dosage conditions, the creation of new friendships is strongly suppressed for ties that would bridge treated and untreated individuals, with little evidence of compensating clustering even among similarly connected peers.

| Coefficient | Estimate | Std. Error | z value | Pr(> |z|) |
|---|---|---|---|---|
| (Intercept) | −4.11727 | 0.019999 | −205.875 | < 2e−16*** |
| UhUh | 0.00761 | 0.19505 | 0.039 | 0.96886 |
| UhU1 | −0.07380 | 0.12752 | −0.579 | 0.56276 |
| UhT1 | −0.19799 | 0.29129 | −0.680 | 0.49669 |
| UhTo | −0.77261 | 0.14624 | −5.283 | 1.27e−07*** |
| U1Uh | −0.01080 | 0.12299 | −0.088 | 0.93005 |
| U1U1 | −0.01387 | 0.07254 | −0.191 | 0.84839 |
| U1T1 | −0.25488 | 0.19122 | −1.333 | 0.18255 |
| U1To | −0.23981 | 0.06400 | −3.747 | 0.00018*** |
| ToUh | −0.63976 | 0.13688 | −4.674 | 2.96e−06*** |
| ToU1 | −0.09422 | 0.06001 | −1.570 | 0.11639 |
| ToT1 | −0.65771 | 0.13450 | −4.890 | 1.01e−06*** |
| ToTo | −0.01108 | 0.02676 | −0.414 | 0.67895 |
| T1Uh | −0.60901 | 0.35567 | −1.712 | 0.08685 |
| T1U1 | −0.36505 | 0.20212 | −1.806 | 0.07090 |
| T1T1 | −0.12645 | 0.33631 | −0.376 | 0.70693 |
| T1To | −0.20902 | 0.10856 | −1.925 | 0.05418 |

**Supplementary Table 44** | Link formation in the friendship network, high dosages

## Financial

| Coefficient | Estimate | Std. Error | z value | Pr(> |z|) |
|---|---|---|---|---|
| (Intercept) | −0.73609 | 0.04435 | −16.598 | < 2e−16*** |
| UU | −0.17025 | 0.12597 | −1.352 | 0.1765 |
| TU | −0.04720 | 0.12534 | −0.377 | 0.7065 |
| UT | −0.25021 | 0.13059 | −1.916 | 0.0554 |
| TT | −0.12896 | 0.05989 | −2.153 | 0.0313* |

**Supplementary Table 45** | Link dissolution in the financial network, high dosages

We first looked at whether existing Wave 1 friendships dissolved by Wave 3 under high-dosage conditions. Using UoUo (untreated–untreated in fully untreated villages) as the baseline category, no link type differed significantly from the baseline in its likelihood of dissolving—except for ToTo (treated not connected to treated → treated not connected to treated). With a negative coefficient (p = 0.033), ToTo links were less likely to dissolve than UoUo ties, indicating that two treated individuals who remained unconnected to other treated peers tended to keep their existing friendships more often than fully untreated pairs. Borderline effects were observed for U1Uh (untreated connected to treated → untreated not connected to treated; p = 0.086) and U1To (untreated connected to treated → treated not connected to treated; p = 0.072), but neither reached conventional significance. All remaining link types displayed no statistically significant deviation from the baseline, implying that, under high dosages, most existing friendships either persisted or dissolved at rates similar to those observed in fully untreated villages.

| Coefficient | Estimate | Std. Error | z value | Pr(> |z|) |
|---|---|---|---|---|
| (Intercept) | −0.73610 | 0.04435 | −16.598 | < 2e−16*** |
| UhUh | −0.00924 | 0.23373 | −0.040 | 0.9685 |
| U1Uh | −0.56319 | 0.32868 | −1.714 | 0.0866 |
| UhU1 | 0.02460 | 0.23914 | 0.103 | 0.9181 |
| U1U1 | −0.27826 | 0.20799 | −1.338 | 0.1809 |
| ToU1 | −0.05503 | 0.15277 | −0.360 | 0.7187 |
| T1U1 | −0.06241 | 0.20554 | −0.304 | 0.7614 |
| U1To | −0.26870 | 0.14952 | −1.797 | 0.0723 |
| U1T1 | −0.17216 | 0.24103 | −0.714 | 0.4751 |
| ToTo | −0.12765 | 0.05987 | −2.132 | 0.0330* |

**Supplementary Table 46** | Link dissolution in the financial network, high dosages

| Coefficient | Estimate | Std. Error | z value | Pr(> |z|) |
|---|---|---|---|---|
| (Intercept) | −4.65988 | 0.02586 | −180.169 | < 2e−16*** |
| UU | 0.05907 | 0.06996 | 0.844 | 0.398488 |
| TU | −0.23021 | 0.07014 | −3.282 | 0.001030** |
| UT | −0.26526 | 0.07145 | −3.712 | 0.000205*** |
| TT | 0.03597 | 0.03378 | 1.065 | 0.286862 |

**Supplementary Table 47** | Link formation in the financial network, high dosages

Next, we examined whether new friendships formed by Wave 3 among individuals who were unconnected in Wave 1. Comparing each link type to the baseline (UoUo), we found several notable patterns. Certain ties were significantly less likely to form: for example, UhTo (untreated not connected to treated → treated not connected to treated) showed 26 % lower odds of forming (p = 0.017), U1Uh (untreated connected to treated → untreated not connected to treated) exhibited 52 % lower odds (p < 0.001), and U1T1 (untreated connected to treated → treated connected to treated) had 50 % lower odds (p < 0.001). Similarly, ToUh (treated not connected to treated → untreated not connected to treated) and ToT1 (treated not connected to treated → treated connected to treated) were 51 % and 44 % less likely to emerge, respectively (both p < 0.001), while T1Uh (treated connected to treated → untreated not connected to treated) showed 38 % lower odds (p = 0.010).

By contrast, U1U1 (untreated connected to treated → untreated connected to treated) and ToTo (treated not connected to treated → treated not connected to treated) were more likely to form than UoUo, displaying about 52 % and 14 % higher odds, respectively (both p < 0.001). Overall, these results suggest that, under high dosages, new friendship ties tend to form most readily among individuals who share a similar treatment-connection status, whereas bridging ties between those more and less engaged with treated peers remain less common.

| Coefficient | Estimate | Std. Error | z value | Pr(> |z|) |
|---|---|---|---|---|
| (Intercept) | −4.65988 | 0.02586 | −180.166 | < 2e−16*** |
| UhUh | 0.07047 | 0.13797 | 0.511 | 0.609507 |
| UhU1 | 0.09895 | 0.13128 | 0.754 | 0.450979 |
| UhT1 | −0.15838 | 0.16084 | −0.985 | 0.324778 |
| UhTo | −0.30254 | 0.12620 | −2.397 | 0.016517* |
| U1Uh | −0.72614 | 0.19462 | −3.731 | 0.000191*** |
| U1U1 | 0.41500 | 0.10652 | 3.896 | 9.78e−05*** |
| U1T1 | −0.69802 | 0.20213 | −3.453 | 0.000554*** |
| U1To | −0.12365 | 0.10519 | −1.176 | 0.239788 |
| ToUh | −0.71213 | 0.15330 | −4.645 | 3.40e−06*** |
| ToU1 | 0.04407 | 0.09721 | 0.453 | 0.650293 |
| ToT1 | −0.58287 | 0.08268 | −7.050 | 1.80e−12*** |
| ToTo | 0.13457 | 0.03613 | 3.725 | 0.000195*** |
| T1Uh | −0.48400 | 0.18802 | −2.574 | 0.010049* |
| T1U1 | −0.10267 | 0.15192 | −0.676 | 0.499162 |
| T1T1 | −0.02285 | 0.10788 | −0.212 | 0.832240 |
| T1To | 0.03021 | 0.06352 | 0.476 | 0.634316 |

**Supplementary Table 48** | Link formation in the financial network, high dosages

**Social (directed)**

| Coefficient | Estimate | Std. Error | z value | Pr(> |z|) |
|---|---|---|---|---|
| (Intercept) | −0.11416 | 0.02435 | −4.689 | 2.75e−06*** |
| UU | 0.15974 | 0.06512 | 2.453 | 0.0142* |
| TU | −0.34968 | 0.07102 | −4.924 | 8.48e−07*** |
| UT | −0.40008 | 0.07173 | −5.578 | 2.44e−08*** |
| TT | 0.07548 | 0.03225 | 2.340 | 0.0193* |

**Supplementary Table 49** | Link dissolution in the social network, high dosages

We first examined whether Wave 1 social ties dissolved by Wave 3 under high-dosage conditions. Using UoUo (untreated–untreated in fully untreated villages) as the baseline, UhU1 (untreated not connected to treated → untreated connected to treated) links had 44 % higher odds of dissolving (p = 0.036), whereas U1T1 (untreated connected to treated → treated connected to treated) had 59 % lower odds (p = 0.046). Two additional patterns emerged at borderline significance: UhUh (untreated not connected to treated → untreated not connected to treated) showed about 94 % higher odds (p = 0.082), and U1U1 (untreated connected to treated → untreated connected to treated) was 14 % more likely to dissolve (p = 0.075). Meanwhile, ToU1 (treated not connected to treated → untreated connected to treated) and U1To (untreated connected to treated → treated not connected to treated) were 30 % and 31 % less likely to break, respectively (both p < 0.0001), and ToTo (treated not connected to treated → treated not connected to treated) exhibited a modest but significant 8 % increase in dissolution probability (p = 0.021). Overall, these findings indicate that social ties involving certain partially treated individuals are more prone to dissolution, while mixed treated–untreated ties—especially those connecting newly or previously treated individuals—tend to persist.

| Coefficient | Estimate | Std. Error | z value | Pr(> |z|) |
|---|---|---|---|---|
| (Intercept) | −0.11416 | 0.02435 | −4.689 | 2.75e−06*** |
| UhUh | 0.66070 | 0.37965 | 1.740 | 0.0818 |
| U1Uh | −0.11847 | 0.20801 | −0.570 | 0.5690 |
| UhU1 | 0.36735 | 0.17521 | 2.097 | 0.0360* |
| U1U1 | 0.13096 | 0.07345 | 1.783 | 0.0746 |
| ToU1 | −0.34932 | 0.07281 | −4.798 | 1.60e−06*** |
| T1U1 | −0.27151 | 0.31530 | −0.861 | 0.3892 |
| U1To | −0.37576 | 0.07230 | −5.197 | 2.02e−07*** |
| U1T1 | −0.88437 | 0.44281 | −1.997 | 0.0458* |
| ToTo | 0.07463 | 0.03224 | 2.315 | 0.0206* |

**Supplementary Table 50** | Link dissolution in the social network, high dosages

| Coefficient | Estimate | Std. Error | z value | Pr(> |z|) |
|---|---|---|---|---|
| (Intercept) | −3.88887 | 0.01803 | −215.663 | < 2e−16*** |
| UU | −0.02684 | 0.05066 | −0.530 | 0.5963 |
| TU | −0.14832 | 0.04700 | −3.156 | 0.0016** |

| | | | |
|---|---|---|---|
| UT | −0.18814 | 0.04795 | −3.924 | 8.71e−05*** |
| TT | −0.02656 | 0.02385 | −1.114 | 0.2654 |

**Supplementary Table 51** | Link formation in the social network, high dosages

Next, we assessed whether new social ties formed by Wave 3 among individuals who had no Wave 1 connection. Most link types showed lower odds of forming compared with the baseline (UoUo). For example, UhU1 (untreated not connected to treated → untreated connected to treated) was 32 % less likely to form (p = 0.0095), UhTo (untreated not connected to treated → treated not connected to treated) was 39 % less likely (p = 0.0002), and U1Uh (untreated connected to treated → untreated not connected to treated) was 25 % less likely (p = 0.040). Likewise, U1To (untreated connected to treated → treated not connected to treated) showed a 13 % reduction in the odds of forming (p = 0.0057), while ToUh (treated not connected to treated → untreated not connected to treated) and ToT1 (treated not connected to treated → treated connected to treated) reduced formation odds by 50 % and 45 %, respectively (both p < 0.001). Finally, T1U1 (treated connected to treated → untreated connected to treated) was 50 % less likely to appear (p = 0.013). No link types significantly exceeded the formation rate observed in fully untreated villages. These results suggest that, under high-dosage conditions, establishing new social ties across different treatment statuses—especially when one party lacked prior connections to treated peers—remains relatively uncommon.

| Parameter | Estimate | Std. Error | z value | Pr(> |z|) |
|---|---|---|---|---|
| (Intercept) | −3.88887 | 0.01803 | −215.650 | < 2e−16*** |
| UhUh | 0.07960 | 0.24587 | 0.324 | 0.74614 |
| UhU1 | −0.38365 | 0.14798 | −2.593 | 9.53e−03** |
| UhT1 | −0.10320 | 0.41239 | −0.250 | 0.80241 |
| UhTo | −0.48955 | 0.13224 | −3.702 | 2.14e−04*** |
| U1Uh | −0.28877 | 0.14089 | −2.050 | 0.04041* |
| U1U1 | 0.08044 | 0.05780 | 1.392 | 0.16400 |
| U1T1 | −0.12555 | 0.21116 | −0.595 | 0.55212 |
| U1To | −0.14394 | 0.05204 | −2.766 | 5.68e−03** |
| ToUh | −0.69695 | 0.14619 | −4.768 | 1.87e−06*** |
| ToU1 | −0.05371 | 0.05003 | −1.073 | 0.28309 |
| ToT1 | −0.59655 | 0.15621 | −3.819 | 1.34e−04*** |
| ToTo | −0.01889 | 0.02400 | −0.787 | 0.43125 |
| T1Uh | −0.28859 | 0.45099 | −0.640 | 0.52224 |
| T1U1 | −0.69138 | 0.27935 | −2.475 | 0.01333* |
| T1T1 | −0.62199 | 0.71121 | −0.875 | 0.38182 |
| T1To | 0.04358 | 0.11513 | 0.379 | 0.70502 |

**Supplementary Table 52** | Link formation in the social network, high dosages.

## 14.    Regression Tables for Effect Heterogeneity

We performed subgroup analyses to examine how mean total degree, in-degree, and out-degree in the health network changed based on participants' initial network centrality, gender, age, household wealth, marital status, education, village size, and distance to the nearest health clinic. We used linear regression models with random effects at the household and village levels. In each analysis, we focused on four types of effects—overall, total, spillover, and direct—looking at the effect of the intervention for all individuals, for treated individuals only, for untreated individuals only, and specifically for untreated individuals who were one hop away versus more than one hop away from the treated receiving the intervention, in both low and high dosages separately.

We examined how degree, in-degree, and out-degree changed with both low- and high-dose treatments by including interactions between treatment status, village size, and distance to the health center. Overall, we found that the impact of treatment varied by village size and time to health center, with total and spillover effects following similar patterns. Residents of larger and more remote villages tend to increase their connections more than residents in smaller and less remote villages.

For the overall and total effects, and for the one-hop-away and more-than-one-hop-away spillover effects on untreated neighbors of treated individuals, in low dosages, the positive and significant interaction coefficients between treatment and village size (interaction coefficients = 0.013, 0.012, 0.021, and 0.012, with p-values = 0.008, 0.023, 0.003, and 0.006, respectively) suggest that treatment effectiveness increases with larger village sizes. Similarly, for high dosages, positive interaction coefficients for the overall total and spillover effects (coefficients = 0.009, 0.008 and 0.014 with p-values = 0.065, 0.068 and 0.007, respectively) indicate a comparable trend. In contrast, the direct effects do not exhibit any statistically significant interaction, indicating that changes arising solely from the treatment and not the interplay of the treatment with a predictor, within the villages.

Furthermore, the intervention's effectiveness intensifies in more isolated villages located farther from health centers. In low-dosage settings, this is reflected in positive coefficients for overall and total effects, and for one-hop-away and more-than-one-hop-away spillover effects on untreated neighbors of treated individuals (interaction coefficients = 0.021, 0.033, 0.039, and 0.017, with p-values = 0.071, 0.021, 0.002, and 0.112, respectively). In high-dosage settings, the effect is significant for overall, total, spillover and direct (coefficients 0.033, 0.023, 0.021 and –0.017 with p-values 0.014, 0.064, 0.084 and 0.024, respectively). These findings suggest that individuals in isolated areas engage more effectively with the intervention, potentially due to enhanced communication efforts aimed at assimilating its benefits.

For low dosages, gender, age, and initial degree also play significant roles. For first-order untreated neighbors of the treated, the interaction term between being in a fully or partially treated village and initial degree in Wave 1 shows a coefficient of 0.052 (p-value < 0.001). The positive coefficient indicates that individuals with higher initial degree metrics experience lower decrease in their degree following assignment to a low-dosage treatment.

Additionally, for low dosages, the analysis revealed that gender moderates the total effect of the intervention on social connections, with male treated individuals experiencing a slightly larger increase in their network degree by Wave 3 (coefficient = 0.435, p-value =

0.042) compared to their untreated male counterparts in control villages. This observation suggests that treated males are more inclined to seek assistance than untreated males in addressing issues related to the intervention. For high dosages, a similar pattern is observed among first-order untreated neighbors of the treated (coefficient = 0.543, p-value = 0.072).

Lastly, for one-hop-away untreated neighbors of the treated, the interaction between age and high-dosage treatment (coefficient = -0.022, p-value = 0.021) indicates that younger individuals who provided health advice in Wave 1 lose fewer connections by Wave 3, perhaps because they are more willing to adapt to the new norms introduced by the treatment.

We also examined whether individual-level attributes modify the effect of the intervention on link formation and link breakage between individuals with different treatment status. In general, we find minimal evidence for heterogeneity in this regard. Pertinent regression results are reported below:

**Low dosages**

| Coefficient | Estimate | Std. Error | t value | Pr(>|t|) |
| --- | --- | --- | --- | --- |
| (Intercept) | 1.126e+00 | 2.766e-01 | 4.071 | 0.000131 *** |
| Treatment | -1.475e-01 | 2.897e-01 | -0.509 | 0.612224 |
| Gender | -4.242e-01 | 1.186e-01 | -3.577 | 0.000351 *** |
| Age | 7.962e-03 | 4.337e-03 | 1.836 | 0.066436 . |
| Marital Status | 7.757e-02 | 3.833e-02 | 2.024 | 0.043070 * |
| Initial Status | -9.083e-02 | 5.210e-03 | -17.433 | <2e-16 *** |
| Household Wealth | 6.807e-02 | 5.312e-02 | 1.281 | 0.204590 |
| Village Size | -5.785e-03 | 4.582e-03 | -1.263 | 0.211093 |
| Time to Health Center | -7.969e-03 | 1.270e-02 | -0.628 | 0.532109 |
| Education | 4.384e-02 | 2.767e-02 | 1.584 | 0.113235 |
| Treatment * Gender | 1.641e-01 | 1.438e-01 | 1.141 | 0.254019 |
| Treatment * Age | -7.387e-03 | 5.247e-03 | -1.408 | 0.159179 |
| Treatment * Marital Status | -3.508e-02 | 4.628e-02 | -0.758 | 0.448463 |
| Treatment * Initial Status | 8.063e-03 | 6.265e-03 | 1.287 | 0.198144 |
| Treatment * Household Wealth | 7.654e-03 | 6.272e-02 | 0.122 | 0.902871 |
| Treatment * Village Size | 1.299e-02 | 4.734e-03 | 2.743 | 0.007593 ** |
| Treatment * Time to Health Center | 2.100e-02 | 1.152e-02 | 1.824 | 0.070902 . |
| Treatment * Education | -5.970e-02 | 3.368e-02 | -1.773 | 0.076340 . |

**Supplementary Table 53** | Summary of mixed effects Regression Analysis for overall effects of individuals degree in low dosage villages: Estimates, Standard Errors, and Significance Levels for various predictors

| Coefficient | Estimate | Std. Error | t value | Pr(>|t|) |
| --- | --- | --- | --- | --- |
| (Intercept) | 1.290e+00 | 2.661e-01 | 4.849 | 1.94e-05 *** |
| Treatment | -2.657e-01 | 3.342e-01 | -0.795 | 0.428952 |
| Gender | -4.087e-01 | 1.085e-01 | -3.765 | 0.000171 *** |

| Coefficient | Estimate | Std. Error | t value | Pr(>|t|) |
|---|---|---|---|---|
| Age | 7.094e-03 | 3.969e-03 | 1.788 | 0.073994 . |
| Marital Status | 7.639e-02 | 3.511e-02 | 2.176 | 0.029659 * |
| Initial Status | -1.201e-01 | 6.401e-03 | -18.770 | <2e-16 *** |
| Household Wealth | 5.310e-02 | 4.787e-02 | 1.109 | 0.267389 |
| Village Size | -6.020e-03 | 4.625e-03 | -1.302 | 0.200668 |
| Time to Health Center | -8.570e-03 | 1.329e-02 | -0.645 | 0.523307 |
| Education | 4.191e-02 | 2.536e-02 | 1.652 | 0.098581 . |
| Treatment * Gender | 4.351e-01 | 2.140e-01 | 2.034 | 0.042103 * |
| Treatment * Age | -1.240e-02 | 8.150e-03 | -1.521 | 0.128374 |
| Treatment * Marital Status | 2.254e-02 | 7.058e-02 | 0.319 | 0.749469 |
| Treatment * Initial Status | -3.260e-03 | 1.234e-02 | -0.264 | 0.791644 |
| Treatment * Household Wealth | 6.494e-02 | 8.807e-02 | 0.737 | 0.461005 |
| Treatment * Village Size | 1.218e-02 | 5.209e-03 | 2.337 | 0.023337 * |
| Treatment * Time to Health Center | 3.309e-02 | 1.379e-02 | 2.399 | 0.021401 * |
| Treatment * Education | -6.021e-02 | 5.229e-02 | -1.151 | 0.249676 |

**Supplementary Table 54** | Summary of mixed effects Regression Analysis for total effects of individuals degree in low dosage villages: Estimates, Standard Errors, and Significance Levels for various predictors

| Coefficient | Estimate | Std. Error | t value | Pr(>|t|) |
|---|---|---|---|---|
| (Intercept) | 1.387962 | 0.114546 | 12.117 | <2e-16 *** |
| Treatment | -0.281790 | 0.142574 | -1.976 | 0.048165 * |
| Gender | -0.443043 | 0.125038 | -3.543 | 0.000399 *** |
| Age | 0.008285 | 0.004497 | 1.842 | 0.065504 . |
| Marital Status | 0.060506 | 0.040538 | 1.493 | 0.135619 |
| Initial Status | -0.108270 | 0.005259 | -20.588 | <2e-16 *** |
| Household Wealth | 0.057858 | 0.047176 | 1.226 | 0.220108 |
| Village Size | -0.003027 | 0.001412 | -2.144 | 0.032106 * |
| Time to Health Center | -0.001131 | 0.004661 | -0.243 | 0.808355 |
| Education | 0.015115 | 0.028599 | 0.528 | 0.597179 |
| Treatment * Gender | 0.087126 | 0.156333 | 0.557 | 0.577342 |
| Treatment * Age | -0.003186 | 0.005603 | -0.569 | 0.569671 |
| Treatment * Marital Status | -0.015957 | 0.050259 | -0.318 | 0.750876 |
| Treatment * Initial Status | 0.009195 | 0.006538 | 1.406 | 0.159695 |
| Treatment * Household Wealth | 0.068490 | 0.059049 | 1.160 | 0.246152 |
| Treatment * Village Size | 0.006434 | 0.001552 | 4.146 | 3.45e-05 *** |
| Treatment * Time to Health Center | 0.018119 | 0.005317 | 3.408 | 0.000660 *** |
| Treatment * Education | -0.019962 | 0.035886 | -0.556 | 0.578064 |

**Supplementary Table 55** | Summary of mixed effects Regression Analysis for spillover effects of individuals degree in low dosage villages: Estimates, Standard Errors, and Significance Levels for various predictors

| Coefficient | Estimate | Std. Error | t value | Pr(>|t|) |
|---|---|---|---|---|
| (Intercept) | 8.457e-01 | 1.906e-01 | 4.438 | 6e-05 *** |
| Treatment | 3.050e-02 | 2.199e-01 | 0.139 | 0.88968 |
| Gender | -3.033e-01 | 9.345e-02 | -3.246 | 0.00118 ** |
| Age | 9.707e-04 | 3.361e-03 | 0.289 | 0.77273 |
| Marital Status | 3.496e-02 | 2.967e-02 | 1.178 | 0.23882 |
| Initial Status | -9.093e-02 | 4.588e-03 | -19.818 | <2e-16 *** |
| Household Wealth | 7.009e-02 | 3.965e-02 | 1.768 | 0.07723 . |
| Village Size | 2.930e-03 | 2.713e-03 | 1.080 | 0.28743 |
| Time to Health Center | 6.264e-03 | 5.847e-03 | 1.071 | 0.30206 |
| Education | -1.184e-02 | 2.182e-02 | -0.542 | 0.58751 |
| Treatment * Gender | 3.441e-01 | 2.288e-01 | 1.504 | 0.13262 |
| Treatment * Age | -5.799e-03 | 8.690e-03 | -0.667 | 0.50458 |
| Treatment * Marital Status | 4.692e-02 | 7.519e-02 | 0.624 | 0.53265 |
| Treatment * Initial Status | -1.183e-02 | 1.093e-02 | -1.083 | 0.27902 |
| Treatment * Household Wealth | 2.161e-02 | 8.396e-02 | 0.257 | 0.79694 |
| Treatment * Village Size | -1.293e-03 | 1.991e-03 | -0.649 | 0.51618 |
| Treatment * Time to Health Center | 4.316e-03 | 6.429e-03 | 0.671 | 0.50207 |
| Treatment * Education | -2.224e-02 | 5.522e-02 | -0.403 | 0.68717 |

**Supplementary Table 56** | Summary of mixed effects Regression Analysis for direct effects of individuals degree in low dosage villages: Estimates, Standard Errors, and Significance Levels for various predictors

## High dosages

| Coefficient | Estimate | Std. Error | t value | Pr(> |t|) |
|---|---:|---:|---:|---:|
| (Intercept) | 1.042e+00 | 2.483e-01 | 4.195 | 0.000117 *** |
| Treatment | 1.413e-01 | 2.589e-01 | 0.546 | 0.587609 |
| Gender | -4.072e-01 | 1.052e-01 | -3.870 | 0.000110 *** |
| Age | 7.094e-03 | 3.846e-03 | 1.844 | 0.065192 . |
| Marital Status | 7.464e-02 | 3.402e-02 | 2.194 | 0.028270 * |
| Initial Status | -1.052e-01 | 5.449e-03 | -19.309 | <2e-16 *** |
| Household Wealth | 5.786e-02 | 4.608e-02 | 1.256 | 0.209340 |
| Village Size | -8.770e-03 | 4.089e-03 | -2.144 | 0.037625 * |
| Time to Health Center | -2.056e-02 | 1.254e-02 | -1.639 | 0.109048 |
| Education | 4.154e-02 | 2.456e-02 | 1.691 | 0.090860 . |
| Treatment × Gender | 1.075e-01 | 1.297e-01 | 0.828 | 0.407431 |
| Treatment × Age | -3.437e-03 | 4.690e-03 | -0.733 | 0.463635 |
| Treatment × Marital Status | -1.399e-02 | 4.198e-02 | -0.333 | 0.739025 |
| Treatment × Initial Status | 4.279e-03 | 6.733e-03 | 0.635 | 0.525139 |
| Treatment × Household Wealth | 5.936e-03 | 5.572e-02 | 0.107 | 0.915160 |
| Treatment × Village Size | 8.860e-03 | 4.691e-03 | 1.889 | 0.065241 . |
| Treatment × Time to Health Center | 3.329e-02 | 1.297e-02 | 2.567 | 0.013323 * |
| Treatment × Education | -2.081e-02 | 3.018e-02 | -0.689 | 0.490649 |

**Supplementary Table 57** | Summary of mixed effects Regression Analysis for overall effects of individuals degree in high dosage villages: Estimates, Standard Errors, and Significance Levels for various predictors

| Coefficient | Estimate | Std. Error | t value | Pr(> |t|) |
|---|---:|---:|---:|---:|
| (Intercept) | 1.089e+00 | 2.328e-01 | 4.678 | <0.001 *** |
| Treatment | 2.236e-01 | 2.393e-01 | 0.934 | 0.354038 |
| Gender | -4.058e-01 | 1.046e-01 | -3.878 | <0.001 *** |
| Age | 7.070e-03 | 3.824e-03 | 1.849 | 0.06453 . |

| | Estimate | Std. Error | t value | Pr(>\|t\|) |
|---|---|---|---|---|
| Marital Status | 7.414e-02 | 3.384e-02 | 2.191 | 0.02852 * |
| Initial Status | -1.100e-01 | 5.657e-03 | -19.452 | <0.001 *** |
| Household Wealth | 6.203e-02 | 4.546e-02 | 1.364 | 0.17256 |
| Village Size | -8.082e-03 | 3.714e-03 | -2.176 | 0.03488 * |
| Time to Health Center | -1.605e-02 | 1.174e-02 | -1.367 | 0.17901 |
| Education | 4.190e-02 | 2.442e-02 | 1.716 | 0.08624 . |
| Treatment × Gender | 2.504e-02 | 1.338e-01 | 0.187 | 0.85157 |
| Treatment × Age | -3.151e-03 | 4.820e-03 | -0.654 | 0.51335 |
| Treatment × Marital Status | -3.298e-03 | 4.341e-02 | -0.076 | 0.93943 |
| Treatment × Initial Status | 5.806e-05 | 7.245e-03 | 0.008 | 0.99361 |
| Treatment × Household Wealth | 7.339e-03 | 5.693e-02 | 0.129 | 0.89743 |
| Treatment × Village Size | 7.969e-03 | 4.280e-03 | 1.862 | 0.06874 . |
| Treatment × Time to Health Center | 2.268e-02 | 1.195e-02 | 1.897 | 0.06358 . |
| Treatment × Education | -1.967e-02 | 3.108e-02 | -0.633 | 0.52701 |

**Supplementary Table 58** | Summary of mixed effects Regression Analysis for total effects of individuals degree in high dosage villages: Estimates, Standard Errors, and Significance Levels for various predictors

| Coefficient | Estimate | Std. Error | t value | Pr(>\|t\|) |
|---|---|---|---|---|
| (Intercept) | 1.083e+00 | 2.860e-01 | 3.789 | 0.000354 *** |
| Treatment | -1.037e-01 | 2.997e-01 | -0.346 | 0.730535 |
| Gender | -4.234e-01 | 1.190e-01 | -3.558 | 0.000377 *** |
| Age | 7.949e-03 | 4.351e-03 | 1.827 | 0.067787 . |
| Marital Status | 7.682e-02 | 3.845e-02 | 1.998 | 0.045802 * |
| Initial Status | -9.058e-02 | 5.228e-03 | -17.325 | <2e-16 *** |
| Household Wealth | 6.953e-02 | 5.611e-02 | 1.239 | 0.224692 |
| Village Size | -6.187e-03 | 4.731e-03 | -1.308 | 0.195678 |
| Time to Health Center | -8.886e-03 | 1.309e-02 | -0.679 | 0.499541 |
| Education | 4.367e-02 | 2.777e-02 | 1.573 | 0.115841 |
| Treatment * Gender | 1.123e-01 | 1.488e-01 | 0.755 | 0.450314 |
| Treatment * Age | -6.800e-03 | 5.410e-03 | -1.257 | 0.208817 |
| Treatment * Marital Status | -4.298e-02 | 4.777e-02 | -0.900 | 0.368260 |
| Treatment * Initial Status | 9.950e-03 | 6.484e-03 | 1.535 | 0.124972 |
| Treatment * Household Wealth | 7.667e-03 | 6.505e-02 | 0.118 | 0.906172 |
| Treatment * Village Size | 1.367e-02 | 4.894e-03 | 2.794 | 0.006706 ** |
| Treatment * Time to Health Center | 2.075e-02 | 1.191e-02 | 1.743 | 0.084288 . |
| Treatment * Education | -5.939e-02 | 3.475e-02 | -1.709 | 0.087527 . |

**Supplementary Table 59** | Summary of mixed effects Regression Analysis for spillover effects of individuals degree in high dosage villages: Estimates, Standard Errors, and Significance Levels for various predictors.

| Coefficient | Estimate | Std. Error | t values | Pr(>\|t\|) |
|---|---|---|---|---|
| (Intercept) | 7.594e-01 | 2.507e-01 | 3.030 | 0.00359 ** |
| Treatment | 4.868e-01 | 1.911e-01 | 2.547 | 0.01091 * |
| Gender | 6.901e-02 | 1.810e-01 | 0.381 | 0.70299 |
| Age | 4.155e-03 | 6.509e-03 | 0.638 | 0.52332 |
| Marital Status | 2.069e-02 | 5.699e-02 | 0.363 | 0.71656 |
| Initial Status | -9.326e-02 | 1.035e-02 | -9.007 | <2e-16 *** |
| Household Wealth | 7.440e-02 | 6.992e-02 | 1.064 | 0.28740 |
| Village Size | -2.853e-03 | 2.930e-03 | -0.974 | 0.33600 |
| Time to Health Center | 3.015e-02 | 1.102e-02 | 2.736 | 0.00901 ** |
| Education | 2.485e-02 | 4.123e-02 | 0.603 | 0.54678 |
| Treatment × Gender | -4.175e-01 | 1.994e-01 | -2.094 | 0.03635 * |
| Treatment × Age | -1.338e-03 | 7.120e-03 | -0.188 | 0.85096 |
| Treatment × Marital Status | 3.858e-02 | 6.313e-02 | 0.611 | 0.54119 |
| Treatment × Initial Status | -2.008e-02 | 1.137e-02 | -1.765 | 0.07761 |
| Treatment × Household Wealth | -1.125e-02 | 7.608e-02 | -0.148 | 0.88249 |
| Treatment × Village Size | 2.277e-03 | 2.017e-03 | 1.129 | 0.25905 |
| Treatment × Time to Health Center | -1.660e-02 | 7.348e-03 | -2.260 | 0.02394 * |
| Treatment × Education | -3.330e-03 | 4.528e-02 | -0.074 | 0.94139 |

**Supplementary Table 60** | Summary of mixed effects Regression Analysis for direct effects of individuals degree in high dosage villages: Estimates, Standard Errors, and Significance Levels for Various Predictors

## 15. Network Spillover Analysis

In wave 1, treated villages contain 661 first-degree neighbors to the treated in villages where a low dosage of the treatment was applied, and 314 first-degree neighbors to the treated where a high dosage was applied. Additionally, there are 2,280 higher-degree neighbors exclusively connected to untreated individuals in low-dosage villages, compared to 268 in high-dosage villages.

Using permutation tests comparing these groups to control individuals, we assessed the extent of spillover effects on different individuals defined by their connection to other treated or untreated individuals. In particular, we conduct two comparative analyses:

(a) First-order spillover effect on the untreated: comparison of untreated individuals who, in wave 1, are directly connected (1st-order neighbors) to treated individuals in treated villages with untreated individuals in untreated villages.

b) Higher-order spillover effect on the untreated: comparison of untreated individuals who, in wave 1, are indirectly connected (higher-order neighbors) to treated individuals in treated villages but not directly connected to any treated individual with untreated individuals in untreated villages.

The results of the analysis reveal substantial disparities in the outcomes of untreated individuals who are one-hop-away neighbors to those treated, in comparison to untreated individuals without connections to treated individuals. This distinction becomes more pronounced when juxtaposed with untreated individuals residing in untreated villages, thereby accentuating the crucial role of network relationships in mediating the treatment's spillover effects. These findings highlight the necessity of incorporating network structures into the analysis to fully comprehend the implications of treatments in interconnected environments.

**Low dosages**

|  | Higher order | First order |
|---|---|---|
| Degree | 9.11% (0.002) | -54.09% (< 0.001) |
| In Degree | 10.17% (0.006) | -57.32% (< 0.001) |
| Out Degree | 8.05% (0.026) | -50.85% (< 0.001) |
| Betweenness | 36.20% (< 0.001) | 6.29% (0.611) |
| Closeness | -8.05% (0.009) | -16.46% (< 0.001) |
| Clustering | 18.60% (0.045) | -16.38% (0.182) |

**Supplementary Table 61** | Higher-order and first-order spillover effect on the untreated for the health network. Each cell presents the percentage change and p-values in untreated individuals in treated low-dosage villages who are high- or first-order neighbors of treated individuals relative to control individuals in untreated villages for the health network, including intra-household connections, calculated using a difference- in- differences estimator on means.

|  | Higher order | First order |
|---|---|---|
| Degree | 8.14% (< 0.001) | -17.83% (< 0.001) |
| In Degree | 9.13% (0.002) | -18.77% (< 0.001) |
| Out Degree | 7.15% (0.009) | -16.89% (< 0.001) |
| Betweenness | 9.01% (0.059) | -6.21% (0.283) |
| Closeness | 1.44% (0.376) | -6.82% (< 0.001) |
| Clustering | 7.08% (0.173) | 8.06% (0.117) |

**Supplementary Table 62** | Higher-order and first-order spillover effect on the untreated for the friendship network. Each cell presents the percentage change and p-values in untreated individuals in treated low-dosage villages who are high- or first-order neighbors of treated individuals relative to control individuals in untreated villages for the friendship network, including intra-household connections, calculated using a difference- in- differences test statistics.

|  | Higher order | First order |
|---|---|---|
| Degree | 6.79% (0.018) | -39.34% (< 0.001) |
| In Degree | 6.74% (0.060) | -40.03% (< 0.001) |
| Out Degree | 6.84% (0.044) | -38.65% (< 0.001) |
| Betweenness | -5.75% (0.385) | -39.88% (< 0.001) |
| Closeness | 1.70% (0.665) | -13.68% (0.006) |
| Clustering | 21.20% (0.021) | -31.18% (0.014) |

**Supplementary Table 63** | Higher-order and first-order spillover effect on the untreated for the financial network. Each cell presents the percentage change and p-values in untreated individuals in treated low-dosage villages who are high- or first-order neighbors of treated individuals relative to control individuals in untreated villages for the financial network, including intra-household connections, calculated using a difference- in- differences test statistics.

|  | Higher order | First order |
|---|---|---|
| Degree | 7.50% (< 0.001) | -20.40% (< 0.001) |
| In Degree | 7.50% (< 0.001) | -20.40% (< 0.001) |
| Out Degree | 7.50% (< 0.001) | -20.40% (< 0.001) |
| Betweenness | 9.04% (0.061) | 1.02% (0.851) |
| Closeness | 0.62% (0.557) | -7.87% (< 0.001) |
| Clustering | 0.47% (0.896) | 3.97% (0.208) |

**Supplementary Table 64** | Higher-order and first-order spillover effect on the untreated for the social network. Each cell presents the percentage change and p-values in untreated individuals in treated low-dosage villages who are high- or first-order neighbors of treated individuals relative to control individuals in untreated villages for the aggregated network, including intra-household connections, calculated using a difference-in-differences test statistics.

## High dosages

|  | Higher order | First order |
|---|---|---|
| Degree | 30.60% (< 0.001) | −19.05% (0.002) |
| In Degree | 23.80% (0.004) | −20.06% (0.018) |
| Out Degree | 37.40% (< 0.001) | −18.04% (0.011) |
| Betweenness | 49.34% (0.004) | 9.01% (0.605) |
| Closeness | 20.44% (0.001) | −14.90% (0.012) |
| Clustering | 48.03% (0.014) | 30.09% (0.087) |

**Supplementary Table 65** | Higher-order and first-order spillover effect on the untreated for the health network. Each cell presents the percentage change and p-values in untreated individuals in treated high-dosage villages who are high- or first-order neighbors of treated individuals relative to control individuals in untreated villages for the health network, including intra-household connections, calculated using permutation tests with the difference-in-differences test statistics.

|  | Higher order | First order |
|---|---|---|
| Degree | 22.48% (< 0.001) | −9.09% (0.005) |
| In Degree | 24.21% (< 0.001) | −7.94% (0.0695) |
| Out Degree | 20.75% (0.003) | −10.24% (0.017) |
| Betweenness | 28.48% (0.024) | −11.42% (0.178) |
| Closeness | 17.54% (< 0.001) | −7.05% (0.0045) |
| Clustering | 7.54% (0.573) | 2.68% (0.727) |

**Supplementary Table 66** | Higher-order and first-order spillover effect on the untreated for the friendship network. Each cell presents the percentage change and p-values in untreated individuals in treated high-dosage villages who are high- or first-order neighbors of treated individuals relative to control individuals in untreated villages for the friendship network, including intra-household connections, calculated using permutation tests with the difference-in-differences test statistics.

|  | Higher order | First order |
|---|---|---|

|               | Higher order         | First order          |
|---------------|----------------------|----------------------|
| Degree        | 23.92% (< 0.001)     | −20.93% (< 0.001)    |
| In Degree     | 16.54% (0.0215)      | −11.64% (0.1185)     |
| Out Degree    | 31.31% (< 0.001)     | −30.21% (< 0.001)    |
| Betweenness   | 15.38% (0.2960)      | −51.80% (< 0.001)    |
| Closeness     | −0.34% (0.9635)      | −12.13% (0.0975)     |
| Clustering    | 44.18% (0.0170)      | −12.51% (0.4755)     |

**Supplementary Table 67** | Higher-order and first-order spillover effect on the untreated for the financial network. Each cell presents the percentage change and p-values in untreated individuals in treated high-dosage villages who are high- or first-order neighbors of treated individuals relative to control individuals in untreated villages a for the financial network, including intra-household connections, calculated using permutation tests with the difference-in-differences test statistics.

|               | Higher order         | First order          |
|---------------|----------------------|----------------------|
| Degree        | 23.67% (< 0.001)     | −6.38% (0.0345)      |
| In Degree     | 23.67% (< 0.001)     | −6.38% (0.0320)      |
| Out Degree    | 23.67% (< 0.001)     | −6.38% (0.0245)      |
| Betweenness   | 37.42% (0.0155)      | −23.88% (0.0050)     |
| Closeness     | 21.57% (< 0.001)     | −0.40% (0.7615)      |
| Clustering    | 5.09% (0.6115)       | 6.93% (0.1365)       |

**Supplementary Table 68** | Higher-order and first-order spillover effect on the untreated for the social network. Each cell presents the percentage change and p-values in untreated individuals in treated high-dosage villages who are high- or first-order neighbors of treated individuals relative to control individuals in untreated villages for the social network, including intra-household connections, calculated using permutation tests with the difference-in-differences test statistics

## 16. Robustness Checks

We conducted several kinds of robustness checks.

First, in the foregoing analyses, we conducted a robustness check for the network analysis, by adding individuals who moved in another household in wave 3, resulting in a dataset of 9944 individuals, which is 1613 more individuals than the original analysis. With the updated sample, we reexamine the network centralities (degree, in-degree, and out-degree) for the aggregated, health, friendship and financial networks, heterogeneous effects, dyad regressions and spatial effects. By conducting this robustness check, we confidently verified that the observed effects in the aggregated, health, and friendship networks remain consistent and reliable, even after accounting for changes in the dataset's composition.

|             | Overall Individuals | Treated Individuals | Untreated Individuals | Control Individuals |
|-------------|---------------------|---------------------|-----------------------|---------------------|
| All Dosages | 9944                | 3800                | 4181                  | 1963                |
| Low Dosages | 6128                | 674                 | 3491                  | 1963                |
| High Dosages | 5779 | 3126 | 690 | 1963 |

**Supplementary Table 69** | Number of individuals in control and treated villages for all, low and high dosages.

| | Overall Individuals | Overall Effect | Total Effect | Spillover Effect |
|---|---|---|---|---|
| All Dosages | 9944 | x | + | - |
| Low Dosages | 5471 | - | - | - |
| High Dosages | 5779 | - | - | - |

**Supplementary Table 70** | With + when the degree of individuals in treated villages is reduced less than the degree of individuals in control villages. With - when the degree of individuals in treated villages is reduced more than individuals in control villages.

## Low dosages

| | Overall | Total | Spillover | Direct |
|---|---|---|---|---|
| Degree | -5.27% (0.075) | -10.11% (0.025) | -4.34% (0.144) | -5.64% (0.193) |
| In Degree | -5.27% (0.196) | -6.72% (0.279) | -4.99% (0.230) | -1.70% (0.799) |
| Out Degree | -5.27% (0.089) | -13.50% (0.009) | -3.68% (0.247) | -9.56% (0.040) |
| Betweenness | 17.04% (0.001) | 12.15% (0.239) | 17.99% (0.003) | -6.03% (0.494) |
| Closeness | -9.16% (<0.001) | -14.29% (<0.001) | -8.17% (0.002) | -6.31% (0.121) |
| Clustering | 4.12% (0.561) | -14.62% (0.200) | 7.74% (0.294) | -21.08% (0.042) |

**Supplementary Table 71** | Each cell presents the percentage change and p-values in treated groups (all individuals in fully and partially treated low-dosage villages for the overall effect, treated individuals in fully and partially treated low-dosage villages for the total and direct effects, untreated individuals in in fully and partially treated low-dosage villages for the spillover effects) relative to control groups (all individuals in untreated villages for the overall, total and spillover effects, and untreated individuals in fully and partially low-dosage treated villages for the direct effects) at wave 3 within the health low-dosage network, including intra-household connections, calculated using permutation tests with the difference-in-differences test statistics.

| | Overall | Total | Spillover | Direct |
|---|---|---|---|---|
| Degree | -0.76% (0.656) | -1.58% (0.564) | -0.60% (0.741) | -0.97% (0.710) |
| In Degree | -0.76% (0.740) | -1.78% (0.622) | -0.56% (0.814) | -1.21% (0.742) |
| Out Degree | -0.76% (0.719) | -1.38% (0.724) | -0.64% (0.766) | -0.73% (0.840) |
| Betweenness | 6.05% (0.121) | 2.67% (0.673) | 6.70% (0.111) | -4.10% (0.484) |
| Closeness | -2.77% (0.013) | -2.00% (0.265) | -2.91% (0.009) | 0.94% (0.619) |
| Clustering | 10.74% (0.006) | 1.99% (0.767) | 12.43% (0.002) | -9.62% (0.083) |

**Supplementary Table 72** | Each cell presents the percentage change and p-values in treated groups (all individuals in fully and partially treated low-dosage villages for the overall effect, treated individuals in fully and partially treated low-dosage villages for the total and direct effects, untreated individuals in in fully and partially treated low-dosage villages for the spillover effects) relative to control groups (all individuals in untreated villages for the overall, total and spillover effects, and untreated individuals in fully and partially low-dosage treated villages for the direct effects) at wave 3 within the friendship low-dosage network, including intra-household connections, calculated using permutation tests with the difference-in-

differences test statistics.

|  | **Overall** | **Total** | **Spillover** | **Direct** |
| --- | --- | --- | --- | --- |
| Degree | -3.47% (0.187) | -8.32% (0.044) | -2.54% (0.351) | -5.74% (0.165) |
| In Degree | -3.47% (0.297) | -7.53% (0.139) | -2.69% (0.447) | -4.83% (0.348) |
| Out Degree | -3.47% (0.241) | -9.11% (0.046) | -2.38% (0.431) | -6.64% (0.164) |
| Betweenness | -6.31% (0.261) | 5.22% (0.577) | -8.53% (0.136) | 17.28% (0.090) |
| Closeness | -3.48% (0.243) | -13.07% (0.009) | -1.63% (0.616) | -11.29% (0.015) |
| Clustering | 18.57% (0.012) | 17.39% (0.171) | 18.80% (0.017) | -1.22% (0.903) |

**Supplementary Table 73** | Each cell presents the percentage change and p-values in treated groups (all individuals in fully and partially treated low-dosage villages for the overall effect, treated individuals in fully and partially treated low-dosage villages for the total and direct effects, untreated individuals in in fully and partially treated low-dosage villages for the spillover effects) relative to control groups (all individuals in untreated villages for the overall, total and spillover effects, and untreated individuals in fully and partially low-dosage treated villages for the direct effects) at wave 3 within the financial low-dosage network, including intra-household connections, calculated using permutation tests with the difference-in-differences test statistics.

|  | **Overall** | **Total** | **Spillover** | **Direct** |
| --- | --- | --- | --- | --- |
| Degree | -4.23% (0.011) | -6.19% (0.021) | -3.85% (0.025) | -2.42% (0.369) |
| Betweenness | 5.33% (0.169) | 4.91% (0.442) | 5.41% (0.181) | -0.50% (0.936) |
| Closeness | -3.50% (<0.001) | -5.13% (<0.001) | -3.18% (<0.001) | -2.04% (0.083) |
| Clustering | 1.54% (0.541) | -2.69% (0.476) | 2.36% (0.348) | -5.08% (0.175) |

**Supplementary Table 74** | Each cell presents the percentage change and p-values in treated groups (all individuals in fully and partially treated low-dosage villages for the overall effect, treated individuals in fully and partially treated low-dosage villages for the total and direct effects, untreated individuals in in fully and partially treated low-dosage villages for the spillover effects) relative to control groups (all individuals in untreated villages for the overall, total and spillover effects, and untreated individuals in fully and partially low-dosage treated villages for the direct effects) at wave 3 within the social low-dosage network, including intra-household connections, calculated using permutation tests with the difference-in-differences test statistics.

**High dosages**

|  | **Overall** | **Total** | **Spillover** | **Direct** |
| --- | --- | --- | --- | --- |
| Degree | 8.14% (<0.001) | 8.60% (0.003) | 6.05% (0.167) | 2.69% (0.522) |
| In Degree | 8.14% (0.025) | 9.39% (0.011) | 2.47% (0.679) | 7.44% (0.197) |
| Out Degree | 8.14% (0.005) | 7.81% (0.009) | 9.62% (0.049) | -1.87% (0.676) |
| Betweenness | 15.67% (0.015) | 15.35% (0.025) | 17.12% (0.092) | -1.77% (0.872) |
| Closeness | -3.93% (0.118) | -3.86% (0.118) | -4.26% (0.301) | 0.41% (0.916) |
| Clustering | 24.19% (0.001) | 22.55% (0.002) | 31.63% (0.005) | -9.05% (0.389) |

**Supplementary Table 75** | Each cell presents the percentage change and p-values in treated groups (all individuals in fully and partially treated high-dosage villages for the overall effect, treated individuals in fully and partially treated high-dosage villages for the total and direct effects, untreated individuals in in fully and partially treated high-dosage villages for the spillover effects) relative to control groups (all individuals in untreated villages for the overall, total and spillover effects, and untreated individuals in fully and partially treated high-dosage villages for the direct effects) at wave 3 within the health high-dosage net work, including intra-household connections, calculated using permutation tests with the difference-in-differences test statistics

|  | Overall | Total | Spillover | Direct |
|---|---|---|---|---|
| Degree | 4.97% (0.004) | 5.33% (0.002) | 3.31% (0.238) | 2.21% (0.426) |
| In Degree | 4.97% (0.024) | 5.22% (0.032) | 3.80% (0.310) | 1.55% (0.657) |
| Out Degree | 4.97% (0.022) | 5.44% (0.020) | 2.83% (0.436) | 2.87% (0.424) |
| Betweenness | -7.40% (0.092) | -7.51% (0.097) | -6.89% (0.304) | -0.58% (0.937) |
| Closeness | 4.02% (<0.001) | 3.96% (0.001) | 4.28% (0.024) | -0.33% (0.858) |
| Clustering | 12.32% (0.002) | 12.05% (0.007) | 13.57% (0.036) | -1.50% (0.800) |

**Supplementary Table 76** | Each cell presents the percentage change and p-values in treated groups (all individuals in fully and partially treated high-dosage villages for the overall effect, treated individuals in fully and partially treated high-dosage villages for the total and direct effects, untreated individuals in in fully and partially treated high-dosage villages for the spillover effects) relative to control groups (all individuals in untreated villages for the overall, total and spillover effects, and untreated individuals in fully and partially treated high-dosage villages for the direct effects) at wave 3 within the friendship low-dosage network, including intra-household connections, calculated using permutation tests with the difference-in-differences test statistics.

|  | Overall | Total | Spillover | Direct |
|---|---|---|---|---|
| Degree | 6.56% (0.008) | 7.37% (0.005) | 2.91% (0.488) | 4.98% (0.240) |
| In Degree | 6.56% (0.036) | 7.36% (0.033) | 2.96% (0.542) | 4.93% (0.356) |
| Out Degree | 6.56% (0.021) | 7.38% (0.011) | 2.87% (0.544) | 5.02% (0.312) |
| Betweenness | 1.78% (0.770) | 2.81% (0.640) | -2.89% (0.771) | 7.18% (0.514) |
| Closeness | -6.05% (0.051) | -5.29% (0.107) | -9.49% (0.065) | 4.10% (0.404) |
| Clustering | 18.04% (0.015) | 16.56% (0.045) | 24.72% (0.047) | -7.97% (0.507) |

**Supplementary Table 77** | Each cell presents the percentage change and p-values in treated groups (all individuals in fully and partially treated high-dosage villages for the overall effect, treated individuals in fully and partially treated high-dosage villages for the total and direct effects, untreated individuals in in fully and partially treated high-dosage villages for the spillover effects) relative to control groups (all individuals in untreated villages for the overall, total and spillover effects, and untreated individuals in fully and partially treated high-dosage villages for the direct effects) at wave 3 within the financial low-dosage network, including intra-household connections, calculated using permutation tests with the difference-in-differences test statistics.

|  | Overall | Total | Spillover | Direct |
|---|---|---|---|---|
| Degree | 2.37% (0.122) | 2.64% (0.095) | 1.15% (0.664) | 1.70% (0.509) |
| Betweenness | -9.52% (0.027) | -8.15% (0.046) | -15.75% (0.018) | 7.70% (0.247) |
| Closeness | 3.87% (<0.001) | 3.83% (<0.001) | 4.09% (<0.001) | -0.27% (0.842) |
| Clustering | 3.52% (0.144) | 2.39% (0.343) | 8.66% (0.029) | -6.48% (0.094) |

**Supplementary Table 78** | Each cell presents the percentage change and p-values in treated groups (all individuals in fully and partially treated high-dosage villages for the overall effect, treated individuals in fully and partially treated high-dosage villages for the total and direct effects, untreated individuals in in fully and partially treated high-dosage villages for the spillover effects) relative to control groups (all individuals in untreated villages for the overall, total and spillover effects, and untreated individuals in fully and partially treated high-dosage villages for the direct effects) at wave 3 within the social high-dosage network, including intra-household connections, calculated using permutation tests with the difference-in-differences test statistics.

As a second robustness check, we also separately examine the residual friendship and financial networks in wave 1 and 3, excluding health ties, for both wave 1 and wave 3.

*Supplementary Figure 8* and *Supplementary Figure 9* show that these networks also change over time. Interestingly, we do not observe any changes at low dosages for the residual friendship or the financial networks, suggesting that the reduction in links seen in both the friendship and financial networks was primarily driven by health care changes. However, at high dosages, financial residual networks exhibit changes, reinforcing the robustness our findings. The collaboration initiated by the health care intervention in the villages led to new financial ties of treated individuals.

In the residual financial network, for high dosages, the overall degree, in-degree, and out-degree increase by 8.71% (p-values = 0.009, 0.054, and 0.036, respectively). Treated individuals show increases of 9.94% in degree, 9.12% in in-degree, and 10.76% in out-degree (p-values = 0.006, 0.061, and 0.009, respectively). Unexpectedly, the health treatment led to a significant increase in financial relationships within high-dosage villages (in the "residual" financial network).

We report the residual friendship and financial networks in Wave 1 and Wave 3 (*Supplementary Table 79* to *Supplementary Figure 82*, excluding health ties, to assess whether changes in social and financial relationships persist independently of health-related interactions. *Supplementary Figure 8* and *Supplementary Figure 9* illustrate that these networks evolve over time, with notable differences between low- and high-dosage conditions. At low dosages, we do not observe significant changes in the residual friendship or financial networks, suggesting that the reductions in links in the main networks were primarily driven by health care interactions. However, at high dosages, financial networks exhibit significant structural shifts, reinforcing the robustness of our findings. These results indicate that the health care intervention facilitated new financial connections, improving financial exchange among treated individuals.

**Low dosages**

**Friendship**

|  | Overall | Total | Spillover | Direct |
|---|---|---|---|---|
| Degree | -2.02% (0.403) | -5.07% (0.158) | -1.42% (0.551) | -3.73% (0.304) |
| In Degree | -2.02% (0.535) | -7.98% (0.117) | -0.86% (0.774) | -7.28% (0.141) |
| Out Degree | -2.02% (0.514) | -2.16% (0.642) | -1.99% (0.541) | -0.17% (0.976) |
| Betweenness | 5.76% (0.246) | -0.57% (0.947) | 7.00% (0.193) | -7.86% (0.340) |
| Closeness | -3.21% (0.152) | -0.62% (0.854) | -3.71% (0.108) | 3.14% (0.373) |
| Clustering | 12.46% (0.043) | 19.89% (0.055) | 11.02% (0.089) | 8.77% (0.323) |

**Supplementary Table 79** | Each cell presents the percentage change and p-values in treated groups all individuals in fully and partially treated low-dosage villages for the overall effect, treated individuals in fully and partially low-dosage treated villages for the total and direct effects, untreated individuals in in fully and partially treated low-dosage villages for the spillover effects) relative to control groups (all individuals in untreated villages for the overall, total and spillover effects, and untreated individuals in fully and partially treated low-dosage villages for the direct effects) at wave 3 within the residual friendship low-dosage network, including intra-household connections, calculated using permutation tests with the difference-in-differences test statistics.

**Financial**

|  | Overall | Total | Spillover | Direct |
|---|---|---|---|---|
| Degree | 2.16% (0.540) | -2.25% (0.682) | 3.01% (0.407) | -5.39% (0.328) |
| In Degree | 2.16% (0.650) | -1.93% (0.803) | 2.95% (0.538) | -5.02% (0.520) |
| Out Degree | 2.16% (0.631) | -2.57% (0.700) | 3.08% (0.482) | -5.75% (0.385) |
| Betweenness | -23.00% (0.012) | -33.04% (0.033) | -21.05% (0.033) | -15.81% (0.414) |
| Closeness | 8.76% (0.033) | -2.97% (0.683) | 11.04% (0.011) | -13.24% (0.034) |
| Clustering | 17.63% (0.161) | 1.09% (0.959) | 20.85% (0.103) | -19.47% (0.300) |

**Supplementary Table 80** | Each cell presents the percentage change and p-values in treated groups (all individuals in fully and partially treated low-dosage villages for the overall effect, treated individuals in fully and partially low-dosage treated villages for the total and direct effects, untreated individuals in in fully and partially treated low-dosage villages for the spillover effects) relative to control groups (all individuals in untreated villages for the overall, total and spillover effects, and untreated individuals in fully and partially treated low-dosage villages for the direct effects) at wave 3 within the residual financial low-dosage network, including intra-household connections, calculated using permutation tests with the difference-in-differences test statistics.

**High dosages**

**Friendship**

|  | Overall | Total | Spillover | Direct |
|---|---|---|---|---|
| Degree | 0.13% (0.964) | 0.32% (0.903) | -0.72% (0.846) | 1.23% (0.729) |
| In Degree | 0.13% (0.967) | -0.24% (0.938) | 1.75% (0.719) | -2.30% (0.676) |
| Out Degree | 0.13% (0.974) | 0.88% (0.793) | -3.20% (0.507) | 4.87% (0.398) |
| Betweenness | 19.25% (<0.001) | 17.60% (<0.001) | 26.56% (0.004) | -7.17% (0.313) |
| Closeness | -10.89% (<0.001) | -9.12% (<0.001) | -18.72% (<0.001) | 11.47% (0.013) |
| Clustering | 13.50% (0.035) | 12.51% (0.054) | 17.90% (0.088) | -5.90% (0.573) |

**Supplementary Table 81** | Each cell presents the percentage change and p-values in treated groups ((all individuals in fully and partially treated high-dosage villages for the overall effect, treated individuals in fully and partially treated high-dosage villages for the total and direct effects, untreated individuals in in fully and partially treated high-dosage villages for the spillover effects) relative to control groups (all individuals in untreated villages for the overall, total and spillover effects, and untreated individuals in fully and partially treated high-dosage villages for the direct effects) at wave 3 within the residual friendship high-dosage network, including intra-household connections, calculated using permutation tests with the difference-in-differences test statistics.

**Financial**

|  | Overall | Total | Spillover | Direct |
|---|---|---|---|---|
| Degree | 8.71% (0.009) | 9.94% (0.006) | 3.28% (0.556) | 8.32% (0.193) |
| In Degree | 8.71% (0.054) | 9.12% (0.061) | 6.93% (0.333) | 2.69% (0.741) |
| Out Degree | 8.71% (0.036) | 10.76% (0.009) | -0.36% (0.952) | 14.17% (0.073) |
| Betweenness | 21.47% (0.019) | 29.56% (<0.001) | -14.33% (0.328) | 74.31% (<0.001) |
| Closeness | -3.75% (0.197) | -4.13% (0.248) | -2.06% (0.271) | -2.06% (0.764) |
| Clustering | 14.63% (0.249) | 15.67% (0.220) | 10.00% (0.643) | 7.30% (0.752) |

**Supplementary Table 82** | Each cell presents the percentage change and p-values in treated groups (all individuals in fully and partially treated high-dosage villages for the overall effect, treated individuals in fully and partially treated high-dosage villages for the total and direct effects, untreated individuals in in fully and partially treated high-dosage villages for the spillover effects) relative to control groups (all individuals in untreated villages for the overall, total and spillover effects, and untreated individuals in fully and partially treated high-dosage villages for the direct effects) at wave 3 within the residual financial high-dosage network, excluding intra-household connections, calculated using permutation tests with the difference-in-differences test statistics.

## 17. R commands

The R commands employed in this study used the igraph package to compute key network measures. Normalized betweenness centrality was calculated using betweenness(,normalized = TRUE) on an directed graph with normalization enabled. Normalized closeness centrality was derived from the undirected version using closeness(,normalized = TRUE). Undirected network was used for closeness to avoid NAs because the health, friendship and financial networks are sparse. The local clustering coefficient for each vertex was computed using transitivity() with the 'local' option, where vertex identifiers were provided via an a matrix conversion of the vertex names and isolates were handled as specified. These commands enable accurate assessment of centrality and clustering in the networks analyzed.

## 18. Link Evolution

**Causal Effects for the Link evolution analysis**

Let $Y_{ijv}(t_{iv}, t_{jv}, \mathbf{t}_{-ijv})$ be the potential link from individual i to j in village v under individual treatment status $T_{iv} = t_{iv}$ for individual i and $T_{jv} = t_{jv}$ for individual j and treatment status in the village v excluding i and j equal to $\mathbf{t}_{-ijv}$. The dyadic average potential outcome and its population average under dyadic treatment status $T_{iv} = t_{iv}$ and $T_{jv} = t_{jv}$ and dosage α are then given by $Y_{ijv}(t_{iv}, t_{jv}, \alpha) = \sum_{\mathbf{t}_v} Y_{ijv}(t_{iv}, t_{jv}, \mathbf{t}_{-ijv}) \cdot \Pr_\alpha (\mathbf{T}_{-ijv} = \mathbf{t}_{-ijv} \mid T_{iv} = t_{iv}, T_{jv} = t_{jv})$, and $Y(t_{iv}, t_{jv}, \alpha) = \frac{1}{V} \sum_v \frac{1}{N_v(N_v-1)} \sum_i \overline{Y}_{ijv}(t_{iv}, t_{jv}, \alpha)$, respectively.

Then, the coefficients of dyadic logistic regression correspond to the following comparisons:

- **β_UU** represents the comparison in the probability of link between untreated-untreated dyads in partially treated villages (UU) and untreated-untreated dyads in control villages (UoUo), that is, the comparison in the logit scale between $Y(0, 0, \alpha)$ for $\alpha > 0$ and $Y(0, 0, 0)$;

- **β_UT** represents the comparison in the probability of link from an untreated individual to a treated one in partially treated villages (UU) and untreated-untreated dyads in control villages (UoUo), that is, the comparison in the logit scale between $Y(0, 1, \alpha)$ for $\alpha > 0$ and $Y(0, 0, 0)$;

- **β_TU** represents the comparison in the probability of link from a treated individual to an untreated one in partially treated villages (UU) and untreated-untreated dyads in control villages (UoUo), that is, the comparison in the logit scale between $Y(1, 0, \alpha)$ for $\alpha > 0$ and $Y(0, 0, 0)$;

- **β_TT** represents the comparison in the probability of link between treated-treated dyads in partially or fully treated villages (TT) and untreated-untreated dyads in control villages (UoUo), that is, the comparison in the logit scale between $Y(1, 1, \alpha)$ for $\alpha > 0$ and $Y(0, 0, 0)$.

Similar causal effects could have been derived using permutation tests to assess causal effects on node-level test statistics.

Let $\overline{Y}_{iv}^{outT}(t_{iv}, \alpha) = \frac{1}{|N_{iv}^{out}|} \sum_{j \in N_{iv}^{out}} \overline{Y}_{ijv}(t_{iv}, 1, \alpha)$ be the outdegree to treated individuals of individual i in village v under individual treatment status $t_{iv}$ and dosage $\alpha$, with $N_{iv}^{out}$ being its out-neighborhood. Similarly let $\overline{Y}_{iv}^{outU}(t_{iv}, \alpha)$ be the outdegree to untreated individuals of individual i in village v. For indegree we let $\overline{Y}_{jv}^{inT}(t_{iv}, \alpha) = \frac{1}{|N_{jv}^{in}|} \sum_{i \in N_{jv}^{in}} \overline{Y}_{ijv}(1, t_{jv}, \alpha)$ be the indegree from treated individuals of individual j in village v under individual treatment status $t_{jv}$ and dosage $\alpha$, with $N_{jv}^{in}$ being its in-neighborhood. Similarly, let $\overline{Y}_{iv}^{inU}(t_{iv}, \alpha)$ be the indegree from untreated individuals of individual j in village v. Population averages $\overline{Y}^{outT}(t_{iv}, \alpha)$, $\overline{Y}^{outU}(t_{iv}, \alpha)$, $\overline{Y}^{inT}(t_{iv}, \alpha)$, and $\overline{Y}^{inU}(t_{iv}, \alpha)$, are then defined by taking the average across the sample.

There is then a one-to-one correspondence between dyadid causal effects estimated with the dyadic logistic regression and causal effects on node-level properties that can be estimated with permutation tests.

- $\beta_{UU}$ corresponds to the spillover effect on the indegree from the untreated, i.e., $\bar{Y}^{inU}(0, \alpha)$ vs $\bar{Y}^{inU}(0,0)$, or the spillover effect on the outdegree to the untreated, i.e., $\bar{Y}^{outU}(0, \alpha)$ vs $\bar{Y}^{outU}(0,0)$;

- $\beta_{UT}$ corresponds to the total effect on the indegree from the untreated, i.e., $\bar{Y}^{inU}(1, \alpha)$ vs $\bar{Y}^{inU}(0,0)$;

- $\beta_{TU}$ corresponds to the total effect on the outdegree to the untreated, i.e., $\bar{Y}^{outU}(1, \alpha)$ vs $\bar{Y}^{outU}(0,0)$;

- $\beta_{TT}$ corresponds to the comparison between the indegree from the treated for the treated under dosage α, i.e., $\bar{Y}^{inT}(1, \alpha)$, and the indegree from the untreated for the untreated in control villages, i.e., $\bar{Y}^{inU}(0,0)$.

For the latter, the test statistics for the permutation tests use a different outcome for the treated group and the control group.

Here we have reported results from the dyadic logistic regressions as they may be more intuitive for the reader. We have also conducted corresponding permutation tests for the node-level analysis. The two approaches show similar results. However, dyadic potential outcomes are weighted differently for node-level and dyadic level estimands, and the logistic regression makes an additional assumption of independence across dyads.

Due to space limitations, we have not included the full set of detailed tables. For completeness, readers interested in the complete results (not data) can request them or access them via [GitHub/MariosPapamix].

## 19. Explanation of Higher Order Untreated Respondent Increase in Degree Centralities

In *Supplementary Tables 61* and *65* for the health network (as well as in *Tables 62-64* and *66-68* for the other networks), we observe that untreated nodes at distance ≥ 2 from treated nodes ("higher-order untreated neighbors") exhibit increases in degree centrality compared to control villages. This pattern arises because, as observed in *Supplementary Section 15*, untreated–untreated (UU) ties persist across waves, and by definition of untreated nodes at distance ≥ 2 from treated nodes are only directly connected to untreated individuals. This explains the observed lower reduction in degree centrality for these higher-order untreated nodes compared to control villages, while allowing them to form new direct connections with treated nodes to which they were not linked in wave 1.

To evaluate whether this result is entirely explained by the stability of the untreated–untreated (UU) ties or it can also be interpretated as a spillover effect of being indirectly connected to treated nodes, we performed a robustness check by defining the higher-order spillover effect on the untreated as the contrast between higher-order untreated neighbors in treated villages—those whose shortest path to any treated node is $\geq 2$ (regardless of their first-order connections to other nodes), and untreated individuals in fully untreated (control) villages.

At baseline (wave 1), we identified 2312 higher-order untreated neighbors in low-dosage villages versus 459 in high-dosage villages. We then used permutation tests to compare changes in degree centrality across waves between these higher-order untreated neighbors and control individuals. Contrary to the gross increase reported in *Supplementary Section 16*, the permutation analysis reveals a statistically significant decrease in degree centrality among higher-order untreated neighbors relative to controls (see *Supplementary Table 65*).

This finding confirms that the apparent increase in Section 16 is not a direct causal effect of the health intervention but rather an indirect artifact of the network's evolving topology.

|  | **Low Dosages** | **High Dosages** |
|---|---|---|
| Degree | -27.50% (<0.001) | -18.97% (<0.001) |
| In Degree | -25.14% (<0.001) | -20.33% (0.0015) |
| Out Degree | -29.86% (<0.001) | -17.62% (<0.001) |
| Betweenness | -20.08% (0.002) | -20.40% (0.010) |
| Closeness | 1.49% (0.629) | 8.16% (0.173) |
| Clustering | -26.67% (0.005) | 10.59% (0.504) |

**Supplementary Table 83** | Higher-order spillover effects on untreated nodes indirectly connected in wave 1 to treated nodes, regardless of their first-order connections, Each cell presents the percentage change and p-values in untreated individuals in treated low- or high-dosage villages who are high-order neighbors of treated individuals, including those whoo are also directly connected to the treated, relative to control individuals in untreated villages, for the health network, excluding intra-household connections, calculated using permutation tests with the difference-in-differences test statistics.